\documentclass{jfm}

\usepackage{graphicx}
\usepackage{newtxtext}
\usepackage{newtxmath}
\usepackage{natbib}

\usepackage{booktabs}
\usepackage{multirow}
\usepackage{color}
\usepackage{pifont}

\usepackage{hyperref}
\hypersetup{
    colorlinks = true,
    linkcolor  = blue,
    urlcolor   = blue,
    citecolor  = blue,
    hypertexnames = false, 
    plainpages = false
}

\renewcommand{\theequation}{\thesection.\arabic{equation}}
\makeatletter
\@addtoreset{equation}{section}
\makeatother

\newcommand{\RomanNumeralCaps}[1]

\usepackage{tikz}
\usetikzlibrary{shapes.geometric}

\title{Mapping-based exact-integral formulation of skin-friction transformations for zero-pressure-gradient compressible turbulent boundary layers}

\author{Xuke Zhu\aff{1},
  Xiaoshuo Yang\aff{1},
  Yongchao Ji\aff{1},
  Shiyi Chen\aff{2},
  \and Zhenhua Xia\aff{1}
  \corresp{\email{xiazh1006@163.com}}
 }
\affiliation{\aff{1}State Key Laboratory of Fluid Power and Mechatronic Systems and Department of Engineering Mechanics, Zhejiang University, Hangzhou 310027, PR China
             \aff{2}Eastern Institute for Advanced Study, Eastern Institute of Technology, Ningbo 315200, PR China}

\begin{document}
\maketitle

\begin{abstract}
A long-standing route to efficient surface-drag prediction in zero-pressure-gradient compressible turbulent boundary layers is to map the skin-friction coefficient $C_f$ and momentum-thickness Reynolds number $Re_\theta$ onto their `incompressible' counterparts, for which established skin-friction scalings apply. Reassessment against an extensive direct numerical simulation database shows, however, that existing formulations do not consistently recover the reference incompressible skin-friction behaviour, even when transformed compressible data exhibit improved collapse. We therefore define the mapped `incompressible' state as a constant-property counterpart of the physical compressible boundary layer and derive the associated transformation factors from prescribed mean-velocity and wall-normal-coordinate mappings. This definition-first approach ties the resulting skin-friction scaling to the full-layer accuracy of the underlying velocity transformation and exposes inherited outer-layer mapping errors. On this basis, van Driest's skin-friction theory is recast within a finite-Reynolds-number exact-integral formulation, with the classical van Driest~I and II transformations emerging as leading-order asymptotic reductions. We further demonstrate that these reductions are not uniformly reliable over practical parameter ranges, and the historical success of the classical van Driest~II transformation is traced to a fortuitous cancellation of finite-Reynolds-number truncation errors. The exact-integral formulation is then used to construct modified skin-friction transformations, which are assessed through both \emph{a priori} scaling and standalone \emph{a posteriori} prediction of $C_f$ from prescribed macroscopic flow and wall-thermal inputs. Among the formulations considered, the modified transformation incorporating an established velocity mapping (Volpiani \emph{et al.}, 2020, \emph{Phys. Rev. Fluids}, vol.~5, 052602) gives the best overall performance. For database cases spanning free-stream Mach numbers $0.30 \leq M_\infty \leq 13.64$ and wall-diabatic parameters $-0.55 \leq \varTheta \leq 2.85$, its \emph{a posteriori} prediction errors remain below $11\%$, with a mean error of $3.07\%$. Overall, the analysis places skin-friction transformations on a mapping-based, exact-integral footing, clarifying them as direct consequences of prescribed mean-flow mappings rather than empirical adjustments, while exposing and avoiding the leading-order asymptotic truncations that limit classical van Driest skin-friction theory at finite Reynolds numbers.
\end{abstract}



\section{Introduction}
\label{sec:introduction}
The efficient and accurate prediction of surface drag and heat transfer in high-speed turbulent boundary layers is essential for reliable aerospace design \citep{bradshaw1977improved, Lee2019recent,cheng2024progress}. Although practical configurations typically involve complex geometries and pressure gradients, the zero-pressure-gradient (ZPG) case remains the fundamental model for elucidating the underlying physics.

In the limiting case of incompressible turbulent boundary layers (ITBLs) with constant fluid properties, this predictive problem is well-established. The skin-friction coefficient, $C_{f,i} = 2 \bar \tau_w / (\bar \rho_w U_\infty^2)$, exhibits excellent data collapse when plotted against the Reynolds number based on the momentum thickness $\theta_i$, i.e., $Re_{\theta,i} = \bar \rho_w U_\infty \theta_i / \bar \mu_w$. Here, $\bar{\tau}_w$ represents the mean wall shear stress, $U_\infty$ is the free-stream velocity, and $\bar{\rho}$ and $\bar{\mu}$ denote the (constant) mean density and dynamic viscosity, respectively. Specifically, the low-Reynolds-number regime is well described by the power-law correlation proposed by \citet{smits1983low},
\begin{equation}
    C_{f,i} = 0.024 Re_{\theta,i}^{-1/4},
    \label{eq:Smits relation}
\end{equation}
whereas the high-Reynolds-number behaviour ($Re_{\theta,i} \gtrsim 5000$), associated with the emergence of outer-layer self-similarity \citep{appelbaum2025onset}, is accurately captured by the Coles--Fernholz (hereafter CF) relation \citep{Fernholz1996incompressible,nagib2007approach},
\begin{equation}
    \sqrt{\frac{2}{C_{f,i}}} = \frac{1}{\kappa_\theta} \ln Re_{\theta,i} + C_\theta,
    \label{eq:Coles-Fernholz relation}
\end{equation}
where $\kappa_\theta \approx 0.384$ and $C_\theta \approx 4.127$. These scalings allow for a direct prediction of the skin-friction coefficient at a prescribed momentum-thickness Reynolds number.

However, the situation becomes considerably more complex in compressible turbulent boundary layers (CTBLs), characterized by strong variations in mean fluid properties due to intense viscous heating and heat transfer, and an intricate interplay between thermodynamics and hydrodynamics \citep{bradshaw1977compressible,lele1994compressibility,yu2024statistics}. Consequently, the skin-friction coefficient significantly deviates from the predictions of \eqref{eq:Smits relation} and \eqref{eq:Coles-Fernholz relation} when fluid properties are evaluated at the wall as suggested by \citet{vonKarman1936problem}. To extend the well-established incompressible scalings to high-speed regimes, a widely adopted approach is the use of compressibility transformations, which map the physical quantities in the compressible state to their corresponding `incompressible' counterparts by accounting for both the Mach-number and wall-temperature effects. Over the past few decades, this concept has proven to be powerful, particularly in the development of compressible laws of the wall for the mean velocity \citep{vanDriest1951turbulent,zhang2012mach,trettel2016mean,patel2016influence,volpiani2020data,griffin2021velocity,hasan2023incorporating,younes2023mean,lee2023compressible,zhu2024velocity,danis2024accuracy,Zhang2026transformation} and mean temperature \citep{patel2017scalar,chen2022unified,huang2023velocity,cheng2024mean,zhu2025enhancing,zhu2025unified,xu2026temperature,wan2026enthalpy,liang2026composite,Zhang2026transformation}, with important implications for near-wall modelling in Reynolds-averaged Navier-Stokes (RANS) \citep{huang1994turbulence,catris2000density,oteroRodriguez2018turbulence,chen2024improved,hu2025viscous,chen2025mean,hasan2025variable} and wall-modelled large-eddy simulation (WMLES) \citep{yang2018semi,hendrickson2022improving,hendrickson2023applying,griffin2023near,xu2025flux,kumar2025improved} frameworks.

Parallel to the well-known van Driest (vD) transformation \citep{vanDriest1951turbulent} for the mean velocity, substantial theoretical efforts were devoted during the 1950s and 1960s to the development of compressibility transformations for the skin-friction coefficient. The underlying idea is intuitive: the compressible skin-friction coefficient, $C_f = 2\bar{\tau}_w / (\rho_\infty u_\infty^2)$, and the Reynolds number based on the momentum thickness $\theta$, $Re_\theta = \rho_\infty u_\infty \theta / \mu_\infty$, are `stretched' by the transformation factors $F_C$ and $F_\theta$, respectively, such that they are mapped onto their corresponding `incompressible' counterparts, $C_{f,i}$ and $Re_{\theta,i}$, i.e. \citep{spalding1964drag,hopkins1971evaluation},
\begin{equation}
    C_{f,i} = F_C C_f, \quad Re_{\theta,i} = F_\theta Re_\theta.
    \label{eq:skin friction transformation}
\end{equation}
Here, $u_\infty$ denotes the free-stream velocity of the compressible flow, introduced to distinguish it from its `incompressible' analogue $U_\infty$, while $\rho_\infty$ and $\mu_\infty$ denote the free-stream density and dynamic viscosity, respectively. Since $C_{f,i}$ and $Re_{\theta,i}$ satisfy the incompressible scaling laws given by \eqref{eq:Smits relation} and \eqref{eq:Coles-Fernholz relation}, their compressible extensions can be readily obtained through \eqref{eq:skin friction transformation}. For example, the compressible CF relation can be written as
\begin{equation}
    \sqrt{\frac{2}{F_C C_f}} = \frac{1}{\kappa_\theta} \ln (F_\theta Re_{\theta} ) + C_\theta.
    \label{eq:Coles-Fernholz relation (compressible extension)}
\end{equation}
The central challenge therefore lies in modelling the compressibility effects embedded in $F_C$ and $F_\theta$. In practice, these factors are commonly expressed as functions of the free-stream Mach number $M_\infty$ and wall-thermal parameters, including the wall-to-recovery-temperature ratio $\bar{T}_w / T_r$, the wall-diabatic parameter $\varTheta = (\bar{T}_w - T_\infty)/(T_r - T_\infty)$ \citep{zhang2014generalized}, or the Eckert number $Ec = (\gamma - 1) M_\infty^2 T_\infty /(T_r - \bar{T}_w)$ \citep{Wenzel2022about}. Here, the recovery temperature is defined as $T_r = [1 + r (\gamma - 1) M_\infty^2/2 ] T_\infty$, where $T_\infty$ is the free-stream temperature, $r = Pr^{1/3}$ denotes the recovery factor, and $Pr$ is the molecular Prandtl number. Once these dependencies are specified, the compressible skin-friction coefficient can be conditionally predicted from the prescribed free-stream conditions, wall-thermal parameters, and momentum-thickness Reynolds number $Re_\theta$.

While not the earliest attempt, a seminal advance was made by \citet{vanDriest1951turbulent}, who semi-analytically derived a transformation commonly referred to as van Driest I (vD~I). A key contribution of this formulation is the explicit treatment of wall-normal variations in mean fluid properties, achieved by combining the Prandtl mixing length hypothesis and the momentum integral equation with a set of mean thermodynamic relations. Subsequently, \citet{vanDriest1956problem} refined this approach by using the von Kármán mixing length hypothesis instead, leading to the so-called van Driest II (vD~II) transformation, which can be regarded as an extension of the earlier quasi-adiabatic formulation of \citet{wilson1950turbulent}. This modification was shown to improve agreement with experimental data at high Mach numbers and in the presence of wall heat transfer \citep{bradshaw1977compressible}, and vD~II has therefore generally been preferred over vD~I. Building on these foundational developments, \citet{spalding1964drag} systematically categorized and quantitatively assessed twenty available theories at the time and proposed a new formulation, commonly referred to as the Spalding--Chi (SC) transformation. A distinctive feature of this approach is its hybrid nature: while the transformation factor $F_C$ retains the same mixing-length-based expression as that employed in vD~I and vD~II, $F_\theta$ was determined entirely empirically to achieve optimal agreement with experimental data. The comparative predictive performance of the SC and vD~II transformations has been widely investigated. While some studies have suggested that the SC transformation may be more suitable for high-Mach-number, high-enthalpy, and very cold-wall conditions \citep{carybertram1974engineering,goyne2003skin}, the vD~II transformation remains widely regarded as the `most popular and best of its type ever developed' \citep{white2006viscous}. Based on flat-plate experimental data spanning $1.5 \leq M_\infty \leq 7.4$ and $0.14 \leq \bar T_w / T_r \leq 1.00$, \citet{hopkins1971evaluation} concluded that the vD~II transformation provides the most reliable performance among the available theories for $\bar T_w / T_r \gtrsim 0.3$, whereas the SC transformation generally leads to systematic underpredictions. The review by \citet{bradshaw1977compressible} subsequently reinforced this preference, while also noting that neither transformation provides satisfactory predictive performance under extremely cold-wall conditions with $\bar T_w / T_r \lesssim 0.2$. More recently, these historical observations have been further validated by \citet{huang2022direct} using an extensive direct numerical simulation (DNS) database covering $2.5 \leq M_\infty \leq 13.64$ and $0.2 \leq \bar T_w / T_r \leq 1.00$. 

Despite their historical success and practical utility, skin-friction transformations have been subject to persistent theoretical scrutiny. A major point of contention is the inconsistency between the comparatively successful vD~II transformation and its underlying compressible law of the wall \citep{coakley1992turbulence,he1994asymptotic}. Consequently, these transformation methods have been criticized for relying on arbitrary adjustments of transformation factors to correlate data \citep{bradshaw1977improved} and for lacking a rigorous physical foundation \citep{huang1993skin}. After a prolonged period of diminishing prominence and limited theoretical development, this classical paradigm has recently been revisited by \citet{zhao2025revisiting}. By defining a nominal momentum thickness intended to isolate Mach-number and wall-heat-transfer effects associated with the mean velocity profile, they proposed a skin-friction scaling based on the GFM velocity transformation \citep{griffin2021velocity}. This formulation was shown to provide a substantially improved \emph{a priori} collapse relative to the classical vD~II transformation, particularly for flows with wall heat transfer. Nevertheless, three issues remain. First, the physical interpretation of the nominal momentum thickness is not fully explicit, since the transformed `incompressible' velocity is integrated together with the physical wall-normal density variation of the compressible flow. Consequently, it is not self-evident whether the resulting quantity genuinely isolates compressibility effects. Second, the empirical logarithmic scaling fitted to the transformed data appears to deviate from the established ZPG ITBL reference behaviour represented by \eqref{eq:Smits relation} and \eqref{eq:Coles-Fernholz relation}. Thus, an improved collapse of transformed compressible data does not guarantee consistency with the incompressible-limit skin-friction scaling. Third, the associated nominal momentum-thickness Reynolds number cannot be evaluated directly from prescribed macroscopic flow parameters. The formulation is therefore primarily suited to \emph{a priori} scaling assessments based on available mean-flow data and is not readily closed as a standalone procedure for the \emph{a posteriori} prediction of $C_f$.

In contrast, the present study interprets $C_{f,i}$ and $Re_{\theta,i}$ as the standard skin-friction coefficient and momentum-thickness Reynolds number of a mapped ZPG ITBL with constant fluid properties, so that the transformation factors $F_C$ and $F_\theta$ follow directly from the prescribed mean-velocity and wall-normal-coordinate mappings. This definition-first construction differs from the nominal-thickness construction by specifying the target mapped state before deriving the associated transformation factors. Building on these relations, van Driest's classical theory is reformulated semi-analytically in an exact-integral form, thereby avoiding the conventional leading-order asymptotic approximations whose finite-Reynolds-number errors are non-negligible over the range of practical interest. The resulting formulation enables the conditional \emph{a posteriori} prediction of $C_f$ from prescribed macroscopic flow parameters, without requiring the available mean-flow profiles.

The remainder of the paper is organized as follows. In~\textsection\ref{sec:section 2}, existing skin-friction transformations are reassessed using an extensive DNS database to assess how well the transformed skin-friction data recover the reference behaviour of ZPG ITBLs over a broad range of Mach numbers and wall-thermal conditions, and to identify the limitations revealed by their departures from this reference. In~\textsection\ref{sec:section 3}, a mapping-based formulation is developed, in which the transformation factors follow as an exact consequence of a prescribed transformed `incompressible' state and the associated mean-velocity and wall-normal-coordinate mappings. Its direct application to existing velocity transformations is then assessed to examine the inherited outer-layer limitations, and its conceptual distinction from the nominal-thickness construction of \citet{zhao2025revisiting} is clarified. In~\textsection\ref{sec:section 4}, van Driest’s classical theory is reformulated within the mapping-based formulation, yielding exact-integral expressions for the transformation factors and clarifying the associated leading-order asymptotic reductions and their finite-Reynolds-number error structure. In~\textsection\ref{sec:section 5}, the resulting modified transformations are evaluated through both \emph{a priori} scaling assessments and standalone \emph{a posteriori} skin-friction predictions. Finally, concluding remarks are provided in~\textsection\ref{sec:conclusions}.

\section{Reassessment of existing skin-friction transformations}
\label{sec:section 2}
Although classical skin-friction transformations such as vD~I, vD~II and SC have been extensively studied, a definitive consensus regarding their accuracy across wide ranges of Mach number and wall-thermal conditions remains elusive. Moreover, the potential extension of the GFM skin-friction transformation recently proposed by \citet{zhao2025revisiting} to other mean velocity scalings has not yet been discussed and evaluated. In this section, we reassess these existing skin-friction transformation approaches using a comprehensive DNS database of ZPG CTBLs, by examining whether the transformed data recover the corresponding incompressible skin-friction behaviour. The datasets span a broad range of free-stream Mach numbers, $0.30 \leq M_\infty \leq 13.64$, and wall-thermal conditions, characterized by wall-to-recovery-temperature ratios $0.18 \leq \bar T_w / T_r \leq 1.89$ and diabatic parameters $-0.55 \leq \varTheta \leq 2.85$. Here, the recovery temperature is consistently evaluated as $T_r = T_\infty [1+r(\gamma-1)M_\infty^2/2]$, with $\gamma=1.4$ and $r=Pr^{1/3}$ based on the Prandtl number listed for each case. Accordingly, cases with $\varTheta \simeq 1$ are classified as quasi-adiabatic, whereas $\varTheta>1$ and $\varTheta<1$ correspond to heated and cooled walls, respectively. For cooled walls, following \citet{gibis2024heat}, the regime is further subdivided into weakly cooled ($0.44 \lesssim \varTheta < 1$), moderately cooled ($0 \lesssim \varTheta \lesssim 0.44$) and strongly cooled ($\varTheta \lesssim 0$) conditions. Since the present database is restricted to $Pr \approx 0.70$--$0.72$, the corresponding recovery factor varies only slightly, $r\approx0.888$--$0.896$, consistent with the $r\approx0.9$ convention used in this classification. \hyperref[tab:the ZPG CTBLs database 1]{Tables~\ref{tab:the ZPG CTBLs database 1}} and \ref{tab:the ZPG CTBLs database 2} summarize the basic flow parameters of all DNS cases considered.

\begin{table}
  \begin{center}
    \def~{\hphantom{0}}
    \begin{tabular}{lcccccccc}
      Source & $M_\infty$ & $Re_\tau$ & $Re_\theta$  & $\bar T_w / T_r$ & $\varTheta$ & $Pr$ & $T_\infty\,(\mathrm{K})$ & $C_f \times10^3$ \\ [3pt]

      AGPW   & 2.00       & 362--1468 & 1711--8288   & 1.00             & 0.99        & 0.71 & 288.15                  & 1.925--2.797     \\[3pt]

      \multirow{3}{*}{BP}
             & 2.00       & 203--1113 & 905--6325    & 1.00             & 1.00        & 0.72 & 169.40                  & 2.105--3.422     \\
             & 3.00       & 399--506  & 3047--4053   & 1.00             & 1.00        & 0.72 & 169.40                  & 1.856--2.008     \\
             & 4.00       & 398--505  & 4811--6051   & 1.00             & 1.00        & 0.72 & 169.40                  & 1.322--1.369     \\[3pt]

      \multirow{7}{*}{WSKR}
             & 0.30       & 253--826  & 626--2528    & 1.00             & 0.99        & 0.71 & 288.15                  & 3.297--4.737     \\
             & 0.50       & 250--834  & 643--2607    & 1.00             & 0.99        & 0.71 & 288.15                  & 3.240--4.628     \\
             & 0.70       & 251--835  & 680--2751    & 1.00             & 0.99        & 0.71 & 288.15                  & 3.127--4.474     \\
             & 0.85       & 254--666  & 719--2212    & 1.00             & 0.99        & 0.71 & 288.15                  & 3.253--4.342     \\
             & 1.50       & 255--489  & 910--1971    & 1.00             & 0.99        & 0.71 & 288.15                  & 2.988--3.703     \\
             & 2.00       & 252--449  & 1109--2191   & 1.00             & 0.99        & 0.71 & 288.15                  & 2.634--3.181     \\
             & 2.50       & 254--356  & 1378--2079   & 1.00             & 0.99        & 0.71 & 288.15                  & 2.367--2.676     \\[3pt]

      \multirow{7}{*}{ZWLSL}
             & 0.50       & 532--666  & 1420--1875   & 1.00             & 1.00        & 0.72 & 298.15                  & 3.614--3.937     \\
             & 2.00       & 617--781  & 2890--3799   & 1.00             & 1.00        & 0.72 & 220.00                  & 2.425--2.627     \\
             & 2.00       & 621--765  & 1246--1570   & 0.50             & -0.20       & 0.72 & 220.00                  & 3.435--3.680     \\
             & 4.00       & 593--725  & 6250--7603   & 1.00             & 1.00        & 0.72 & 220.00                  & 1.330--1.382     \\
             & 6.00       & 570--683  & 18152--21436 & 1.00             & 1.00        & 0.72 & 55.00                   & 0.774--0.792     \\
             & 8.00       & 574--660  & 31299--36377 & 1.00             & 1.00        & 0.72 & 51.80                   & 0.466--0.486     \\
             & 8.00       & 595--686  & 15120--17340 & 0.50             & 0.46        & 0.72 & 51.80                   & 0.699--0.722     \\[3pt]

      \multirow{5}{*}{GSKW}
             & 2.00       & 445--799  & 3267--6321   & 1.42             & 2.02        & 0.71 & 288.15                  & 1.775--2.096     \\
             & 2.00       & 443--799  & 2164--4284   & 1.00             & 1.00        & 0.71 & 288.15                  & 2.217--2.650     \\
             & 2.00       & 442--800  & 1652--3315   & 0.79             & 0.50        & 0.71 & 288.15                  & 2.488--3.000     \\
             & 2.00       & 445--909  & 1123--2610   & 0.58             & 0.00        & 0.71 & 288.15                  & 2.845--3.550     \\
             & 2.00       & 598--999  & 770--1446    & 0.35             & -0.55       & 0.71 & 288.15                  & 3.522--4.130     \\
    \end{tabular}
    \caption{Summary of DNS datasets for ZPG CTBLs with and without wall heat transfer. The data sources are abbreviated as follows: AGPW for \citet{appelbaum2025onset}, BP for \citet{bernardini2011wall,pirozzoli2011turbulence}, WSKR for \citet{wenzel2018dns}, ZWLSL for \citet{zhang2022wall,zhang2024intrinsic}, and GSKW for \citet{gibis2024heat}. Here, $M_\infty = u_\infty / \sqrt{\gamma R T_\infty}$ denotes the free-stream Mach number, and the friction Reynolds number is defined as $Re_\tau = \left. y^+ \right|_{y = \delta_e} = \bar \rho_w u_\tau \delta_e / \bar \mu_w$, with $y^+ = y / \delta_\nu$ denoting the inner-scaled wall-normal coordinate. For all cases listed in this table, the dynamic viscosity is modelled using Sutherland's law \eqref{eq:mean mu-T relation (Sutherland law)}, with a Sutherland temperature of $T_s=120\,\mathrm{K}$ for the datasets of BP and $T_s=110.4\,\mathrm{K}$ for all other cases.}
    \label{tab:the ZPG CTBLs database 1}
  \end{center}
\end{table}

\begin{table}
  \begin{center}
    \def~{\hphantom{0}}
    \begin{tabular}{lcccccccc}
      Source & $M_\infty$ & $Re_\tau$ & $Re_\theta$ & $\bar T_w / T_r$ & $\varTheta$ & $Pr$ & $T_\infty\,(\mathrm{K})$ & $C_f \times10^3$ \\ [3pt]

      \multirow{13}{*}{CBCBP}
             & 2.00       & 444       & 2057        & 1.00             & 1.00        & 0.72 & 220.00       & 2.788                                       \\
             & 2.00       & 445       & 1790        & 0.90             & 0.75        & 0.72 & 220.00       & 2.979                                       \\
             & 2.00       & 444       & 1548        & 0.79             & 0.50        & 0.72 & 220.00       & 3.166                                       \\
             & 2.00       & 1948      & 8302        & 0.76             & 0.42        & 0.72 & 100.00       & 2.159                                       \\
             & 2.00       & 444       & 1293        & 0.69             & 0.25        & 0.72 & 220.00       & 3.403                                       \\
             & 4.00       & 443       & 3633        & 0.81             & 0.75        & 0.72 & 220.00       & 1.610                                       \\
             & 4.00       & 444       & 2733        & 0.63             & 0.50        & 0.72 & 220.00       & 1.856                                       \\
             & 4.00       & 442       & 1831        & 0.44             & 0.25        & 0.72 & 220.00       & 2.197                                       \\
             & 5.86       & 1950      & 38705       & 0.76             & 0.72        & 0.72 & 100.00       & 0.678                                       \\
             & 6.00       & 446       & 7850        & 1.00             & 1.00        & 0.72 & 220.00       & 0.809                                       \\
             & 6.00       & 444       & 6178        & 0.78             & 0.75        & 0.72 & 220.00       & 0.927                                       \\
             & 6.00       & 445       & 4529        & 0.57             & 0.50        & 0.72 & 220.00       & 1.090                                       \\
             & 6.00       & 443       & 2624        & 0.35             & 0.25        & 0.72 & 220.00       & 1.398                                       \\ [3pt]

      \multirow{5}{*}{ZDC}
             & 2.50       & 505       & 2788        & 0.99             & 0.99        & 0.71 & 270.00       & 2.294                                       \\
             & 5.84       & 450       & 2049        & 0.25             & 0.13        & 0.71 & 55.20        & 1.665                                       \\
             & 5.86       & 456       & 9231        & 0.75             & 0.71        & 0.71 & 55.00        & 0.993                                       \\
             & 7.87       & 480       & 9427        & 0.48             & 0.43        & 0.71 & 51.80        & 0.745                                       \\
             & 13.64      & 641       & 13987       & 0.18             & 0.16        & 0.71 & 47.40        & 0.382                                       \\ [3pt]

      \multirow{8}{*}{VBL}
             & 2.28       & 100       & 879         & 1.89             & 2.85        & 0.70 & --            & 2.704                                       \\
             & 2.28       & 224       & 1053        & 0.99             & 0.98        & 0.70 & --            & 3.107                                       \\
             & 2.28       & 396       & 2003        & 0.99             & 0.99        & 0.70 & --            & 2.622                                       \\
             & 2.28       & 511       & 1245        & 0.50             & -0.04       & 0.70 & --            & 3.292                                       \\
             & 5.00       & 175       & 3938        & 1.86             & 2.06        & 0.70 & --            & 0.993                                       \\
             & 5.00       & 390       & 3838        & 0.79             & 0.74        & 0.70 & --            & 1.342                                       \\
             & 5.00       & 553       & 5497        & 0.79             & 0.74        & 0.70 & --            & 1.254                                       \\
             & 5.00       & 685       & 7043        & 0.79             & 0.75        & 0.70 & --            & 1.160                                       \\
    \end{tabular}
    \caption{Continuation of the DNS database for ZPG CTBLs. The data sources are abbreviated as follows: CBCBP for \citet{cogo2022direct,cogo2023assessment}, ZDC for \citet{zhang2018direct}, and VBL for \citet{volpiani2018effects,volpiani2020effects}. All flow parameters follow the definitions in \hyperref[tab:the ZPG CTBLs database 1]{table~\ref{tab:the ZPG CTBLs database 1}}. The VBL datasets employ a power-law viscosity-temperature model \eqref{eq:mean mu-T relation (power law)} with exponent $n=0.75$, and $T_\infty$ is not reported (denoted by `--'), whereas all other cases adopt Sutherland's law \eqref{eq:mean mu-T relation (Sutherland law)} with $T_s=110.4\,\mathrm{K}$.}
    \label{tab:the ZPG CTBLs database 2}
  \end{center}
\end{table}

Throughout this manuscript, `$\overline{(\cdot)}$' denotes Reynolds averaging, while the subscripts `$w$' and `$\infty$' indicate quantities evaluated at the wall and in the free stream, respectively. The superscript `+' denotes inner scaling, non-dimensionalized by the friction velocity $u_\tau = \sqrt{\bar \tau_w / \bar \rho_w}$ and the viscous length scale $\delta_\nu = \bar \mu_w / (\bar \rho_w u_\tau)$. Similarly, the superscript `*' denotes semi-local scaling, based on the semi-local friction velocity $u_\tau^* = \sqrt{\bar \tau_w / \bar \rho}$ and the semi-local viscous length scale $\delta_\nu^* = \bar \mu / (\bar \rho u_\tau^*)$. Hereafter, transformed `incompressible' variables are distinguished from their compressible counterparts by uppercase letters or the subscript `$i$'. Furthermore, the compressible momentum thickness is defined as
\begin{equation}
    \theta = \int_0^{\delta_e}{\frac{\bar \rho}{\rho_\infty}\frac{\bar u}{u_\infty} \left( 1 - \frac{\bar u}{u_\infty} \right) \, \mathrm{d}y}
    = \int_0^{\delta_e}{\frac{\bar \rho}{\rho_\infty} z (1-z) \, \mathrm{d}y},
    \label{eq:compressible momentum thickness}
\end{equation}
where $z = \bar u / u_\infty$ represents the mean velocity normalized by the free-stream value. Here, $y$ denotes the wall-normal coordinate, and the boundary-layer edge is consistently taken as $\delta_e = \delta_{99}$, defined by $\left. \bar u \right|_{y = \delta_{99}} \approx 0.99 u_\infty$. This practical convention follows standard ZPG CTBL practice and recent skin-friction scaling assessments \citep{wenzel2018dns,zhao2025revisiting}, and provides a uniform treatment of datasets with different outer-domain extents.

\subsection{The classical vD~I, vD~II and SC transformations}
\label{sec:subsection 2.1}
Within the framework of~\eqref{eq:skin friction transformation}, the vD~I transformation is defined as \citep{vanDriest1951turbulent}
\begin{equation}
    ( F_C )_{\mathrm{vD\,I}} = \frac{T_r/T_{\infty}-1}{( \sin^{-1} \alpha_0 +\sin^{-1} \beta_0) ^2},\quad
    ( F_{\theta} )_{\mathrm{vD\,I}}=\sqrt{\frac{T_{\infty}}{\bar{T}_w}}\frac{\mu _{\infty}}{\bar{\mu}_w}.
    \label{eq:vD I skin friction transformation}
\end{equation}
Here, $\alpha_0 = (2A_0^2 - B_0) / (B_0^2 + 4A_0^2)^{1/2}$, $\beta_0 = B_0 / (B_0^2 + 4A_0^2)^{1/2}$, where $A_0^2 =( T_r-T_{\infty} ) /\bar{T}_w$ and $B_0=T_r/\bar{T}_w-1$ are coefficients derived from Walz's equation \citep{walz1966}, namely
\begin{equation}
    \frac{\bar T}{\bar T_w} = 1 + B_0 z - A_0^2 z^2.
    \label{eq:Walz's equation}
\end{equation}

The vD~II transformation shares the same factor $F_C$ as vD~I, but employs a different form for $F_\theta$. It is given by \citep{vanDriest1956problem}:
\begin{equation}
    ( F_C )_{\mathrm{vD\,II}} = ( F_C )_{\mathrm{vD\,I}},\quad
    ( F_{\theta} ) _{\mathrm{vD\,II}}=\frac{\mu _{\infty}}{\bar{\mu}_w}.
    \label{eq:vD II skin friction transformation}
\end{equation}
It is worth noting that the quasi-adiabatic ($\bar T_w /T_r = 1$) form of \eqref{eq:vD II skin friction transformation} was proposed earlier by \citet{wilson1950turbulent}, for which the coefficients reduce to $A_0^2 = 1 - T_\infty / \bar T_w$ and $B_0 = 0$. 

In contrast to the semi-analytically derived transformations~\eqref{eq:vD I skin friction transformation} and~\eqref{eq:vD II skin friction transformation}, the SC transformation adopts an empirically fitted form for $F_\theta$, namely \citep{spalding1964drag}
\begin{equation}
    ( F_C )_{\mathrm{SC}} = ( F_C )_{\mathrm{vD\,I}},\quad
    ( F_{\theta} ) _{\mathrm{SC}}= \left( \frac{T_\infty}{\bar T_w} \right)^{0.702} \left( \frac{T_r}{\bar T_w} \right)^{0.772}.
    \label{eq:SC skin friction transformation}
\end{equation}

\hyperref[fig:original vD I, vD II, SC transformations (logarithmic)]{Figure~\ref{fig:original vD I, vD II, SC transformations (logarithmic)}} shows the distribution of $\sqrt{2/C_{f,i}}$ in the transformed `incompressible' state as a function of $Re_{\theta,i}$ in semi-logarithmic coordinates. Ideally, a perfect compressibility transformation would collapse the mapped data onto a nearly universal curve described by the incompressible scaling laws~\eqref{eq:Smits relation} and \eqref{eq:Coles-Fernholz relation}, as observed for ZPG ITBLs. However, none of the transformations considered here demonstrates consistent performance across different Mach numbers and wall-thermal conditions. Even for quasi-adiabatic cases, a satisfactory collapse is not achieved. Specifically, the vD~I and SC transformations produce widely scattered data, and the logarithmic correlations obtained from linear fits of $\sqrt{2/C_{f,i}}$ against $\ln Re_{\theta,i}$ deviate markedly from both the incompressible reference data and the corresponding scaling laws, with coefficients of determination $R^2 = 0.66$ and $0.72$, respectively. By contrast, the vD~II transformation yields an improved collapse of the data, and the fitted logarithmic law is in much closer agreement with the CF relation. 

\begin{figure}
  \centering
  \includegraphics[width=0.49\textwidth]{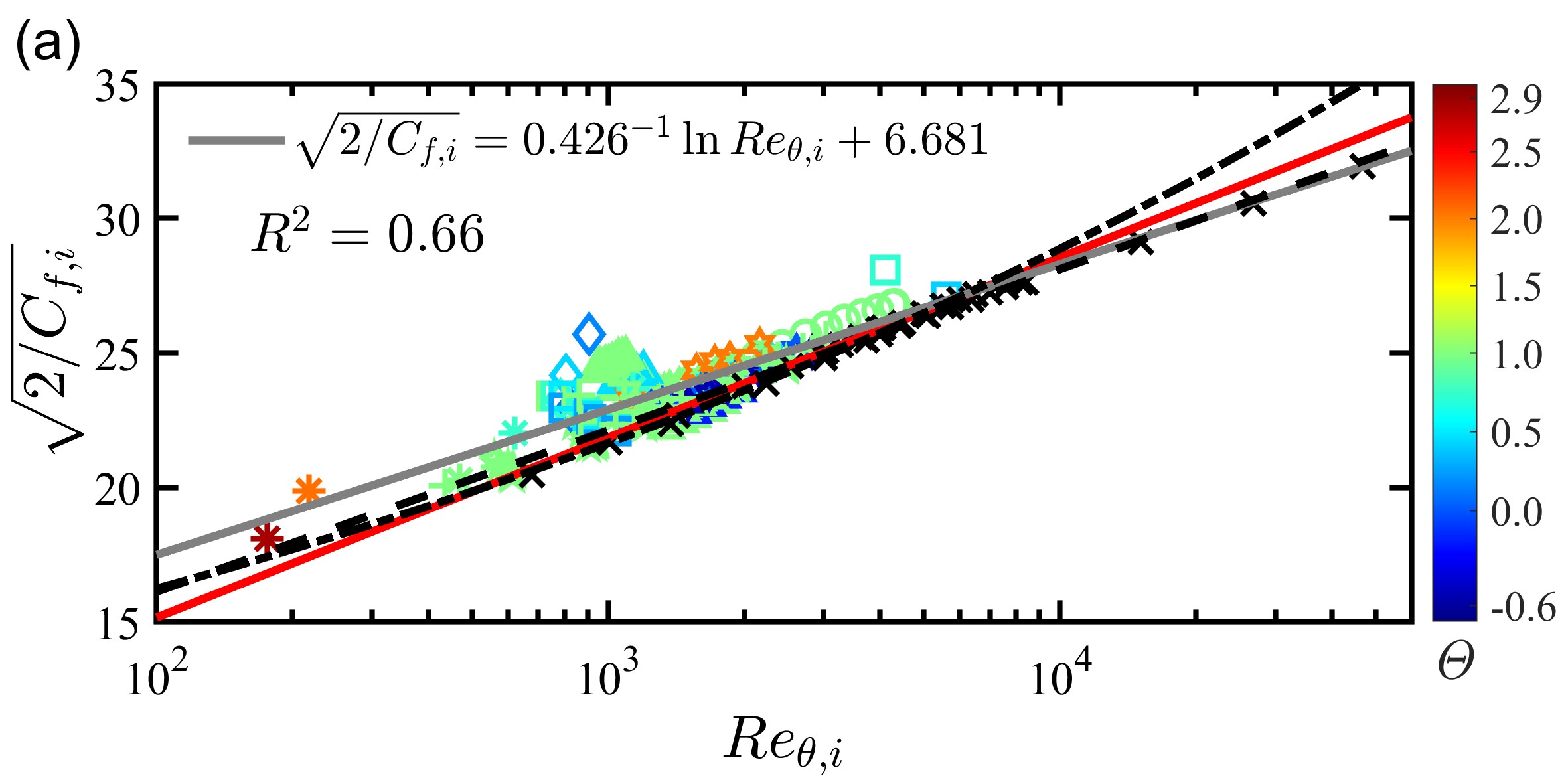}
  \includegraphics[width=0.49\textwidth]{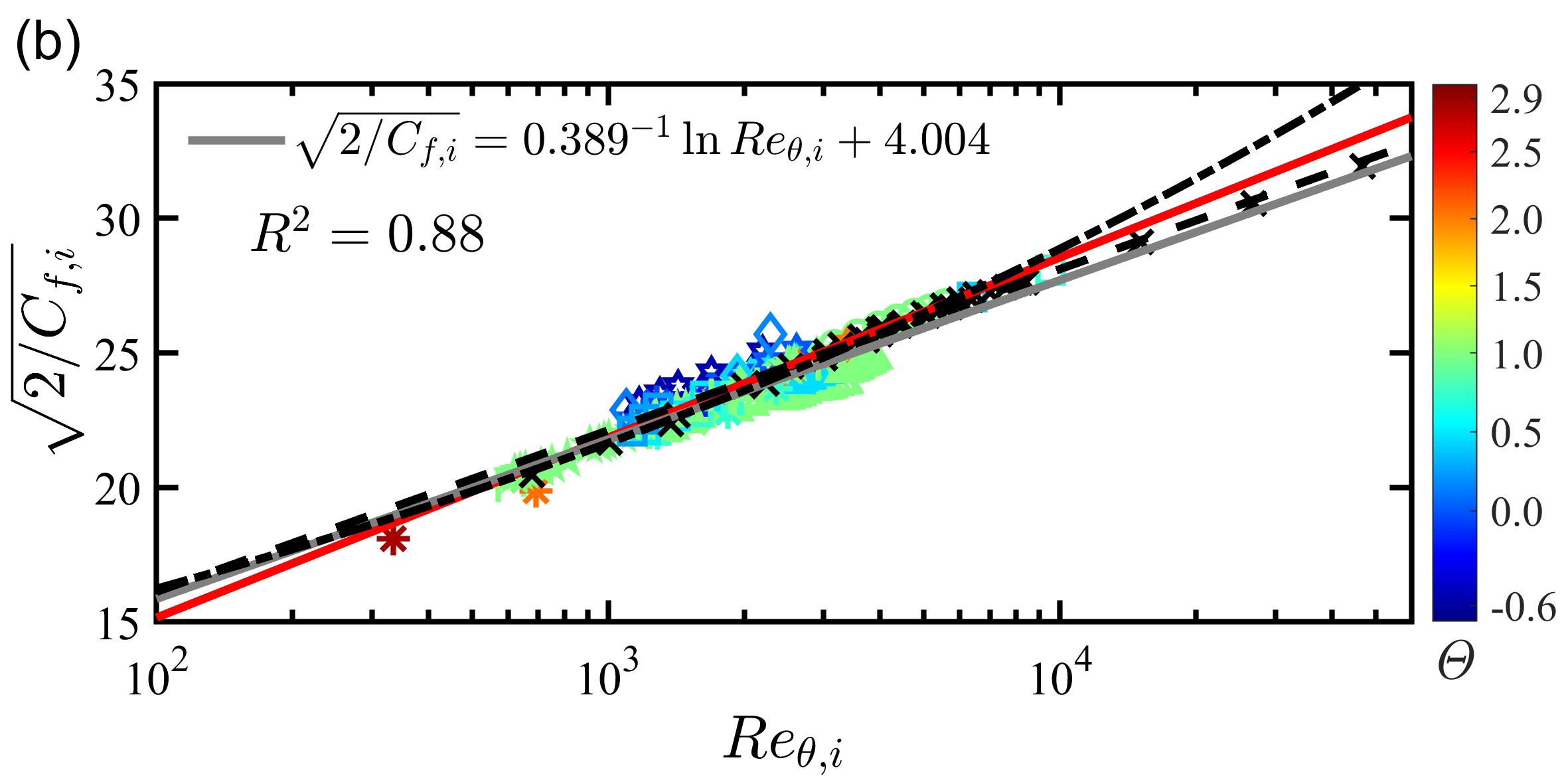}
  \includegraphics[width=0.49\textwidth]{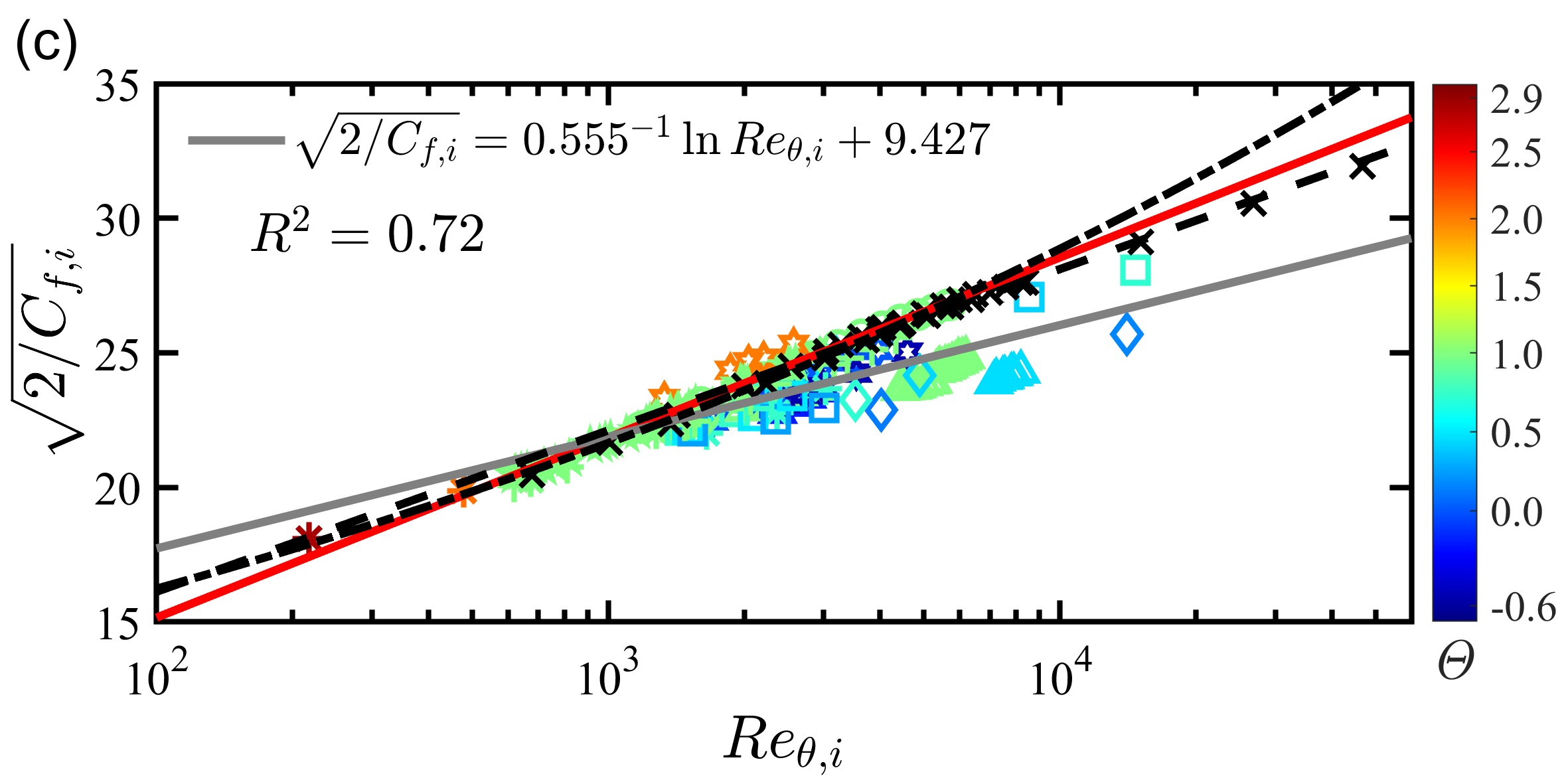}
  \includegraphics[width=0.49\textwidth]{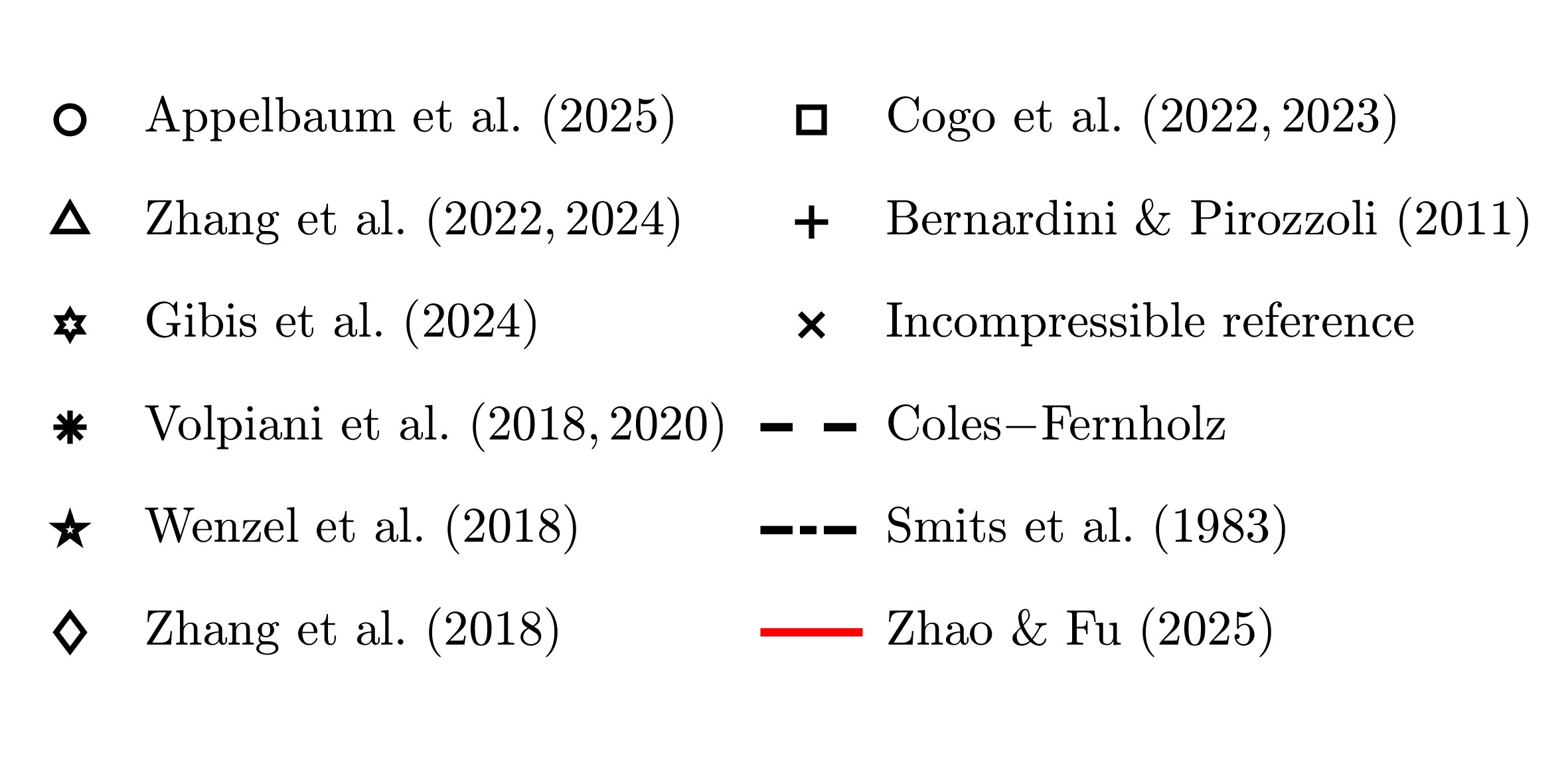}
  \caption{Transformed `incompressible' $\sqrt{2/C_{f,i}}$ versus $Re_{\theta,i}$ in semi-logarithmic coordinates, using the (a) vD~I~\eqref{eq:vD I skin friction transformation}, (b) vD~II~\eqref{eq:vD II skin friction transformation}, and (c) SC~\eqref{eq:SC skin friction transformation} transformations. Hereafter, coloured symbols indicate DNS data of ZPG CTBLs from different sources listed in \hyperref[tab:the ZPG CTBLs database 1]{tables~\ref{tab:the ZPG CTBLs database 1}-\ref{tab:the ZPG CTBLs database 2}}, with colour representing the diabatic parameter $\varTheta$. Black $\times$ symbols denote reference DNS \citep{Schlatter2010assessment,sillero2013one,eitel2014simulation} or experimental data \citep{vallikivi2015turbulent} of ZPG ITBLs, spanning $677 \leq Re_{\theta,i} \leq 234670$. The black dashed line corresponds to the CF relation~\eqref{eq:Coles-Fernholz relation}, the black dash-dotted line to the power-law correlation of \citet{smits1983low}~\eqref{eq:Smits relation}, and the red solid line corresponds to the fitted scaling reported by \citet{zhao2025revisiting} for their proposed skin-friction transformation. The grey solid line in each panel represents a distinct scaling fitted to the transformed compressible DNS data considered in this study, with the corresponding coefficient of determination, $R^2$, reported in the upper-left corner.}
\label{fig:original vD I, vD II, SC transformations (logarithmic)}
\end{figure}

To quantify the deviation of the transformed data from the reference behaviour of ZPG ITBLs, an incompressible skin-friction baseline must be specified. Since the transformed values of $Re_{\theta,i}$ span both low- and high-Reynolds-number regimes, using a single correlation over the entire range would mix the transformation error with the intrinsic Reynolds-number dependence of the incompressible reference. To avoid this ambiguity, the \emph{a priori} error assessment in \textsection~\ref{sec:section 2}--\ref{sec:section 3} is based on the following composite incompressible reference scaling:
\begin{equation}
    C_{f,i}^{\mathrm{ref}}(Re_{\theta,i}) =
    \begin{cases}
    0.024 Re_{\theta,i}^{-1/4}, & Re_{\theta,i} \leq Re_{\theta,i}^{\mathrm{cr}}, \\[3pt]
    2\left(\kappa_\theta^{-1}\ln Re_{\theta,i}+C_\theta\right)^{-2}, & Re_{\theta,i}>Re_{\theta,i}^{\mathrm{cr}},
    \end{cases}
    \label{eq:incompressible skin-friction correlation}
\end{equation}
where the low- and high-Reynolds-number branches correspond, respectively, to the correlation~\eqref{eq:Smits relation} of \citet{smits1983low} and the CF relation~\eqref{eq:Coles-Fernholz relation}. The transition Reynolds number is chosen as their high-Reynolds-number intersection, yielding $Re_{\theta,i}^{\mathrm{cr}} \approx 3887$. This value is close to $Re_{\theta,i} \approx 4125$ recently reported by \citet{appelbaum2025onset} for the onset of outer-layer self-similarity, where the wake-strength parameter approaches an asymptotic constant. The corresponding absolute relative error is then defined as
\begin{equation}
    \mathcal{E}(C_{f,i}) = \frac{\left| C_{f,i} - C_{f,i}^{\mathrm{ref}} \right|}{C_{f,i}^{\mathrm{ref}}} \times 100\%.
    \label{eq:error against composite incompressible reference}
\end{equation}

\hyperref[fig:original vD I, vD II, SC transformations (linear)]{Figure~\ref{fig:original vD I, vD II, SC transformations (linear)}} presents the transformed $C_{f,i}$ as a function of $Re_{\theta,i}$, together with the absolute relative errors $\mathcal{E}(C_{f,i})$. As anticipated, the errors for the reference ZPG ITBL cases remain below 2\% over the full Reynolds-number range, confirming that the composite baseline~\eqref{eq:incompressible skin-friction correlation} provides an accurate description of the expected incompressible behaviour. The vD~I transformation systematically underpredicts $C_{f,i}$ across a substantial portion of the database, yielding a mean error of 9.02\%. The largest deviations occur mainly for moderately cooled high-Mach-number cases at $Re_{\theta,i} \lesssim 2000$, with local errors approaching 30\%. Even for several quasi-adiabatic cases, the relative errors are of order 20\%. In contrast, the SC transformation tends to overpredict $C_{f,i}$ for high-Mach-number quasi-adiabatic cases as well as for cold-wall cases, with local errors reaching approximately 20--30\%, resulting in a mean error of 7.68\%. The vD~II transformation provides the best overall performance among the three classical transformations, reducing the mean error to 3.67\%. Although it tends to underpredict $C_{f,i}$ for moderately and strongly cooled cases while overpredicting some quasi-adiabatic and strongly heated cases, the relative errors remain within 10\% for almost all cases. While these results confirm the relative success of the vD~II transformation, it still falls short of providing a consistently satisfactory mapping to the expected incompressible baseline across the broad parameter space considered here.

\begin{figure}
  \centering
  \includegraphics[width=0.49\textwidth]{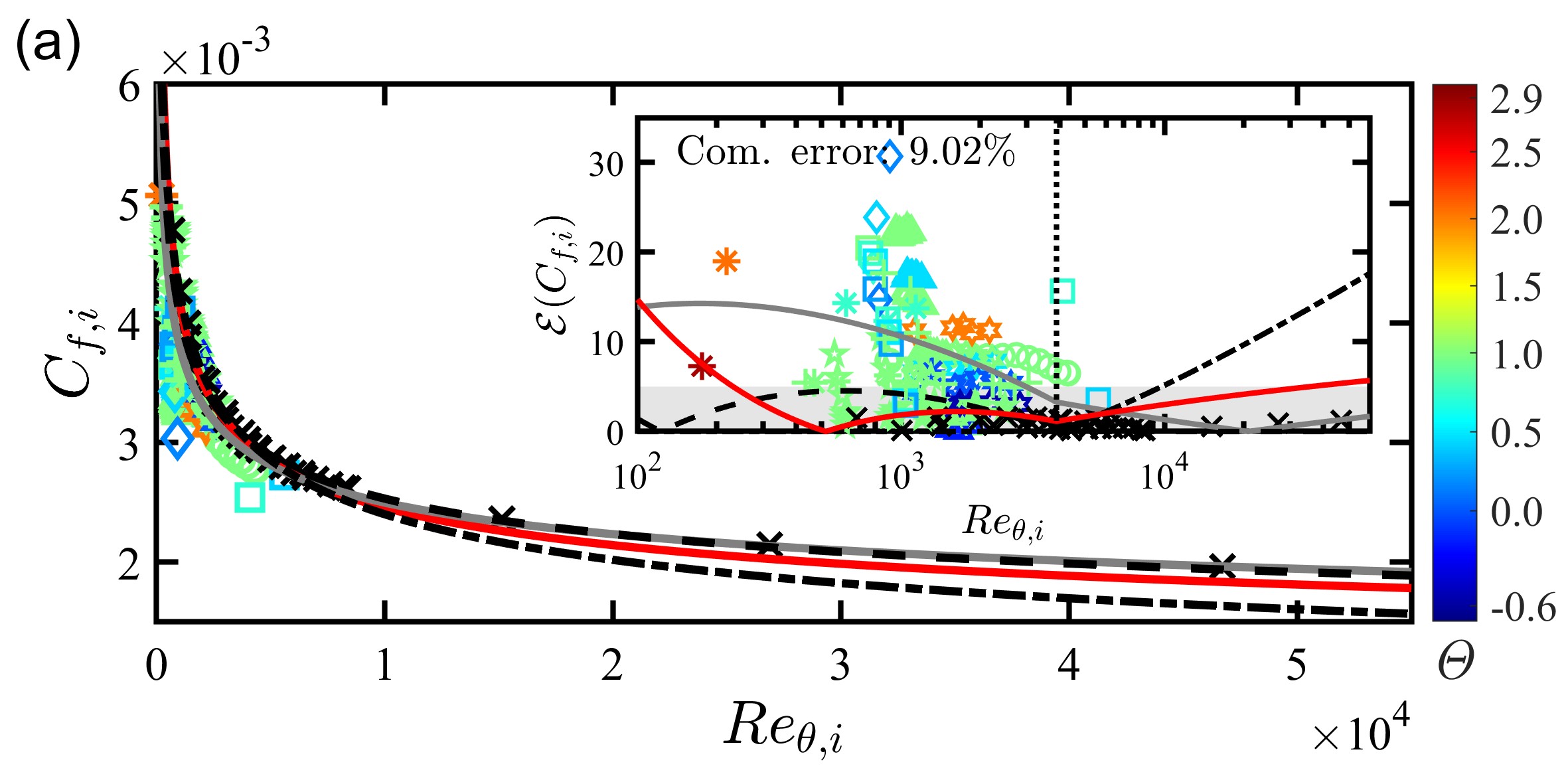}
  \includegraphics[width=0.49\textwidth]{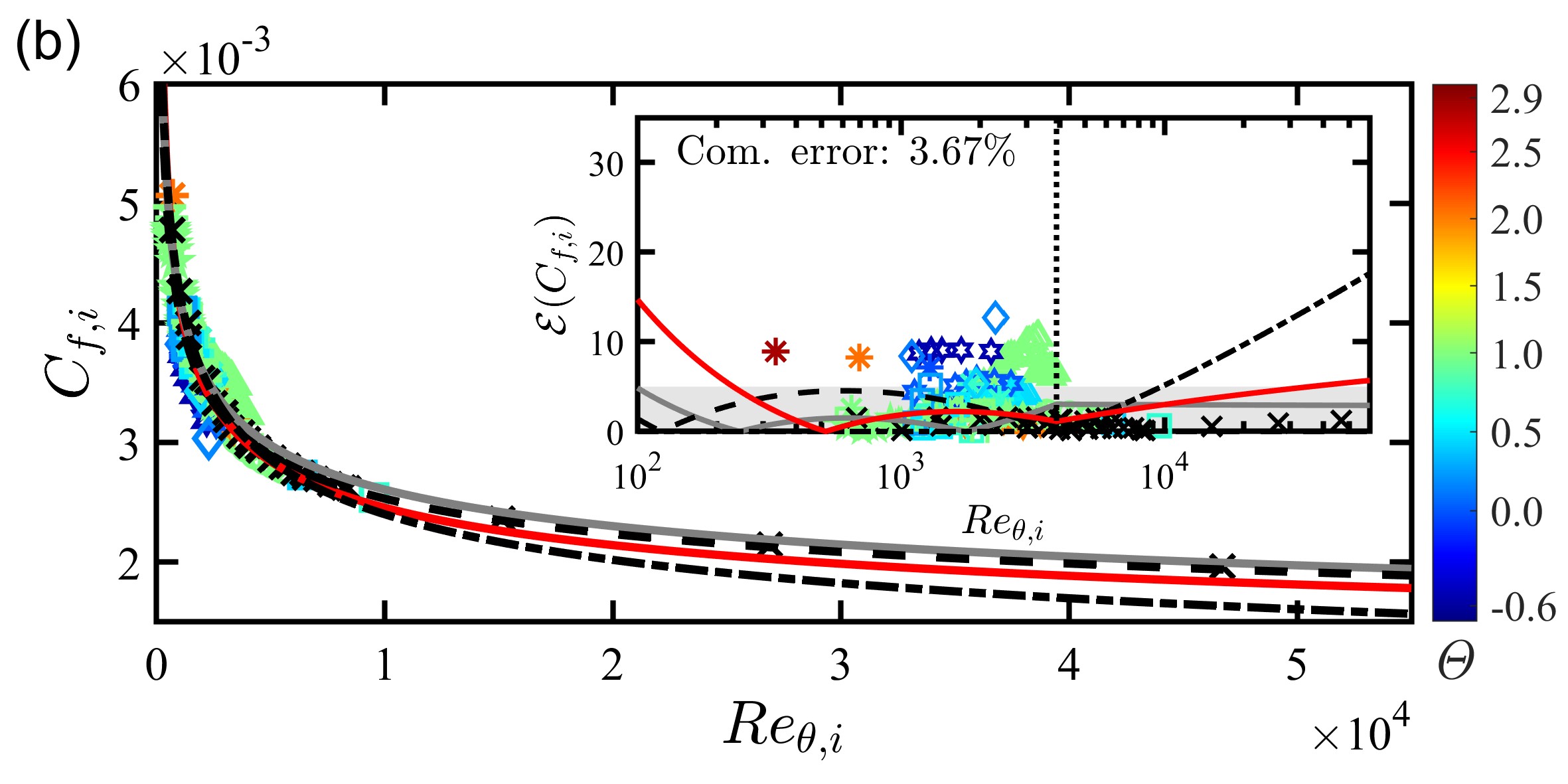}
  \includegraphics[width=0.49\textwidth]{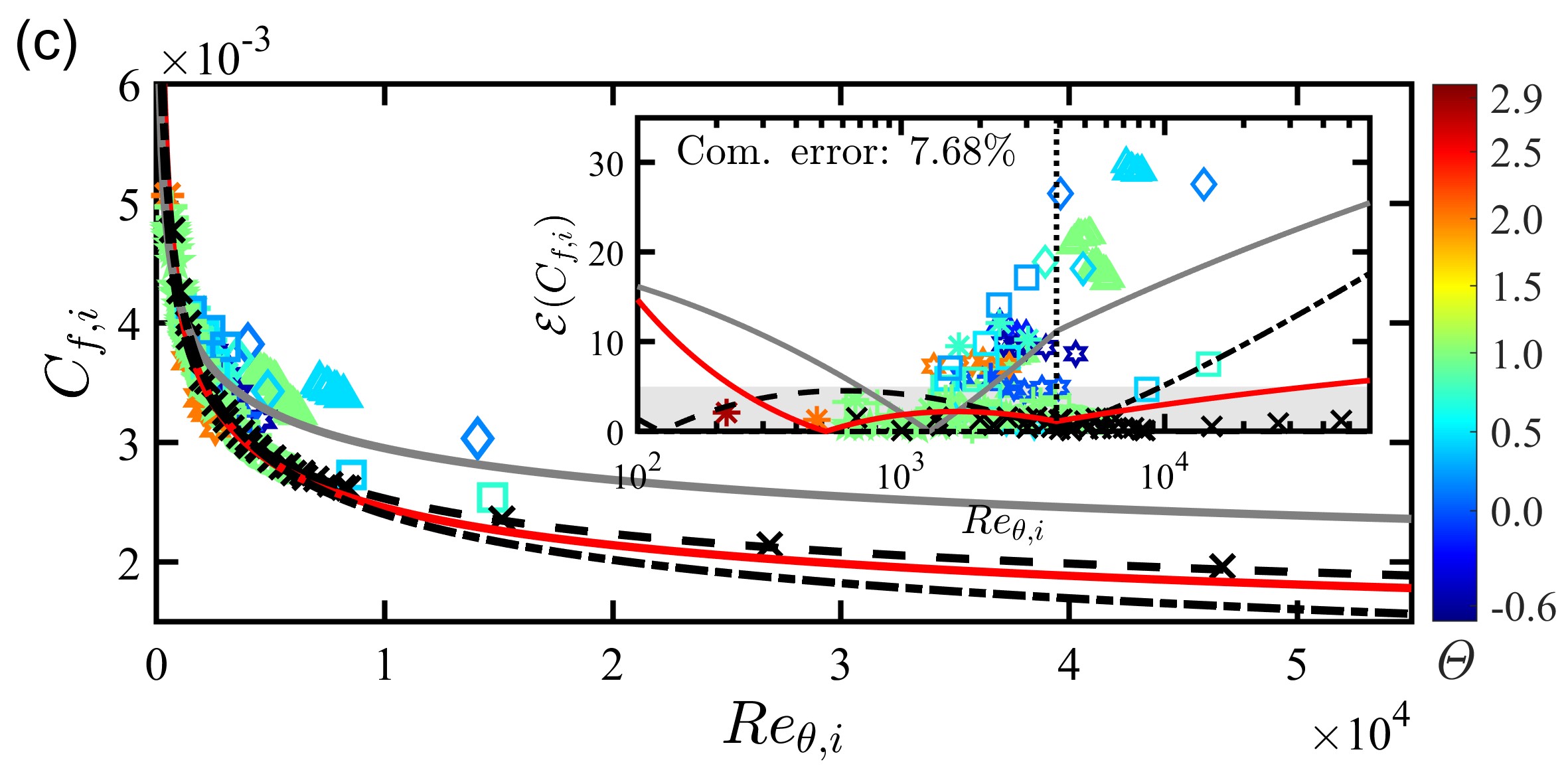}
  \includegraphics[width=0.49\textwidth]{Figure/Legend.jpg}
  \caption{Transformed `incompressible' $C_{f,i}$ versus $Re_{\theta,i}$, using the (a) vD~I~\eqref{eq:vD I skin friction transformation}, (b) vD~II~\eqref{eq:vD II skin friction transformation}, and (c) SC~\eqref{eq:SC skin friction transformation} transformations. Insets show the absolute relative errors $\mathcal{E}(C_{f,i})$ defined by \eqref{eq:error against composite incompressible reference}, with $C_{f,i}^{\mathrm{ref}}$ given by the composite incompressible reference relation~\eqref{eq:incompressible skin-friction correlation}. The vertical dotted line marks the transition Reynolds number $Re_{\theta,i}^{\mathrm{cr}} \approx 3887$. The mean error over all compressible cases is reported in the upper-left corner of each panel, with the grey shaded region highlighting a 5\% error bound. Symbols, colours, and line styles follow those in \hyperref[fig:original vD I, vD II, SC transformations (logarithmic)]{figure~\ref{fig:original vD I, vD II, SC transformations (logarithmic)}}.}
  \label{fig:original vD I, vD II, SC transformations (linear)}
\end{figure}

The present results are consistent with previous findings \citep{hopkins1971evaluation,bradshaw1977compressible,huang2022direct,zhao2025revisiting}, yet the performance of these transformations deviates from their respective theoretical expectations. Following \citet{vanDriest1951turbulent,vanDriest1956problem}, the vD~I and vD~II transformations are based on the Prandtl and von Kármán mixing length hypotheses, respectively. Interestingly, while \eqref{eq:vD II skin friction transformation} clearly outperforms \eqref{eq:vD I skin friction transformation}, the underlying von Kármán mixing length has not been widely accepted as providing improved agreement with compressible data \citep{he1994asymptotic,schetz2011boundary}. The discrepancy between vD~II and its corresponding compressible law of the wall has also been noted previously \citep{coakley1992turbulence,huang1993skin}. Furthermore, the SC transformation is essentially equivalent to vD~II for quasi-adiabatic flows at $M_\infty \leq 5$, where $\bar{\mu} \propto \bar{T}^{0.702}$ approximately holds \citep{bradshaw1977compressible}. However, the correction factor accounting for wall-temperature effects, $(T_r/\bar{T}_w)^{0.772}$, fails to improve performance for flows with wall heat transfer, and instead leads to poorer agreement relative to vD~II. 

\subsection{Recent skin-friction scaling by \citet{zhao2025revisiting}}
\label{sec:subsection 2.2}
When cast into the framework of \eqref{eq:skin friction transformation}, the skin-friction scaling of \citet{zhao2025revisiting} implies:
\begin{equation}
    (F_C)_{\mathrm{ZF}} = \frac{\rho_\infty}{\bar \rho_w} F^{-2}, \quad (F_\theta)_{\mathrm{ZF}} = \frac{\bar \rho_w \mu_\infty}{\rho_\infty \bar \mu_w} F \frac{\hat{\theta}}{\theta}.
    \label{eq:transformation factors of Zhao and Fu}
\end{equation}
Here, $\hat{\theta}$ was interpreted as a momentum thickness intended to remove the Mach-number and wall-heat-transfer effects associated with the mean velocity profile. It is originally defined as
\begin{equation}
    \hat{\theta} = \delta_\nu \hat{\theta}^+ = \delta_\nu \int_0^{Re_{\tau,e}^*}{\frac{\bar \rho}{\rho_\infty} \frac{\bar U_{\mathrm{GFM}}^+}{U_{\mathrm{GFM},\infty}^+} \left( 1 - \frac{\bar U_{\mathrm{GFM}}^+}{U_{\mathrm{GFM},\infty}^+} \right) \, \mathrm{d} y^*},
    \label{eq:definition of theta_hat}
\end{equation}
where $y^* = y / \delta_\nu^*$ denotes the semi-local wall-normal coordinate, and $Re_{\tau,e}^* = \left. y^* \right|_{y = \delta_e}$ is the semi-local friction Reynolds number. The inner-scaled mean velocity in the transformed `incompressible' state is defined via the GFM velocity transformation \citep{griffin2021velocity}:
\begin{equation}
    \bar U_{\mathrm{GFM}}^+(y^*) = \int_0^{\bar u^+}{\frac{\mathrm{d}\bar u^+}{\bar \mu^+ + \frac{\partial \bar u^+}{\partial y^*} - \bar\mu^{+2} \frac{\partial \bar u^+}{\partial y^+}}}.
    \label{eq:GFM transformation}
\end{equation}
Consequently, the ratio between the free-stream velocities after and before the transformation can be expressed as $F = U_{\mathrm{GFM},\infty}^+ / u_\infty^+ = \int_0^1{\left( \bar{\mu}^++\frac{\partial \bar{u}^+}{\partial y^*}-\bar{\mu}^{+2}\frac{\partial \bar{u}^+}{\partial y^+} \right) ^{-1} \, \mathrm{d}z}$.

Although not explicitly reported in \citet{zhao2025revisiting}, \eqref{eq:transformation factors of Zhao and Fu} can be formally applied in conjunction with other appropriate mean velocity scalings by replacing the GFM-transformed variables in \eqref{eq:definition of theta_hat}. In the present study, the VIPL transformation \citep{volpiani2020data} serves as an illustrative extension, as it has also been shown to effectively collapse velocity profiles onto the incompressible law of the wall for ZPG CTBLs. Specifically, $y^*$ and $\bar U_{\mathrm{GFM}}^+$ in \eqref{eq:definition of theta_hat} are replaced by their corresponding VIPL-transformed counterparts, $y_{\mathrm{V}}^+$ and $\bar U_{\mathrm{V}}^+$, defined as
\begin{equation}
    y_{\mathrm{V}}^+ = \int_0^{y^+}{\frac{( \bar \rho^+ )^{1/2}}{( \bar \mu^+ )^{3/2}} \, \mathrm{d}y^+}, \quad 
    \bar U_{\mathrm{V}}^+ (y_{\mathrm{V}}^+) = \int_0^{\bar u^+}{\frac{( \bar \rho^+ )^{1/2}}{( \bar \mu^+ )^{1/2}} \, \mathrm{d}\bar u^+}.
    \label{eq:VIPL transformation}
\end{equation}
Evaluating this mapping at the boundary-layer edge yields the corresponding free-stream velocity ratio, $F = U_{\mathrm{V},\infty}^+ / u_\infty^+ = \int_0^1{(\bar{\rho}^+)^{1/2}(\bar{\mu}^+)^{-1/2} \, \mathrm{d}z}$.

\hyperref[fig:GFM and VIPL transformations (ZF, logarithmic)]{Figure~\ref{fig:GFM and VIPL transformations (ZF, logarithmic)}} shows the transformed $\sqrt{2/C_{f,i}}$ as a function of $Re_{\theta,i}$ in semi-logarithmic coordinates. Consistent with \citet{zhao2025revisiting}, when the transformation factors in \eqref{eq:transformation factors of Zhao and Fu} are evaluated using the GFM velocity transformation~\eqref{eq:GFM transformation}, the transformed data are better organized by a logarithmic trend and exhibit reduced scatter compared with the vD~II transformation. As illustrated in \hyperref[fig:GFM and VIPL transformations (ZF, logarithmic)]{figure~\ref{fig:GFM and VIPL transformations (ZF, logarithmic)}(a)}, the fitted correlation is in close agreement with the original relation of \citet{zhao2025revisiting}, yielding a coefficient of determination of $R^2 = 0.98$. Minor deviations between the two correlations emerge at high Reynolds numbers, which can reasonably be attributed to differences in the compressible DNS datasets employed. However, it is noteworthy that both correlations gradually depart from the well-established CF relation for $Re_{\theta,i} \gtrsim 6000$. Such a trend is not consistent with the behaviour observed in ZPG ITBLs, as indicated by the incompressible reference data. A similar but more pronounced behaviour is observed when the VIPL velocity transformation~\eqref{eq:VIPL transformation} is employed. Although the transformed data remain well organized by an empirical logarithmic fit, again yielding $R^2=0.98$, they deviate markedly from the incompressible reference data.

\begin{figure}
  \centering
  \includegraphics[width=0.49\textwidth]{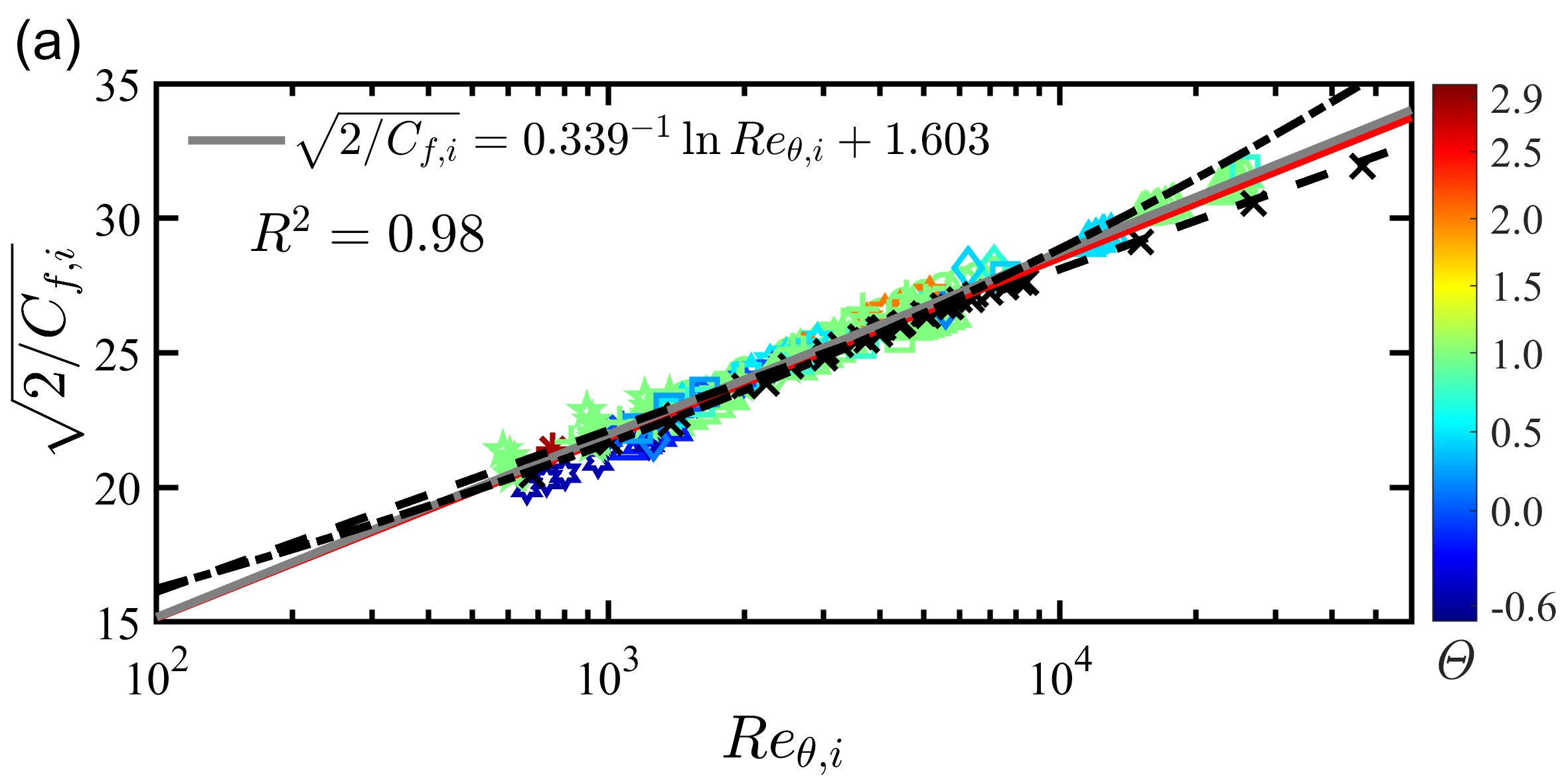}
  \includegraphics[width=0.49\textwidth]{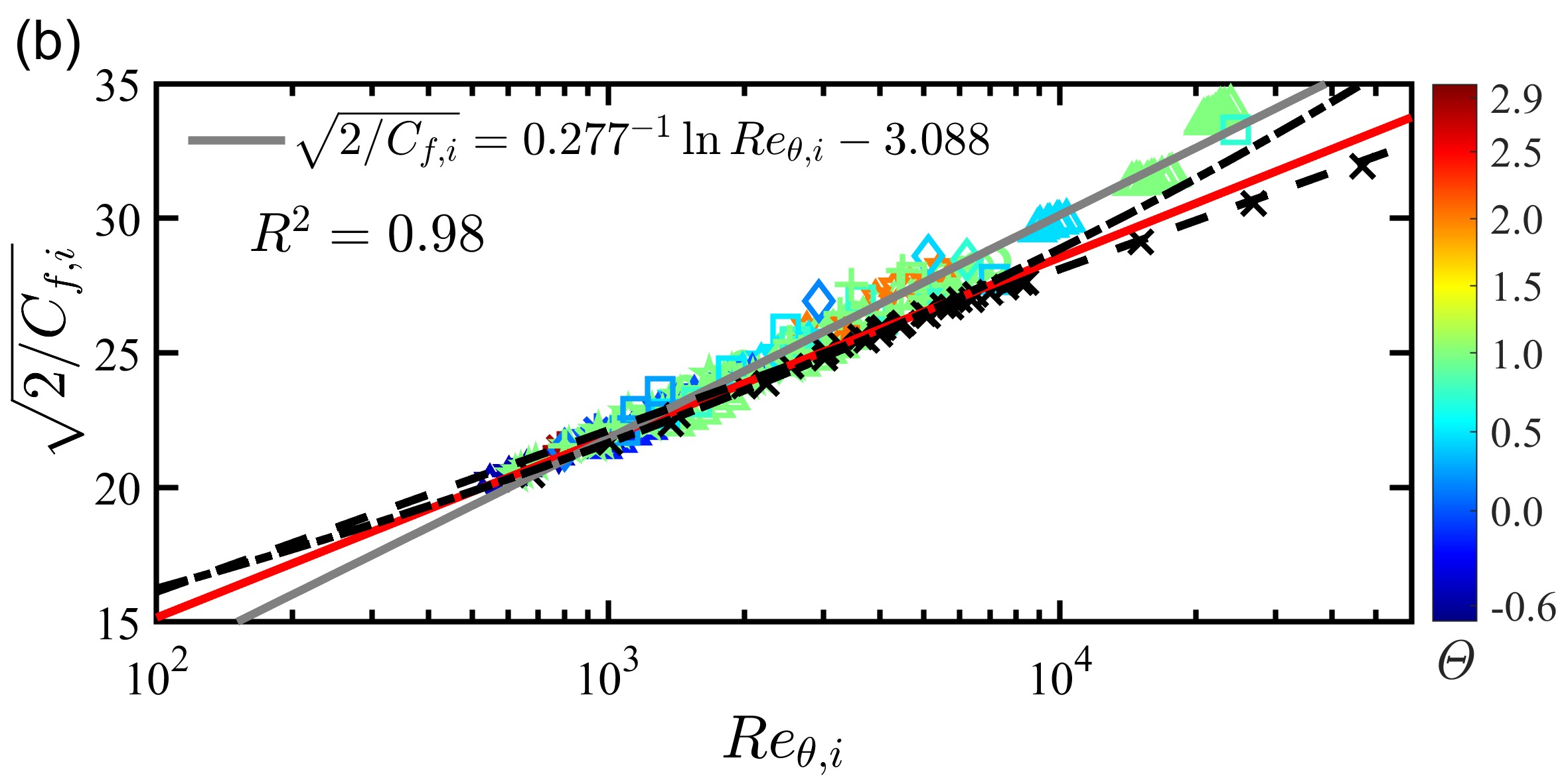}
  \caption{Transformed `incompressible' $\sqrt{2/C_{f,i}}$ versus $Re_{\theta,i}$ in semi-logarithmic coordinates, using the formulation~\eqref{eq:transformation factors of Zhao and Fu} of \citet{zhao2025revisiting} together with the (a) GFM~\eqref{eq:GFM transformation} and (b) VIPL~\eqref{eq:VIPL transformation} transformations. Symbols, colours, and line styles follow those in \hyperref[fig:original vD I, vD II, SC transformations (logarithmic)]{figure~\ref{fig:original vD I, vD II, SC transformations (logarithmic)}}.}
\label{fig:GFM and VIPL transformations (ZF, logarithmic)}
\end{figure}

\hyperref[fig:GFM and VIPL transformations (ZF, linear)]{Figure~\ref{fig:GFM and VIPL transformations (ZF, linear)}} further illustrates the variation of $C_{f,i}$ with $Re_{\theta,i}$, together with the absolute relative errors $\mathcal{E}(C_{f,i})$ evaluated against the composite incompressible reference relation~\eqref{eq:incompressible skin-friction correlation}. When combined with the GFM transformation, the formulation of \citet{zhao2025revisiting} yields a visibly improved collapse of the transformed data, but this reduction in scatter does not translate into a uniformly closer agreement with the incompressible reference. While errors remain within 10\% at moderate-to-high Reynolds numbers, many cases still deviate from the composite incompressible baseline by several per cent, indicating that the data are not tightly centred on the reference relation. At lower Reynolds numbers ($Re_{\theta,i} \lesssim 2000$), more pronounced deviations from the low-Reynolds-number incompressible baseline~\eqref{eq:Smits relation} are observed, particularly for several cases with $M_\infty \geq 1.5$ from the WSKR database \citep{wenzel2018dns}. These deviations contribute to an overall mean error of 4.27\%, which is slightly larger than the corresponding vD~II value of 3.67\% using the same reference baseline. Furthermore, when this formulation is applied in conjunction with the VIPL transformation, the transformed data systematically underpredict $C_{f,i}$ across a substantial portion of the database, especially at $Re_{\theta,i} \gtrsim 2000$, driving the mean error up to 7.23\%.

\begin{figure}
  \centering
  \includegraphics[width=0.49\textwidth]{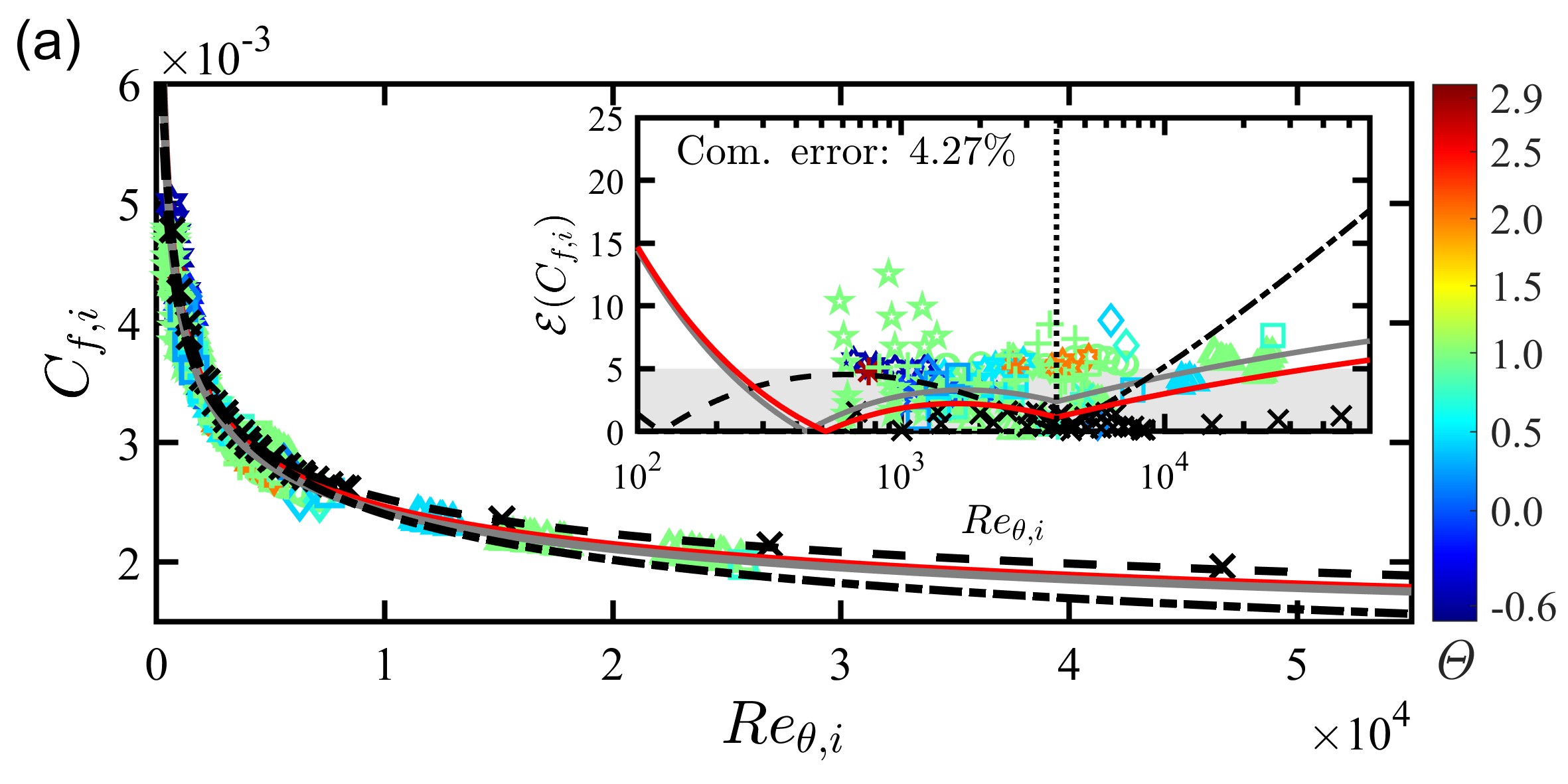}
  \includegraphics[width=0.49\textwidth]{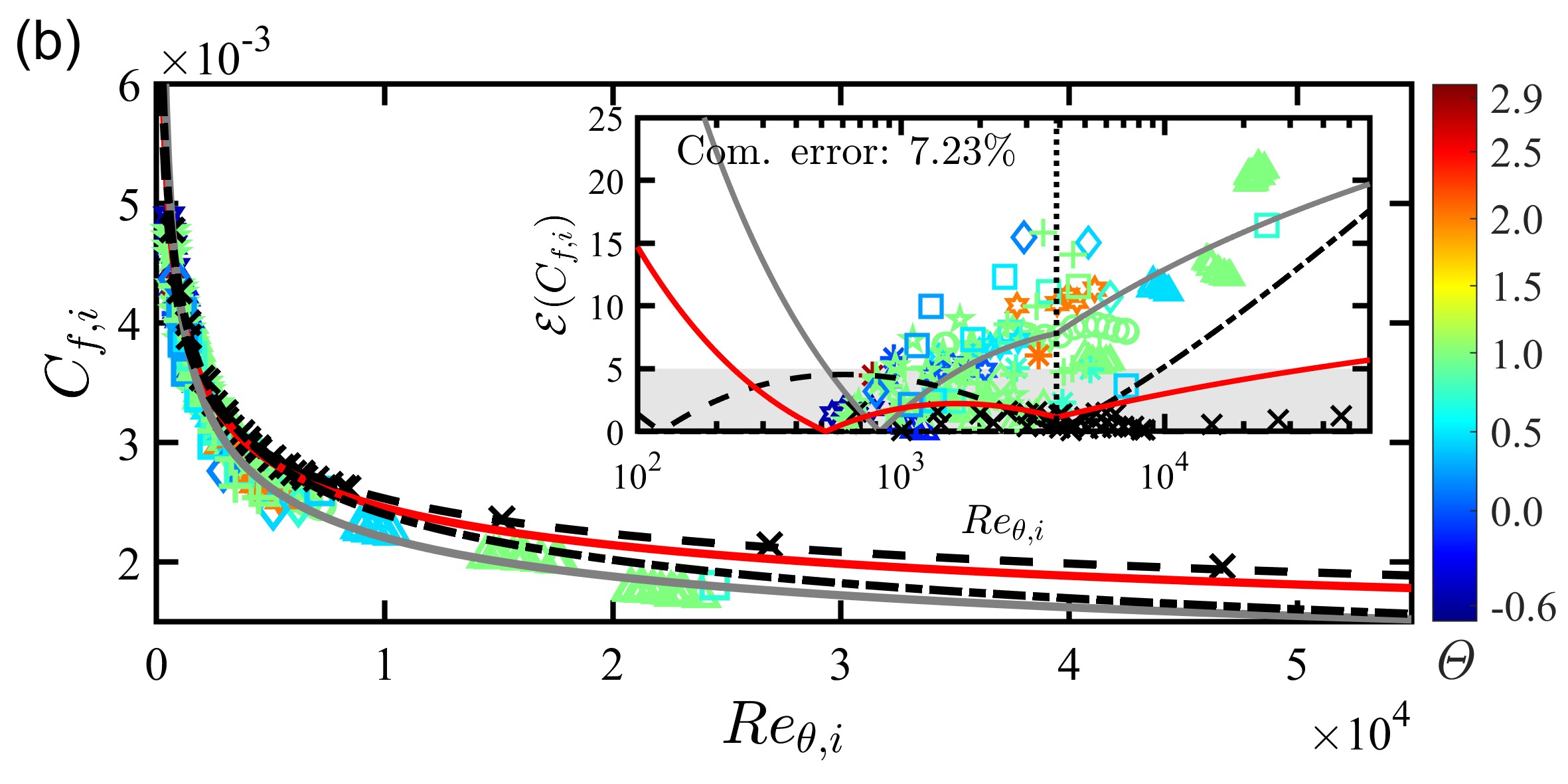}
  \caption{Transformed `incompressible' $C_{f,i}$ versus $Re_{\theta,i}$, using the formulation~\eqref{eq:transformation factors of Zhao and Fu} of \citet{zhao2025revisiting} together with the (a) GFM~\eqref{eq:GFM transformation} and (b) VIPL~\eqref{eq:VIPL transformation} transformations. Insets show the absolute relative errors $\mathcal{E}(C_{f,i})$ defined by \eqref{eq:error against composite incompressible reference}, with the vertical dotted line marking $Re_{\theta,i}^{\mathrm{cr}} \approx 3887$. Symbols, colours, and line styles follow those in \hyperref[fig:original vD I, vD II, SC transformations (logarithmic)]{figure~\ref{fig:original vD I, vD II, SC transformations (logarithmic)}}, while the inset and error-band conventions follow those in \hyperref[fig:original vD I, vD II, SC transformations (linear)]{figure~\ref{fig:original vD I, vD II, SC transformations (linear)}}.}
\label{fig:GFM and VIPL transformations (ZF, linear)}
\end{figure}

The above results confirm that, when combined with the GFM transformation, the formulation of \citet{zhao2025revisiting} yields an improved collapse of the transformed skin-friction data in an \emph{a priori} assessment. Nevertheless, two limitations should be noted. The first is conceptual, concerning whether the transformed momentum-thickness Reynolds number implied by this formulation genuinely removes Mach-number and wall-temperature effects. To see this more clearly, the transformed variables implied by \eqref{eq:transformation factors of Zhao and Fu} are explicitly written as
\begin{equation}
    C_{f,i} = \frac{2}{U_{\mathrm{GFM},\infty}^{+2}}, \quad
    Re_{\theta,i} = U_{\mathrm{GFM},\infty}^{+} \int_0^{Re_{\tau,e}^*} \frac{\bar\rho}{\rho_\infty} \frac{\bar U_{\mathrm{GFM}}^+}{U_{\mathrm{GFM},\infty}^+} \left( 1-\frac{\bar U_{\mathrm{GFM}}^+}{U_{\mathrm{GFM},\infty}^+} \right) \, \mathrm{d}y^* .
    \label{eq:implied transformed variables (Zhao and Fu)}
\end{equation}
Although the GFM transformation is intended to collapse compressible mean velocity profiles onto an `incompressible' counterpart $\bar U_{\mathrm{GFM}}^+(y^*)$, the expression for $Re_{\theta,i}$ explicitly weights this mapped velocity by the physical density ratio $\bar\rho / \rho_\infty$ of the compressible flow. Therefore, even if $\bar U_{\mathrm{GFM}}^+(y^*)$ were perfectly independent of Mach number and wall temperature, residual compressibility effects would be inherently retained in $Re_{\theta,i}$ through this density weighting. In this sense, the improved empirical collapse of $\sqrt{2/C_{f,i}}$ against $\ln Re_{\theta,i}$ does not necessarily imply a genuine recovery of the reference incompressible skin-friction behaviour. The second limitation is practical. Because the quantities $F$ and $\hat{\theta}/\theta$ must be evaluated from the available mean-flow profiles, the transformation factors $(F_C)_{\mathrm{ZF}}$ and $(F_\theta)_{\mathrm{ZF}}$ cannot be deduced solely from prescribed free-stream and wall-thermal inputs, thereby losing a key practical feature of the classical vD and SC approaches. Consequently, while this formulation may serve as a useful tool for \emph{a priori} scaling assessments of existing datasets, it lacks the mathematical closure required to function as a standalone \emph{a posteriori} predictive model for $C_f$ under prescribed macroscopic conditions.

\section{Skin-friction transformations implied by mean-velocity mappings}
\label{sec:section 3}
The reassessment in \textsection~\ref{sec:section 2} shows that the transformed $(C_{f,i},Re_{\theta,i})$ pairs produced by existing skin-friction transformations do not consistently follow the reference ZPG ITBL behaviour. This motivates first specifying the target of the mapping through a clear definition of the transformed `incompressible' state associated with a given physical ZPG CTBL. In the formulation developed below, once this state, together with the mean-velocity and wall-normal-coordinate mappings, is prescribed, $C_{f,i}$ and $Re_{\theta,i}$ are evaluated from their standard incompressible definitions, and the factors $F_C$ and $F_\theta$ follow accordingly. Consequently, these factors are not introduced as additional empirical corrections or nominal-thickness assumptions. This perspective also elucidates how the direct application of these mapping-based relations is inevitably constrained by the inherited outer-layer limitations of existing velocity transformations.

To avoid ambiguity, we first state the notation conventions used below. Lowercase profile and coordinate variables, such as $\bar u$, $y$ and $z=\bar u/u_\infty$, refer to the physical compressible ZPG CTBL, whereas their uppercase counterparts, such as $\bar U$, $Y$ and $Z=\bar U/U_\infty$, refer to the mapped constant-property ZPG ITBL. Global quantities in the mapped state are distinguished from their physical compressible counterparts by the subscript `$i$'. Accordingly, $\theta$, $C_f$ and $Re_\theta$ denote the physical compressible momentum thickness, skin-friction coefficient and momentum-thickness Reynolds number, while $\theta_i$, $C_{f,i}$ and $Re_{\theta,i}$ represent the corresponding quantities obtained from standard incompressible definitions in the mapped state. The nominal thickness $\hat{\theta}$, the semi-local coordinate $y^*$ and the associated $Re_{\tau,e}^*$ retain the definitions introduced in \textsection~\ref{sec:subsection 2.2} for the nominal-thickness construction of \citet{zhao2025revisiting}; these quantities are not part of the mapped-state definitions adopted below.

\subsection{Transformed state and exact transformation factors}
\label{sec:subsection 3.1}
Compressibility transformations have become a well-established approach for mapping a physical compressible mean velocity profile $\bar u^+(y^+)$ onto an `incompressible' analogue $\bar U^+(Y^+)$. In their ideal form, such transformations are intended to establish a one-to-one mapping throughout the entire boundary layer, expressed as \citep{modesti2016reynolds}
\begin{equation}
    Y^+ = \int_0^{y^+}{f_I \mathrm{d} \, y^+}, \quad \bar U^+ = \int_0^{\bar u^+}{g_I \mathrm{d} \, \bar u^+},
    \label{eq:definition of velocity transformation}
\end{equation}
where $f_I$ and $g_I$ denote the kernel functions governing the coordinate and velocity mappings, respectively. Here, the `incompressible' analogue is defined as a hypothetical ZPG ITBL with constant fluid properties. Accordingly, $C_{f,i}$ and $Re_{\theta,i}$ are evaluated from their standard incompressible definitions, and their expected scaling behaviour is represented by the incompressible reference relation~\eqref{eq:incompressible skin-friction correlation}. In the present study, following the wall-referenced convention of \citet{trettel2016mean}, this mapped state is represented using the wall quantities of the corresponding ZPG CTBL, namely $\bar \tau_w$, $\bar \rho_w$, and $\bar \mu_w$. While this choice establishes a common inner normalization for the mapped and physical variables, the derivation below requires only the aforementioned constant-property assumption.

By definition, the skin-friction coefficient in the transformed `incompressible' state is $C_{f,i} = 2 \bar \tau_w / (\bar \rho_w U_\infty^2) = 2 / U_{\infty}^{+2}$, whereas its physical compressible counterpart is $C_f = 2 \bar \tau_w / (\rho_\infty u_\infty^2) = 2 (\bar \rho_w / \rho_\infty) / u_\infty^{+2}$. The transformation factor $F_C$ therefore follows directly as
\begin{equation}
    F_{C}=\frac{C_{f,i}}{C_f}=\frac{{\rho}_{\infty}}{\bar{\rho}_w}F^{-2},
    \label{eq:a general formula for Cf transformation}
\end{equation}
where $F = U_\infty^+ / u_\infty^+ = \int_0^1{g_I \, \mathrm{d}z}$ denotes the free-stream velocity ratio, with $z = \bar u^+ / u_\infty^+$.

The momentum thickness of the transformed constant-property ZPG ITBL is evaluated using the classical incompressible expression,
\begin{equation}
    \theta_i = \delta_\nu \theta_i^+ = \delta_\nu \int_0^{Re_T}{\frac{\bar U^+}{U_\infty^+} \left( 1 - \frac{\bar U^+}{U_\infty^+} \right) \, \mathrm{d}Y^+}
    = \delta_\nu \int_0^{Re_T}{Z(1-Z) \, \mathrm{d}Y^+},
    \label{eq:incompressible momentum thickness}
\end{equation}
where $Z = \bar U^+ / U_\infty^+$ is the normalized mean velocity in the mapped state, and $Re_T = \left. Y^+ \right|_{Y = \Delta_e} = \left. Y^+ \right|_{y = \delta_e}$ denotes the friction Reynolds number evaluated at the transformed boundary-layer edge $Y = \Delta_e$, which corresponds to $y = \delta_e$ in physical space. Importantly, $\theta_i$ is not an arbitrary redefinition of the physical compressible momentum thickness. Rather, it is the conventional momentum thickness of the transformed constant-property ZPG ITBL, for which the density ratio entering the integrand is identically unity. Consequently, the transformation factor $F_\theta$ is directly obtained as the ratio of $Re_{\theta,i} = \bar \rho_w U_\infty \theta_i / \bar \mu_w = U_\infty^+ \theta_i^+$ to $Re_\theta = \rho_\infty u_\infty \theta / \mu_\infty = \left( \rho_{\infty} / \bar{\rho}_w \right) \left( \bar{\mu}_w / \mu_{\infty} \right) u_{\infty}^{+}\theta ^+$:
\begin{equation}
    F_{\theta}=\frac{Re_{\theta,i}}{Re_\theta}=\frac{\bar \rho_w \mu_{\infty}}{\rho_{\infty} \bar \mu_w } F \frac{\theta_i^+}{\theta^+} = \frac{\bar\rho_w \mu_{\infty}}{\rho_{\infty} \bar \mu_w} F \frac{\theta_i}{\theta}.
    \label{eq:a general formula for Re_theta transformation}
\end{equation}
The final equality holds because the wall-referenced convention implies a shared viscous length scale $\delta_\nu$ for both the physical and mapped states. Fundamentally, however, for a prescribed velocity transformation \eqref{eq:definition of velocity transformation}, $C_{f,i} = 2 / U_\infty^{+2}$ and $Re_{\theta,i} = U_\infty^+ \theta_i^+$ are entirely determined by the mapped variables $\bar U^+$ and $Y^+$. Thus, the resulting transformation factors are not tied to the specific dimensional wall quantities ($\bar \tau_w$, $\bar \rho_w$, and $\bar \mu_w$) used to parameterize the hypothetical constant-property target state.

\subsection{Application to existing velocity transformations and inherited limitations}
\label{sec:subsection 3.2}
The transformation factors derived in \textsection~\ref{sec:subsection 3.1} are exact consequences of prescribed mean-velocity and wall-normal-coordinate mappings, and do not involve additional empirical modelling assumptions. In practice, however, no existing velocity transformation can be expected to realize this ideal full-layer mapping perfectly. We therefore examine the direct application of \eqref{eq:a general formula for Cf transformation} and \eqref{eq:a general formula for Re_theta transformation} by substituting $Y^+$ and $\bar U^+$ supplied by two representative modern velocity transformations, namely the GFM~\eqref{eq:GFM transformation} and the VIPL~\eqref{eq:VIPL transformation}. The GFM is selected because it is the baseline mapping adopted in the nominal-thickness construction of \citet{zhao2025revisiting}, whereas the VIPL provides an independent modern velocity mapping. Both transformations have been shown to collapse ZPG CTBL mean velocity profiles reasonably well onto the incompressible law of the wall within the inner layer \citep{larsson2025turbulence}; together, they provide a focused test of how the induced skin-friction transformation depends on the underlying velocity mapping.

\begin{figure}
  \centering
  \includegraphics[width=0.49\textwidth]{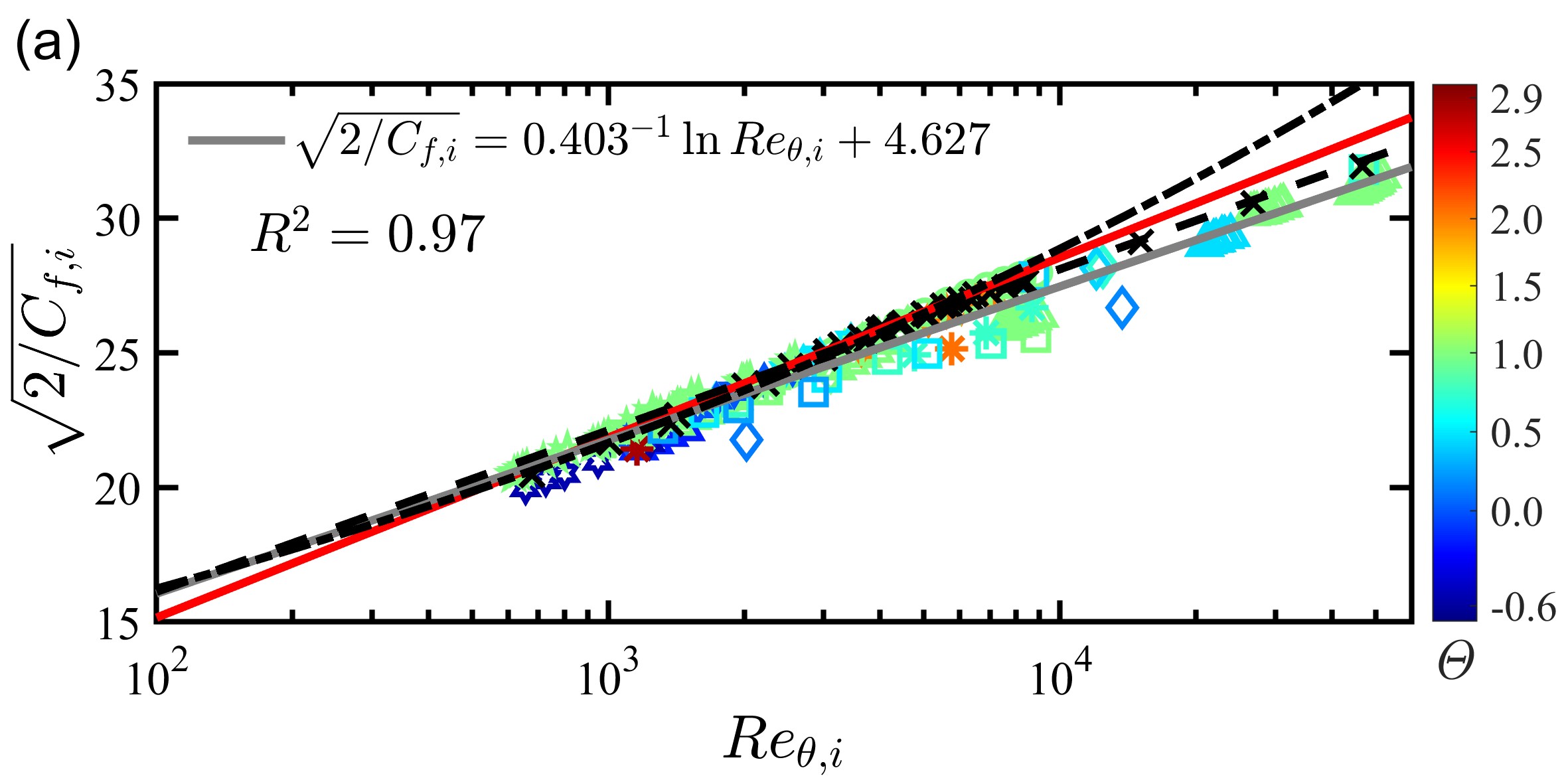}
  \includegraphics[width=0.49\textwidth]{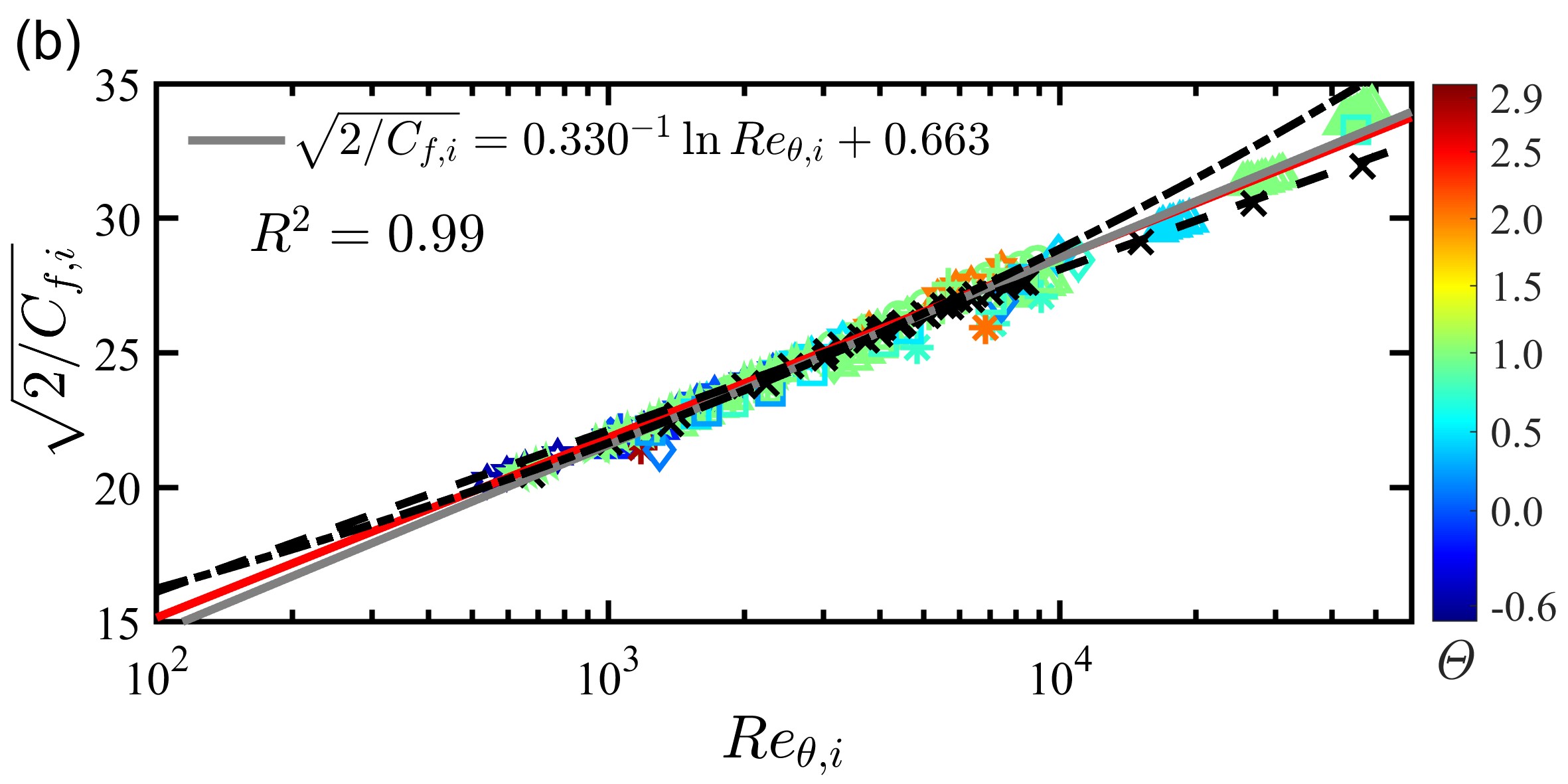}
  \caption{Transformed `incompressible' $\sqrt{2/C_{f,i}}$ versus $Re_{\theta,i}$ in semi-logarithmic coordinates, using the present mapping-based formulation~\eqref{eq:a general formula for Cf transformation} and \eqref{eq:a general formula for Re_theta transformation}, together with the (a) GFM~\eqref{eq:GFM transformation} and (b) VIPL~\eqref{eq:VIPL transformation} transformations. Plotting conventions follow those in \hyperref[fig:GFM and VIPL transformations (ZF, logarithmic)]{figure~\ref{fig:GFM and VIPL transformations (ZF, logarithmic)}}.}
\label{fig:GFM and VIPL transformations (present without modification, logarithmic)}
\end{figure}

\begin{figure}
  \centering
  \includegraphics[width=0.49\textwidth]{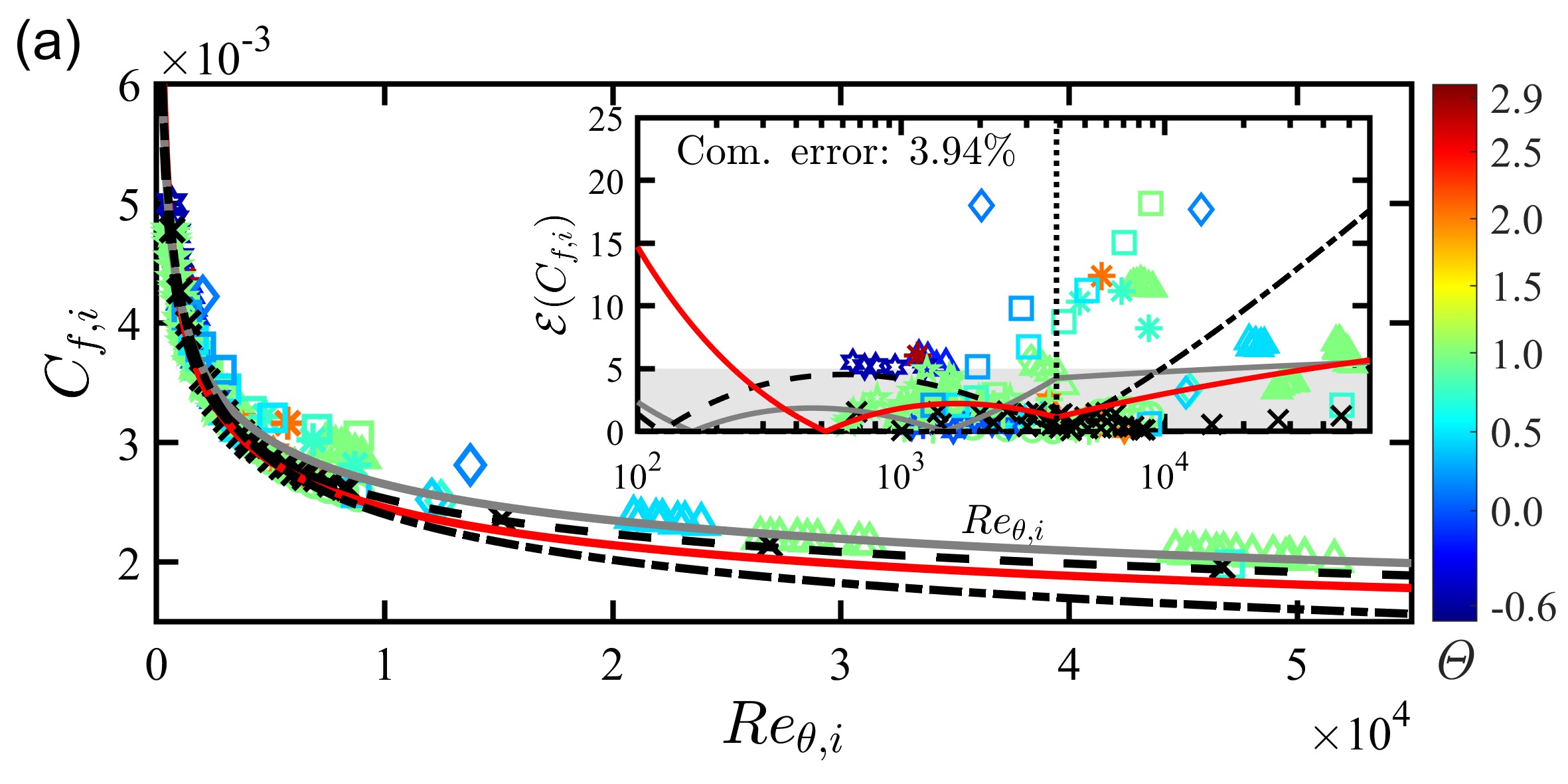}
  \includegraphics[width=0.49\textwidth]{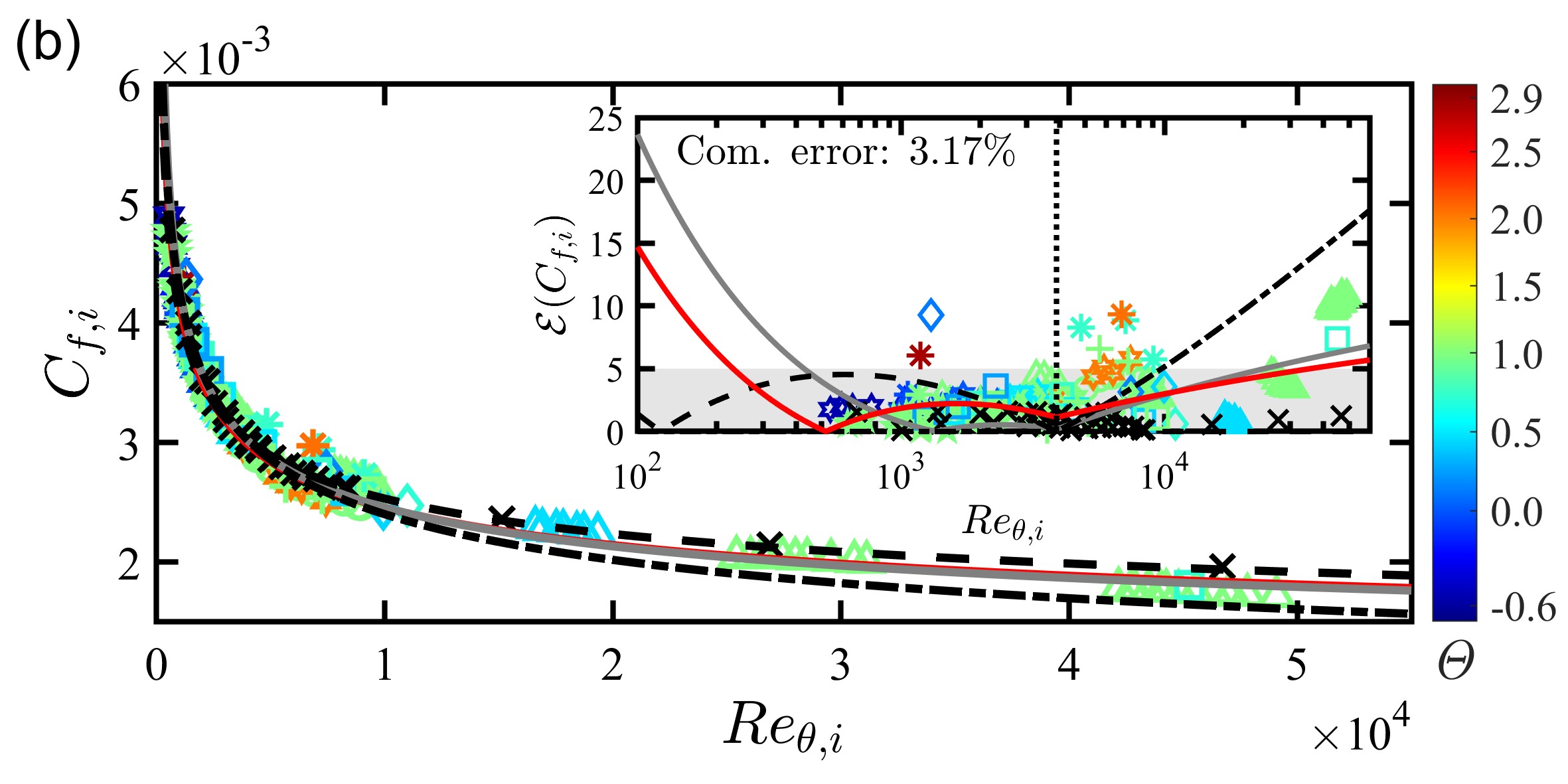}
  \caption{Transformed `incompressible' $C_{f,i}$ versus $Re_{\theta,i}$, using the present mapping-based formulation~\eqref{eq:a general formula for Cf transformation} and \eqref{eq:a general formula for Re_theta transformation}, together with the (a) GFM~\eqref{eq:GFM transformation} and (b) VIPL~\eqref{eq:VIPL transformation} transformations. Insets show the absolute relative errors $\mathcal{E}(C_{f,i})$ defined by \eqref{eq:error against composite incompressible reference}, with the vertical dotted line marking $Re_{\theta,i}^{\mathrm{cr}} \approx 3887$. Plotting, inset, and error-band conventions follow those in \hyperref[fig:GFM and VIPL transformations (ZF, linear)]{figure~\ref{fig:GFM and VIPL transformations (ZF, linear)}}.}
\label{fig:GFM and VIPL transformations (present without modification, linear)}
\end{figure}

The results are presented in \hyperref[fig:GFM and VIPL transformations (present without modification, logarithmic)]{figure~\ref{fig:GFM and VIPL transformations (present without modification, logarithmic)}} and \hyperref[fig:GFM and VIPL transformations (present without modification, linear)]{figure~\ref{fig:GFM and VIPL transformations (present without modification, linear)}}. Compared with the corresponding results obtained using the nominal-thickness construction in \hyperref[fig:GFM and VIPL transformations (ZF, logarithmic)]{figure~\ref{fig:GFM and VIPL transformations (ZF, logarithmic)}} and \hyperref[fig:GFM and VIPL transformations (ZF, linear)]{figure~\ref{fig:GFM and VIPL transformations (ZF, linear)}}, the transformed data exhibit a tighter collapse and closer agreement with the ZPG ITBL reference when the VIPL transformation is employed. This improvement is particularly evident at low-to-moderate $Re_{\theta,i}$, where most transformed data points remain close to the composite incompressible reference relation~\eqref{eq:incompressible skin-friction correlation}. The fitted logarithmic scaling yields $R^2=0.99$, and the mean relative error evaluated with respect to \eqref{eq:incompressible skin-friction correlation} is 3.17\%, with the majority of deviations remaining within 5\%. This error is lower than the 4.27\% value obtained by combining \eqref{eq:transformation factors of Zhao and Fu} with the GFM transformation. Nevertheless, noticeable scatter persists for several diabatic cases, especially those involving high Mach numbers and appreciable wall heat transfer. Moreover, a systematic underprediction of $C_{f,i}$ becomes apparent at high Reynolds numbers ($Re_{\theta,i} \gtrsim 10^4$), indicating that the incompressible reference behaviour is not uniformly recovered. In contrast, the direct use of the GFM transformation within the mapping-based formulation leads to a less satisfactory collapse and larger deviations relative to the incompressible baseline, yielding a lower $R^2=0.97$ and a mean relative error of 3.94\%. While cases with low-to-moderate Mach numbers ($M_\infty \leq 3$) and quasi-adiabatic walls, as well as those with weak cooling or heating, remain in reasonable agreement with the incompressible reference, several cases characterized by high Mach numbers or appreciable wall heat transfer exhibit a pronounced overprediction of $C_{f,i}$.

Although the relations~\eqref{eq:a general formula for Cf transformation} and \eqref{eq:a general formula for Re_theta transformation} follow exactly from the prescribed mappings, their direct application necessarily inherits any inaccuracies in the mapped variables supplied by the chosen velocity transformation. To identify the origin of the deviations observed above, we compare the GFM- and VIPL-transformed mean velocity profiles with the corresponding theoretically expected `incompressible' reference profiles. For a given velocity transformation, the transformed friction Reynolds number $Re_T = \left. Y^+ \right|_{Y=\Delta_e} = \left. Y^+ \right|_{y=\delta_e}$ characterizes the mapped ZPG ITBL. The theoretically expected `incompressible' reference profile $\bar U_{\mathrm{th}}^+(Y^+)$ within $Y \in [0,\Delta_e]$ may be represented by a composite law of the wall and law of the wake as \citep{manzoor2024estimating}
\begin{equation}
    \bar U_{\mathrm{th}}^+ (Y^+) = \int_0^{Y^+}{\frac{\mathrm{d}Y^+}{1+\bar{\mu}_{t,\mathrm{in}}/\bar{\mu}_w}} + \frac{2\Pi_{\mathrm{in}}}{\kappa} \sin ^2\left( \frac{\pi}{2} \frac{Y^+}{Re_T} \right),
    \label{eq:incompressible mean velocity profiles}
\end{equation}
where the `incompressible' eddy viscosity is modelled using the Johnson-King (JK) formulation \citep{johnson1985mathematically,hasan2023incorporating}, $\bar \mu_{t,\mathrm{in}} / \bar \mu_w = \kappa Y^+ [1 - \exp(-Y^+/A^+)]^2$, with $\kappa = 0.41$ and $A^+ = 17$. Although alternative formulations are available \citep{manzoor2024estimating,appelbaum2025onset}, the wake parameter $\Pi_{\mathrm{in}}$ is modelled here using the Cebeci–Smith relation \citep{cebeci1974analysis} recalibrated by \citet{wenzel2018dns},
\begin{equation}
    \Pi_{\mathrm{in}} = 0.66\left[ 1 - \exp (-0.4\sqrt{\xi} - 0.48\xi) \right], \quad \xi = Re_{\theta,i}/1000 .
    \label{eq:modified C-S relation by wenzel}
\end{equation}
The corresponding theoretically expected edge velocity is then obtained by evaluating \eqref{eq:incompressible mean velocity profiles} at $Y^+ = Re_T$, yielding
\begin{equation}
    \bar U_{\mathrm{th},e}^+ = \int_0^{Re_T}{\frac{\mathrm{d}Y^+}{1+\bar{\mu}_{t,\mathrm{in}}/\bar{\mu}_w}} + \frac{2\Pi_{\mathrm{in}}}{\kappa}.
\label{eq:mean velocity at the transformed boundary layer edge}
\end{equation}

\begin{figure}
  \centering
  \includegraphics[width=0.32\textwidth]{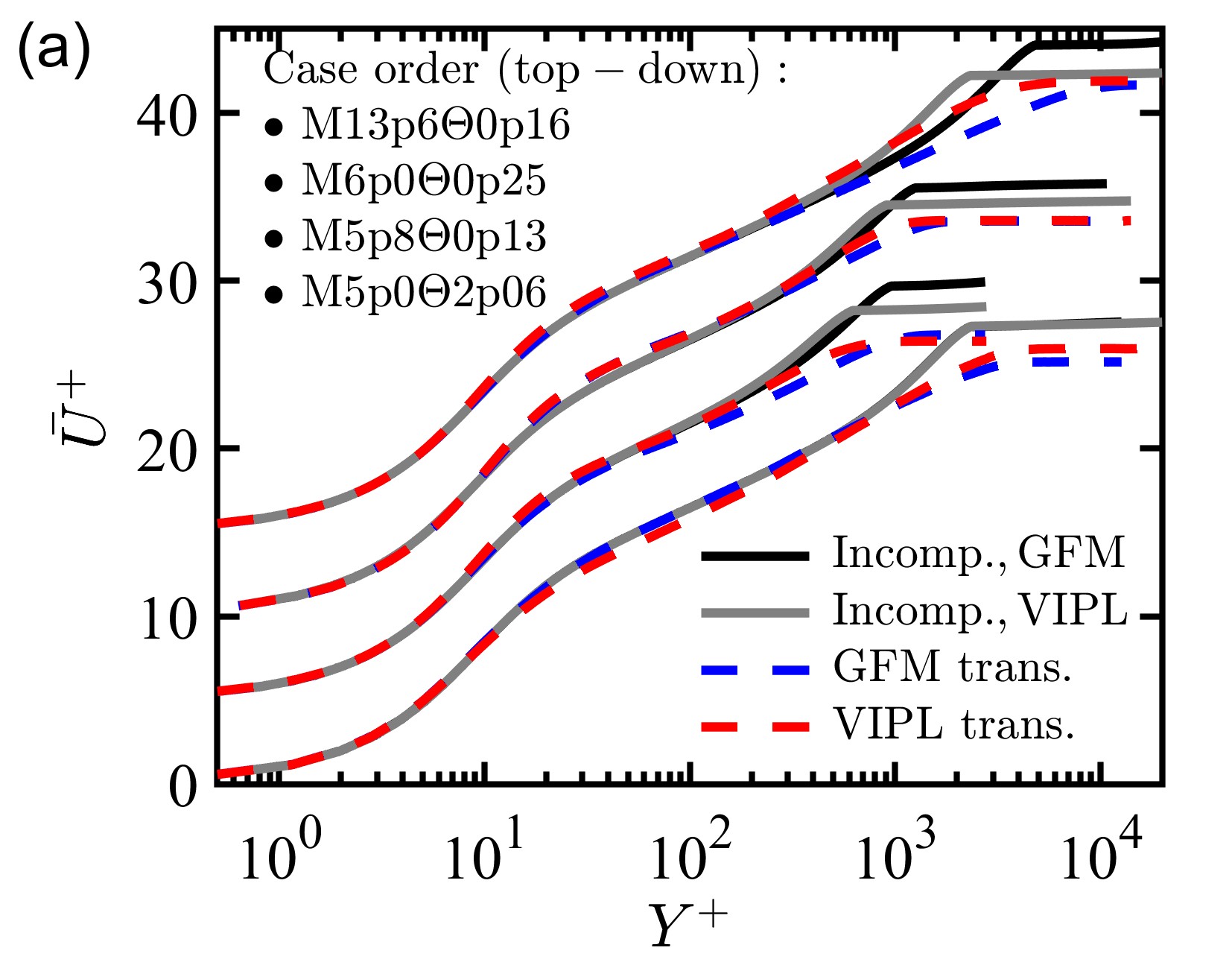}
  \includegraphics[width=0.32\textwidth]{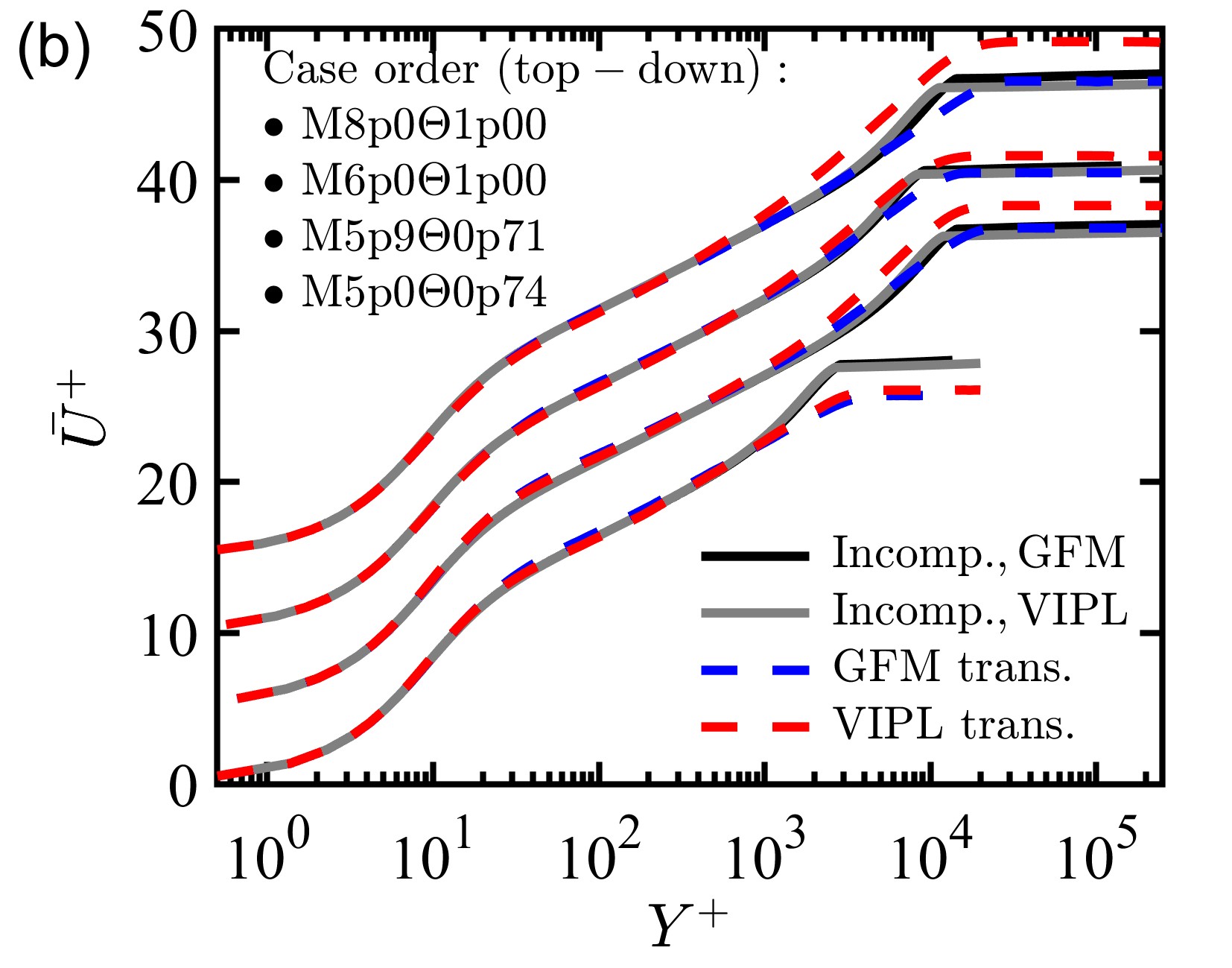}
  \includegraphics[width=0.32\textwidth]{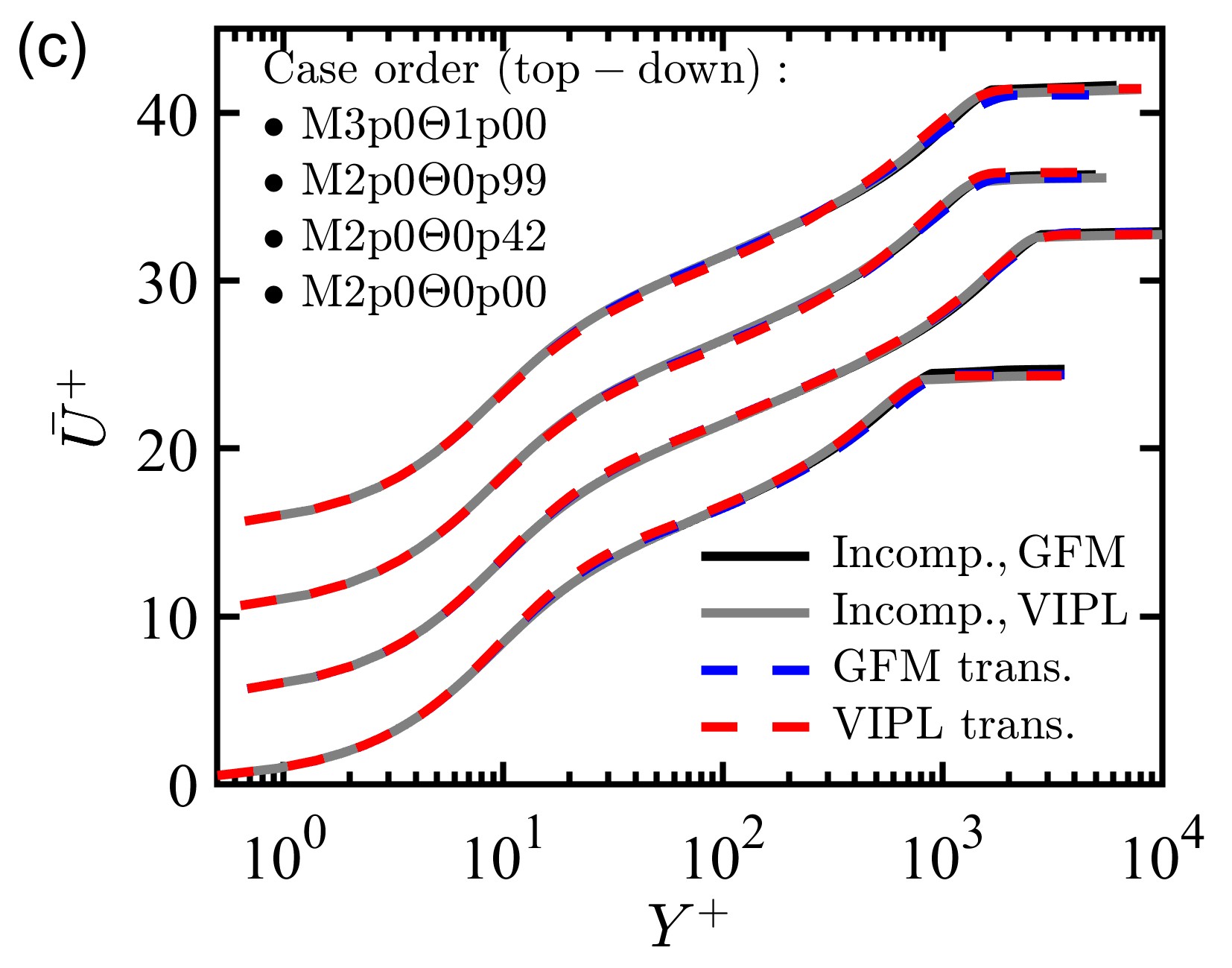}
  \caption{Comparison of transformed mean velocity profiles using the GFM~\eqref{eq:GFM transformation} (blue dashed lines) and VIPL~\eqref{eq:VIPL transformation} (red dashed lines) transformations against their respective `incompressible' references for (a) high-Mach-number cases with moderately cooled and heated walls, (b) high-Mach-number cases with weakly cooled and quasi-adiabatic walls, and (c) low-to-moderate Mach-number cases. The `incompressible' reference profiles over $Y \in [0, \Delta_e]$, shown by black and grey solid lines, are obtained by evaluating \eqref{eq:incompressible mean velocity profiles} with $Y^+ = y^*$ and $Y^+ = y_{\mathrm{V}}^+$, respectively. Outside the boundary layer, the reference profiles are linearly extrapolated from the boundary-layer edge value $\bar U_{\mathrm{th},e}^+$ \eqref{eq:mean velocity at the transformed boundary layer edge} to the free-stream value $U_{\mathrm{th},\infty}^+ = \bar U_{\mathrm{th},e}^+ / 0.99$. The cases shown in each panel are indicated in the upper-left corner (from top to bottom) and are listed in detail in \hyperref[tab:the ZPG CTBLs database 1]{tables~\ref{tab:the ZPG CTBLs database 1}-\ref{tab:the ZPG CTBLs database 2}}.}
\label{fig:GFM and VIPL transformed profiles}
\end{figure}

Existing velocity transformations were developed primarily to collapse compressible mean velocity profiles onto the incompressible law of the wall within the inner layer, whereas their outer-layer behaviour is less tightly constrained. \hyperref[fig:GFM and VIPL transformed profiles]{Figure~\ref{fig:GFM and VIPL transformed profiles}} illustrates this inherited limitation by comparing the GFM- and VIPL-transformed mean velocity profiles for representative cases against their theoretically expected `incompressible' references given by \eqref{eq:incompressible mean velocity profiles} and \eqref{eq:modified C-S relation by wenzel}. Both transformations accurately recover the law of the wall below the logarithmic region. Furthermore, their outer-layer behaviour remains reasonably accurate for low-to-moderate Mach number flows, despite not being explicitly targeted during their development. Substantial discrepancies arise, however, in high-Mach-number flows with appreciable wall heat transfer. Specifically, both transformations exhibit pronounced underprediction in the outer layer for cases with moderate-to-strong wall cooling or heating, with the discrepancy being more severe for the GFM transformation, as shown in panel (a). Similar issues persist in high-Mach-number cases with quasi-adiabatic or weakly cooled walls, as evident in panel (b). These profile-level discrepancies account for the scaling departures observed in \hyperref[fig:GFM and VIPL transformations (present without modification, logarithmic)]{figure~\ref{fig:GFM and VIPL transformations (present without modification, logarithmic)}} and \hyperref[fig:GFM and VIPL transformations (present without modification, linear)]{figure~\ref{fig:GFM and VIPL transformations (present without modification, linear)}}, since the directly transformed skin-friction coefficient is determined by the mapped free-stream velocity, i.e., $C_{f,i} = 2 / U_{\mathrm{GFM},\infty}^{+2}$ or $C_{f,i} = 2 / U_{\mathrm{V},\infty}^{+2}$. For instance, the four cases shown in \hyperref[fig:GFM and VIPL transformed profiles]{figure~\ref{fig:GFM and VIPL transformed profiles}(a)} exhibit a downward shift in $U_\infty^+$ that directly translates to the commensurate downward shift of $\sqrt{2 / C_{f,i}}$ (see \hyperref[fig:GFM and VIPL transformations (present without modification, logarithmic)]{figure~\ref{fig:GFM and VIPL transformations (present without modification, logarithmic)}}), whereas the three high-Reynolds-number cases in \hyperref[fig:GFM and VIPL transformed profiles]{figure~\ref{fig:GFM and VIPL transformed profiles}(b)} display an upward deviation in $U_{V,\infty}^+$ that directly drives the overprediction trends in \hyperref[fig:GFM and VIPL transformations (present without modification, logarithmic)]{figure~\ref{fig:GFM and VIPL transformations (present without modification, logarithmic)}(b)}. In contrast, cases at low-to-moderate Mach numbers exhibit consistently good agreement. The systematic biases observed in the skin-friction plots can thus be traced to the inherited outer-layer inaccuracies of the underlying velocity transformations rather than to the mapping-based relations themselves.

To further isolate the origin of the scaling errors, a controlled assessment is performed. Specifically, the mapped free-stream velocity $U_\infty^+ = \int_0^{u_\infty^+}{g_I \, \mathrm{d}\bar u^+}$ yielded by the velocity transformation is replaced with its theoretically expected counterpart, $U_{\mathrm{th},\infty}^+ = \bar U_{\mathrm{th},e}^+ / 0.99$, where $\bar U_{\mathrm{th},e}^+$ is obtained from \eqref{eq:mean velocity at the transformed boundary layer edge}. The reconstructed quantities are then evaluated as $C_{f,i} = 2 / U_{\mathrm{th},\infty}^{+2}$ and $Re_{\theta,i} = U_{\mathrm{th},\infty}^+ \theta_i^+$. Crucially, $\theta_i^+$ is kept unchanged from the direct transformation, namely it is still computed from the mapped variables $Y^+$ and $Z=\bar U^+/U_{\infty}^+$ in \eqref{eq:incompressible momentum thickness}. This diagnostic replacement therefore separates the scaling errors caused by the mapped free-stream velocity from those inherent to the momentum-thickness integral.

\begin{figure}
  \centering
  \includegraphics[width=0.49\textwidth]{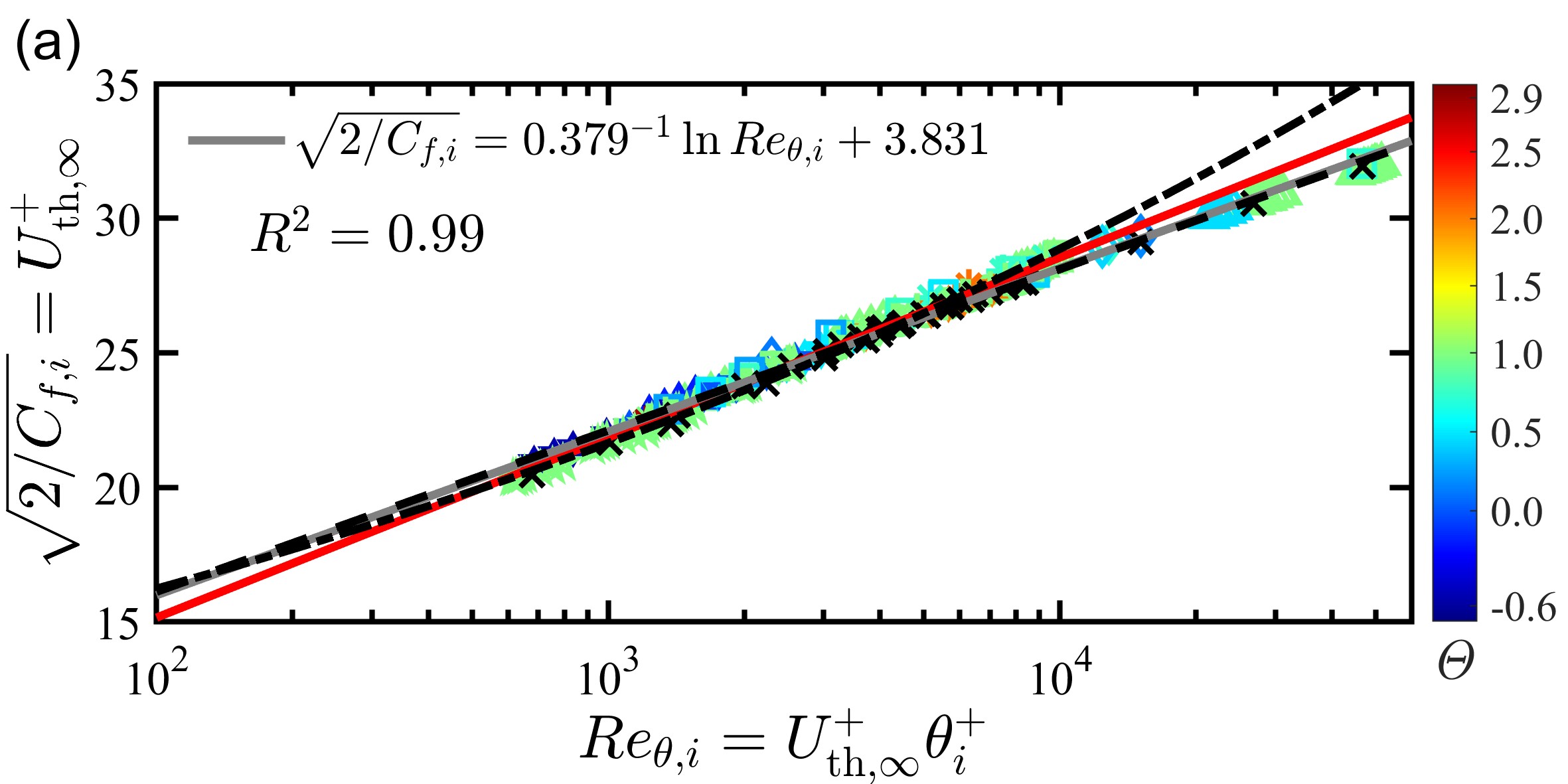}
  \includegraphics[width=0.49\textwidth]{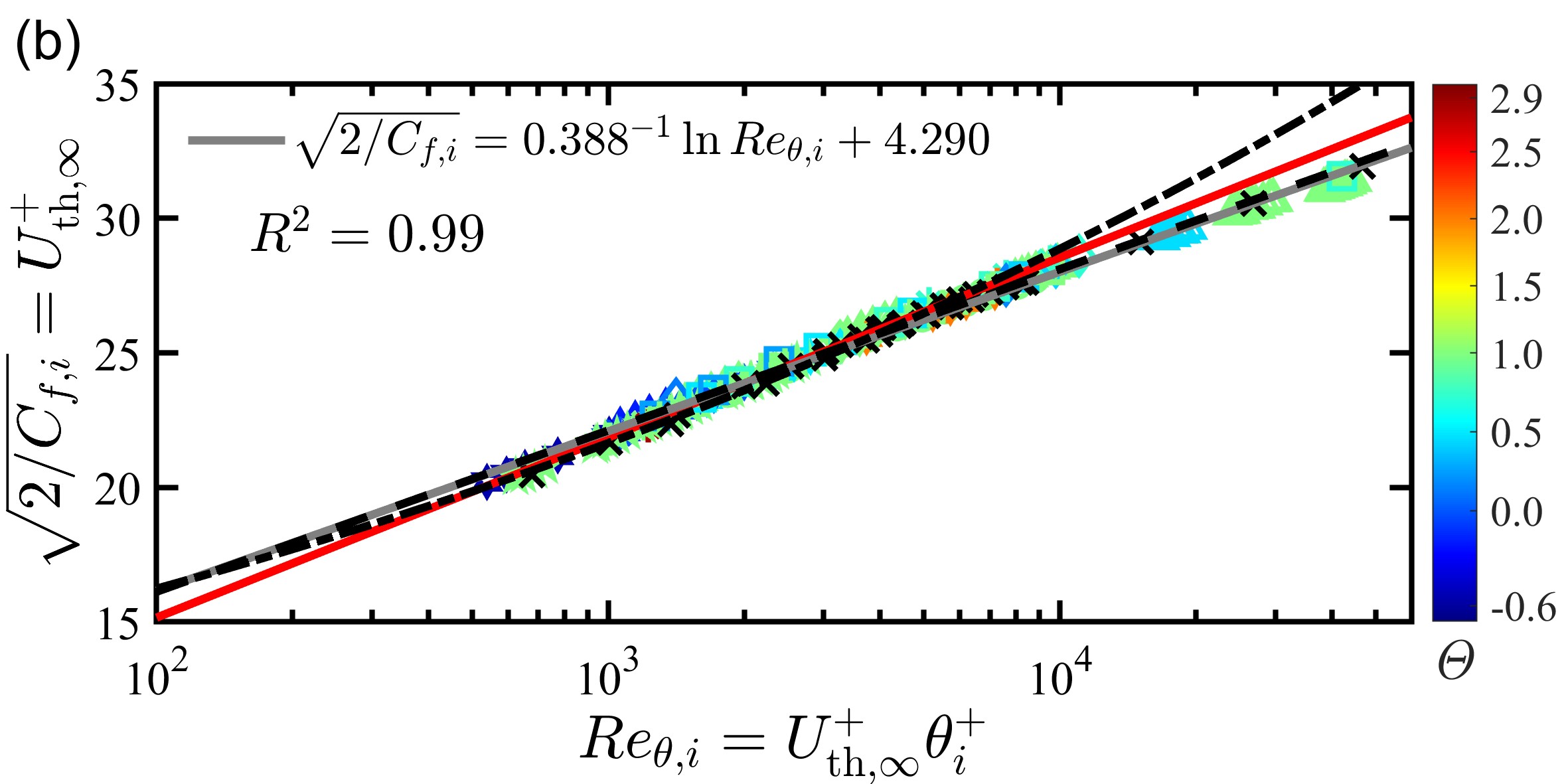}
  \caption{Reconstructed `incompressible' $\sqrt{2/C_{f,i}} = U_{\mathrm{th},\infty}^+$ versus $Re_{\theta,i} = U_{\mathrm{th},\infty}^+ \theta_i^+$ in semi-logarithmic coordinates. The corrected free-stream velocity is prescribed as $U_{\mathrm{th},\infty}^+ = \bar U_{\mathrm{th},e}^+ / 0.99$, with $\bar U_{\mathrm{th},e}^+$ obtained from \eqref{eq:mean velocity at the transformed boundary layer edge}, whereas $\theta_i^+$ is retained from the direct transformation through \eqref{eq:incompressible momentum thickness}. Panels (a) and (b) correspond to the GFM and VIPL transformations, respectively. Plotting conventions follow those in \hyperref[fig:GFM and VIPL transformations (ZF, logarithmic)]{figure~\ref{fig:GFM and VIPL transformations (ZF, logarithmic)}}.}
\label{fig:GFM and VIPL transformations (present with modification, logarithmic)}
\end{figure}

\begin{figure}
  \centering
  \includegraphics[width=0.49\textwidth]{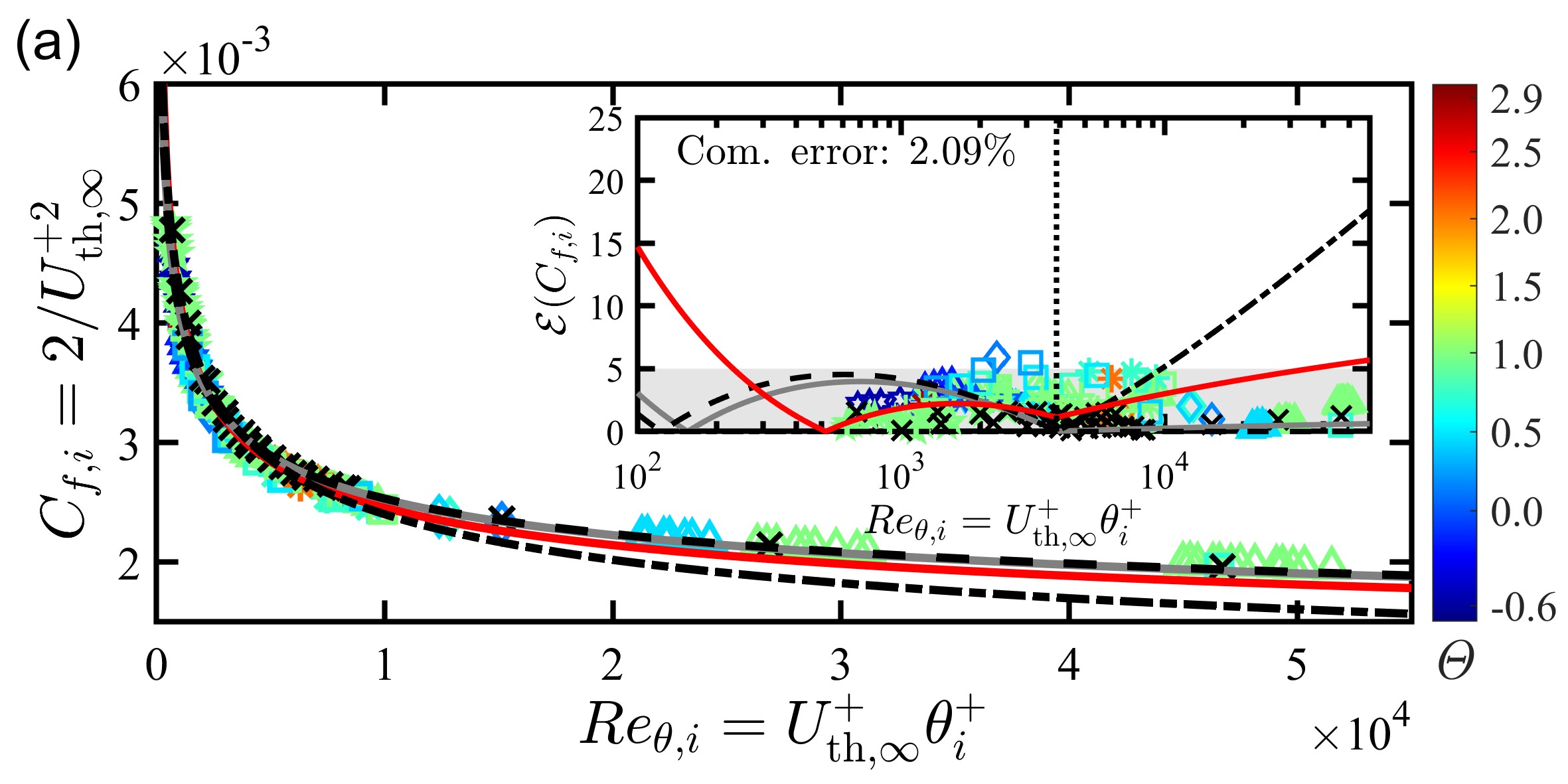}
  \includegraphics[width=0.49\textwidth]{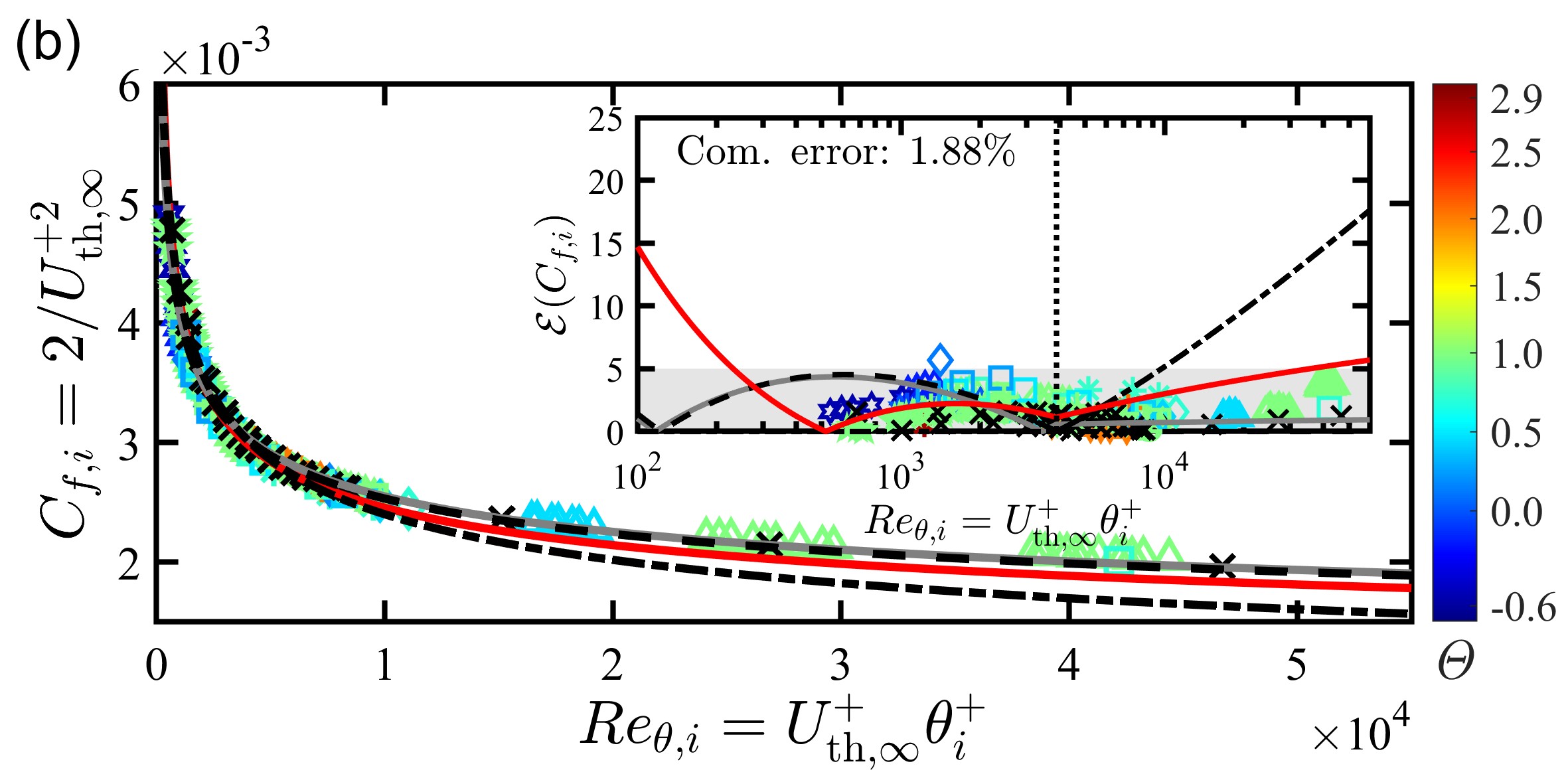}
  \caption{Reconstructed `incompressible' $C_{f,i} = 2 / U_{\mathrm{th},\infty}^{+2}$ versus $Re_{\theta,i} = U_{\mathrm{th},\infty}^+ \theta_i^+$. The quantities $U_{\mathrm{th},\infty}^+$ and $\theta_i^+$ are defined consistently with \hyperref[fig:GFM and VIPL transformations (present with modification, logarithmic)]{figure~\ref{fig:GFM and VIPL transformations (present with modification, logarithmic)}}. Insets show the absolute relative errors $\mathcal{E}(C_{f,i})$ defined by \eqref{eq:error against composite incompressible reference}, with the vertical dotted line marking $Re_{\theta,i}^{\mathrm{cr}} \approx 3887$. Plotting, inset, and error-band conventions follow those in \hyperref[fig:GFM and VIPL transformations (ZF, linear)]{figure~\ref{fig:GFM and VIPL transformations (ZF, linear)}}.}
\label{fig:GFM and VIPL transformations (present with modification, linear)}
\end{figure}

The results of this free-stream-corrected assessment, corresponding to the original evaluations in \hyperref[fig:GFM and VIPL transformations (present without modification, logarithmic)]{figure~\ref{fig:GFM and VIPL transformations (present without modification, logarithmic)}} and \hyperref[fig:GFM and VIPL transformations (present without modification, linear)]{figure~\ref{fig:GFM and VIPL transformations (present without modification, linear)}}, are presented in \hyperref[fig:GFM and VIPL transformations (present with modification, logarithmic)]{figure~\ref{fig:GFM and VIPL transformations (present with modification, logarithmic)}} and \hyperref[fig:GFM and VIPL transformations (present with modification, linear)]{figure~\ref{fig:GFM and VIPL transformations (present with modification, linear)}}, respectively. For both the GFM and VIPL transformations, replacing only the mapped free-stream velocity yields a nearly universal collapse of $\sqrt{2/C_{f,i}}=U_{\mathrm{th},\infty}^+$ against $Re_{\theta,i}=U_{\mathrm{th},\infty}^+\theta_i^+$, with little apparent dependence on Mach number and wall-thermal condition. The fitted logarithmic scalings give $R^2 = 0.99$ in both cases, in close agreement with the CF relation~\eqref{eq:Coles-Fernholz relation}. Consistently, the reconstructed $C_{f,i} = 2 / U_{\mathrm{th},\infty}^{+2}$ decreases monotonically with increasing $Re_{\theta,i}$ and remains in satisfactory agreement with the reference incompressible relation~\eqref{eq:incompressible skin-friction correlation}. As indicated in the insets, most relative errors remain within 5\%, with the mean errors substantially reduced to 2.09\% and 1.88\% for the GFM and VIPL transformations, respectively.

These controlled results confirm that the deviations observed in the direct application arise primarily from inaccuracies in the mapped free-stream velocity $U_\infty^+$, rather than from the mapping-based definitions of $C_{f,i}$ and $Re_{\theta,i}$ themselves. Once $U_\infty^+$ is replaced by $U_{\mathrm{th},\infty}^+$, the inner-scaled momentum thickness $\theta_i^+$ obtained from the GFM and VIPL mapped profiles proves sufficiently accurate, within the present dataset, to recover the incompressible skin-friction scaling. This replacement should nevertheless be interpreted only as a diagnostic test, rather than a closed predictive correction, since $U_{\mathrm{th},\infty}^+$ is prescribed \emph{a priori} from the theoretically expected incompressible reference profile. The practical limitation of the direct transformation therefore lies primarily in the inherited outer-layer inaccuracies of existing velocity transformations, especially under high-Mach-number conditions with appreciable wall heat transfer. This also indicates that future improvements in the outer-layer consistency of velocity transformations would directly improve the induced skin-friction transformation; in the limiting case of an ideal full-layer mapping, the corresponding incompressible skin-friction scaling would be recovered by construction from the definitions of $C_{f,i}$ and $Re_{\theta,i}$.

\subsection{Distinctions from the nominal-thickness construction of \citet{zhao2025revisiting}}
\label{sec:subsection 3.3}
Both the present mapping-based formulation and the nominal-thickness construction of \citet{zhao2025revisiting} aim to use a prescribed velocity transformation to map the compressible pair $(C_f,Re_\theta)$ to transformed `incompressible' counterparts $(C_{f,i},Re_{\theta,i})$, which are expected to be characterized by the well-established skin-friction scaling of ZPG ITBLs. When cast into the framework of \eqref{eq:skin friction transformation}, the resulting skin-friction factor $F_C$ is identical in the two formulations. The essential distinction lies instead in the factor $F_\theta$, or more fundamentally, in how the transformed momentum-thickness Reynolds number $Re_{\theta,i}$ is defined.

The construction of \citet{zhao2025revisiting} was motivated by the observation that a single linear transformation factor, such as those employed in the classical vD and SC theories, cannot fully account for the nonlinear dependence of the compressible momentum thickness $\theta$ on the mean velocity profile. They therefore heuristically modified $\theta$ to form a nominal thickness $\hat{\theta}$~\eqref{eq:definition of theta_hat} by replacing the compressible velocity ratio $z = \bar u / u_\infty$ and wall-normal coordinate $y^+$ in the integrand with their GFM-transformed counterparts $Z_{\mathrm{GFM}} = \bar U_{\mathrm{GFM}}^+/U_{\mathrm{GFM},\infty}^+$ and $y^*$, while retaining the wall-normal density ratio $\bar\rho/\rho_\infty$. The resulting $\hat{\theta}$ is thus a hybrid mathematical construct, mixing transformed `incompressible' kinematics with compressible density weighting, and does not correspond to the standard momentum thickness of either the physical ZPG CTBL or a specified constant-property mapped ZPG ITBL. The transformed momentum-thickness Reynolds number implied by this nominal thickness is $Re_{\theta,i}=U_{\mathrm{GFM},\infty}^+\hat{\theta}^+$, whose explicit form is given in \eqref{eq:implied transformed variables (Zhao and Fu)}. As discussed in \textsection~\ref{sec:subsection 2.2}, it does not guarantee the complete removal of Mach-number and wall-temperature effects.

The present mapping-based formulation proceeds from a different standpoint. Rather than constructing a modified thickness and subsequently invoking an asymptotic argument to deduce its skin-friction scaling, we first specify the target mapped state as a hypothetical ZPG ITBL with constant fluid properties. The quantities $C_{f,i}$ and $Re_{\theta,i}$ are then evaluated by applying the standard incompressible definitions to the mapped kinematics, so that, for an ideal full-layer mapping, the corresponding incompressible skin-friction scaling is recovered by construction. In particular,
\begin{equation}
    C_{f,i} = \frac{2}{U_{\infty}^{+2}}, \quad
    Re_{\theta,i} = U_{\infty}^{+} \int_0^{Re_T} \frac{\bar U^+}{U_{\infty}^+} \left( 1-\frac{\bar U^+}{U_{\infty}^+} \right) \, \mathrm{d}Y^+ .
    \label{eq:implied transformed variables (present)}
\end{equation}
The absence of the density ratio $\bar \rho / \rho_\infty$ in $\theta_i$ is therefore not an arbitrary omission, but follows directly from the constant-property target state, for which this ratio is identically unity. Once this target state and the velocity mapping~\eqref{eq:definition of velocity transformation} are prescribed, the transformation factors $F_C$ and $F_\theta$ follow as exact relations connecting the transformed variables to their physical compressible counterparts. This definition-first construction is consistent with recent developments in compressible velocity transformations \citep{trettel2016mean,zhu2024velocity} and temperature transformations \citep{huang2023velocity,zhu2025enhancing}, in which the target mapped state is specified before the associated transformation relations are derived.

The preceding comparison shows that the conceptual distinction fundamentally alters the physical interpretation of $Re_{\theta,i}$, rather than merely reflecting the presence or absence of a density factor in the momentum-thickness integral. In the present formulation, an ideal full-layer velocity mapping would reproduce the corresponding constant-property ZPG ITBL, ensuring that $C_{f,i}$ and $Re_{\theta,i}$ in \eqref{eq:implied transformed variables (present)} recover the incompressible skin-friction scaling by construction. In the nominal-thickness construction~\eqref{eq:implied transformed variables (Zhao and Fu)}, however, $Re_{\theta,i}$ retains an explicit dependence on the compressible density field even under such an ideal velocity mapping, meaning that such a recovery is no longer guaranteed by construction. Consequently, this construction does not fully isolate the transformed variables from residual compressibility effects associated with the free-stream Mach number and wall heat transfer.

This distinction also helps interpret the contrasting behaviour observed in \hyperref[fig:GFM and VIPL transformations (ZF, logarithmic)]{figure~\ref{fig:GFM and VIPL transformations (ZF, logarithmic)}} and \hyperref[fig:GFM and VIPL transformations (present without modification, logarithmic)]{figure~\ref{fig:GFM and VIPL transformations (present without modification, logarithmic)}}. As established in \textsection~\ref{sec:subsection 3.2}, the present mapping-based formulation provides a transparent diagnostic of the prescribed velocity transformation: since $C_{f,i}=2/U_\infty^{+2}$, any bias in the mapped free-stream velocity $U_\infty^+$ relative to its incompressible reference is directly reflected in the ordinate $\sqrt{2/C_{f,i}}$. The observed deviations therefore reflect the inherited outer-layer limitations of the underlying velocity transformation. In the nominal-thickness construction, however, these mapping-induced deviations are further entangled with the residual density weighting retained in $\hat{\theta}$. For the present database, this weighting generally reduces $\hat{\theta}^+$ relative to $\theta_i^+$, shifting the transformed data towards lower $Re_{\theta,i}$. Such a density-induced horizontal shift can partly compensate for a downward bias in $\sqrt{2/C_{f,i}}$, or even convert it into an apparent overprediction, thereby improving the apparent collapse without necessarily recovering the target-state coordinates $(C_{f,i},Re_{\theta,i})$ of a mapped constant-property ZPG ITBL, as illustrated by the GFM-based comparison in \hyperref[fig:GFM and VIPL transformations (ZF, logarithmic)]{figure~\ref{fig:GFM and VIPL transformations (ZF, logarithmic)}(a)} and \hyperref[fig:GFM and VIPL transformations (present without modification, logarithmic)]{figure~\ref{fig:GFM and VIPL transformations (present without modification, logarithmic)}(a)}. 
Conversely, when the underlying velocity mapping already provides a close full-layer correspondence, the same density-induced shift becomes an additional source of discrepancy, as illustrated by the VIPL-based comparison in \hyperref[fig:GFM and VIPL transformations (ZF, logarithmic)]{figure~\ref{fig:GFM and VIPL transformations (ZF, logarithmic)}(b)} and \hyperref[fig:GFM and VIPL transformations (present without modification, logarithmic)]{figure~\ref{fig:GFM and VIPL transformations (present without modification, logarithmic)}(b)}. The apparent collapse achieved by the nominal-thickness construction should therefore be interpreted with care, since it can reflect a mixture of velocity-mapping errors and residual density-weighting effects, rather than a unique recovery of the corresponding incompressible state.

Finally, beyond these conceptual and phenomenological distinctions, the present work differs further from the study of \citet{zhao2025revisiting} by pursuing the practical closure and predictive use of skin-friction transformations, which forms a central contribution of this paper. Both the nominal-thickness construction of \citet{zhao2025revisiting} and the direct applications considered in \textsection~\ref{sec:subsection 3.2} require the available compressible mean-flow profiles to evaluate the transformation factors $F_C$ and $F_\theta$. They are therefore primarily restricted to profile-based \emph{a priori} scaling assessments of existing DNS or experimental datasets. As shown above, such assessments either diagnose the adopted velocity transformation itself or reflect its interaction with residual density weighting in the nominal-thickness construction; they do not by themselves furnish a closed predictive relation for $C_f$ from prescribed macroscopic inputs, such as $M_\infty$, $Re_\theta$, $\bar T_w/T_r$ (or $\varTheta$), $Pr$, $T_\infty$, and the viscosity law. This limitation forfeits a key practical feature of the classical vD and SC transformations. Whereas the study of \citet{zhao2025revisiting} remains at this profile-based \emph{a priori} scaling-assessment stage, the subsequent development of the present definition-first construction addresses this closure problem directly. Specifically, van Driest's classical theory is semi-analytically reformulated through exact-integral relations, allowing the transformation factors to be modelled while retaining finite-Reynolds-number effects and yielding a generally implicit predictive formulation for $C_f$ based solely on prescribed macroscopic flow and wall-thermal conditions, without requiring DNS-extracted mean profiles.

\section{Exact-integral reformulation and asymptotic analysis}
\label{sec:section 4}
The preceding section established the mapping-based relations for $F_C$ and $F_\theta$, demonstrating that their direct application serves primarily as a transparent diagnostic of the prescribed velocity transformation. To bridge the gap between profile-based \emph{a priori} scaling assessment and practical predictive implementation, this section develops a semi-analytical formulation in which the DNS-extracted mean profiles are replaced by prescribed physical modelling ingredients, including a simplified representation of the transformed `incompressible' mean velocity profile, the inverse form of a selected velocity transformation, a quadratic temperature--velocity (TV) relation, an ideal-gas density closure, and a specified viscosity law. These components provide the mathematical closure required to incorporate the transformation factors into the standalone \emph{a posteriori} prediction procedure for $C_f$ developed in \textsection~\ref{sec:section 5}, based solely on prescribed macroscopic flow parameters and wall-thermal conditions. In this regard, the present treatment follows the strategy of \citet{vanDriest1951turbulent} by treating a closed set of compressible relations in a semi-analytical manner. Its purpose, however, is not to reproduce the original derivation, but to recast this strategy entirely within the mapping-based relations \eqref{eq:a general formula for Cf transformation} and \eqref{eq:a general formula for Re_theta transformation}, such that the `incompressible' and compressible momentum thicknesses form a paired exact-integral problem for a prescribed velocity mapping. This reformulation retains finite-Reynolds-number effects and clarifies how the classical vD~I and vD~II skin-friction transformations arise as leading-order asymptotic reductions.

The present implementation is restricted to canonical smooth-wall ZPG CTBLs of calorically perfect, air-like gases with specified viscosity laws, consistent with the DNS database summarized in \hyperref[tab:the ZPG CTBLs database 1]{tables~\ref{tab:the ZPG CTBLs database 1}--\ref{tab:the ZPG CTBLs database 2}}. Nevertheless, the formulation is inherently modular in structure: the selected velocity transformation, TV relation, density closure and viscosity law can in principle be replaced by alternative closures appropriate to other flow regimes, although such extensions are not assessed here. Within this specified scope, \textsection~\ref{sec:subsection 4.1} derives the exact-integral expressions for $F_C$ and $F_\theta$ once the modelling ingredients are prescribed; \textsection~\ref{sec:subsection 4.2} performs a leading-order asymptotic analysis, clarifying both the recovery of the classical vD~I and vD~II transformations and the limitations of conventional asymptotic approximations at practical Reynolds numbers; and \textsection~\ref{sec:subsection 4.3} quantitatively investigates how the associated asymptotic errors interact, exposing the fortuitous error cancellation behind the historical success of the classical vD~II transformation.

\subsection{Semi-analytical modelling of $F_C$ and $F_\theta$}
\label{sec:subsection 4.1}
We begin by considering the transformed target state, namely a constant-property ZPG ITBL. To evaluate the momentum thickness $\theta_i$ defined in \eqref{eq:incompressible momentum thickness}, the corresponding mean velocity profile $\bar U^+(Y^+)$ must first be prescribed. Following the integral treatments of \citet{vanDriest1951turbulent,vanDriest1956problem} and \citet{zhao2025revisiting}, we assume the logarithmic law, $\bar U^+ = \kappa_U^{-1} \ln Y^+ + C_U$, as an effective representation of the mapped `incompressible' mean velocity profile over the boundary layer. This simplified assumption provides an analytically tractable closure for evaluating $\theta_i$ and facilitating the subsequent asymptotic analysis. Substitution of this profile into \eqref{eq:incompressible momentum thickness} yields
\begin{equation} 
    \theta_i = \frac{\bar{\mu}_w}{\bar{\rho}_w U_{\infty}} \frac{a^2 F^2 I}{\kappa_U E}, \quad 
    I = \int_0^{Z_e}{Z ( 1-Z ) \exp (aFZ) \, \mathrm{d}Z,} 
    \label{eq:original expression for theta_i}
\end{equation}
where $Z_e = \bar U_e^+ / U_\infty^+$ is the normalized velocity at the transformed boundary-layer edge $Y = \Delta_e$, $E = \exp (\kappa_U C_U)$, and $a = \kappa_U u_\infty^+$. Crucially, $\kappa_U$ and $C_U$ should not be conflated with the canonical von Kármán constant and log-law intercept, e.g., $\kappa = 0.41$ and $C = 5.2$. Instead, they act as effective closure parameters for the integral formulation, determined indirectly by enforcing consistency between \eqref{eq:original expression for theta_i} and the CF relation~\eqref{eq:Coles-Fernholz relation}, as will be discussed in \textsection~\ref{sec:subsection 4.2}.

We now proceed to formulate the free-stream velocity ratio $F$ and the compressible momentum thickness $\theta$. For the present semi-analytical implementation, we restrict our attention to a class of velocity transformations in which the kernels $f_I$ and $g_I$ depend solely on the mean density ratio $\bar \rho / \bar \rho_w$ and the mean viscosity ratio $\bar \mu / \bar \mu_w$ \citep{volpiani2020data}. Representative examples include the vD~I~\eqref{eq:vD I velocity transformation}, vD~II~\eqref{eq:vD II velocity transformation}, and VIPL~\eqref{eq:VIPL transformation} transformations. To explicitly relate the mean thermodynamic and velocity fields, we adopt the established quadratic TV relation derived by \citet{zhang2014generalized}, which is consistent with the empirical relation reported by \citet{duan2011direct} for canonical air flows,
\begin{equation}
    \frac{\bar{T}}{\bar{T}_w} = 1 + Bz - A^2 z^2,
    \label{eq:mean T-u relation}
\end{equation}
where $A^2 = (T_r - T_\infty) / \bar T_w + (1 - sPr) (1 - T_r / \bar T_w)$, $B = sPr (T_r / \bar T_w - 1)$, and $s = 2 C_h / C_f$ is the Reynolds-analogy factor through which the wall-heat-flux closure enters the relation, with $C_h = \bar q_w / [\rho_\infty u_\infty c_p(\bar T_w-T_r)]$ denoting the Stanton number. We use $s \approx 1.14$ for canonical air flows \citep{chen2025mean}, while $Pr$ is taken from the corresponding DNS setting. It is worth noting that Walz's equation~\eqref{eq:Walz's equation}, which is employed in the classical transformations~\eqref{eq:vD I skin friction transformation}, \eqref{eq:vD II skin friction transformation} and \eqref{eq:SC skin friction transformation}, is recovered from \eqref{eq:mean T-u relation} when either $sPr=1$ or $T_r/\bar T_w=1$. In the latter quasi-adiabatic limit, $B=0$ and $A^2 = 1 - T_\infty / \bar T_w$. Thus, \eqref{eq:mean T-u relation} retains the $sPr$-dependent wall-heat-transfer correction under general diabatic conditions, while reducing to Walz's equation for quasi-adiabatic walls. To express the mean density ratio explicitly as a function of $z = \bar u^+ / u_\infty^+$, we further invoke the equation of state for a calorically perfect gas, $p = \rho R T$, together with the boundary-layer approximation $\partial \bar{p}/\partial y \approx 0$, yielding
\begin{equation}
    \bar \rho^+ (z) = \frac{\bar \rho}{\bar \rho_w} \approx \frac{\bar T_w}{\bar T} = \frac{1}{1 + Bz - A^2 z^2}.
    \label{eq:mean rho-T relation}
\end{equation}
The viscosity-temperature dependence is modelled using either the power law,
\begin{equation}
    \bar \mu^+ (z) = \frac{\bar \mu}{\bar \mu_w} = \left( \frac{\bar T}{\bar T_w} \right)^n = ( 1 + Bz - A^2 z^2 )^n,
    \label{eq:mean mu-T relation (power law)}
\end{equation}
or Sutherland's law,
\begin{equation}
    \bar \mu^+ (z) =\frac{1+{T}_s/\bar{T}_w}{\bar T/\bar{T}_w+{T}_s/\bar{T}_w} \left( \frac{\bar{T}}{\bar{T}_w} \right) ^{3/2}
    = \frac{( 1+T_s/\bar T_w ) ( 1 + Bz - A^2 z^2 )^{3/2}}{1 + Bz - A^2 z^2+ T_s/\bar T_w}.
    \label{eq:mean mu-T relation (Sutherland law)}
\end{equation}
Equations~\eqref{eq:mean T-u relation}--\eqref{eq:mean mu-T relation (Sutherland law)} are therefore used as necessary closure ingredients for implementing the subsequent exact-integral formulation for calorically perfect, air-like gases, consistent with the flow conditions covered by the DNS database in \hyperref[tab:the ZPG CTBLs database 1]{tables~\ref{tab:the ZPG CTBLs database 1}--\ref{tab:the ZPG CTBLs database 2}}. Their role is to close the thermodynamic and transport-property dependence of the formulation, rather than to introduce new models for these quantities. This closure strategy is also consistent with that commonly adopted in recent mean-flow prediction approaches for ZPG CTBLs \citep{kumar2022modular,manzoor2024estimating,ying2025general}. Applications to flows exhibiting non-air-like viscosity behaviour, supercritical thermophysical properties or strong real-gas effects would require the corresponding closures to be replaced.

With the aid of \eqref{eq:mean T-u relation}–\eqref{eq:mean mu-T relation (Sutherland law)}, the coordinate and velocity transformation kernels can be explicitly reformulated as functions of $z$, expressed as $f_I = \partial Y^+ / \partial y^+ = \mathcal{F}(z)$ and $g_I = \partial \bar U^+ / \partial \bar u^+ = \mathcal{G}(z)$. The velocity mapping~\eqref{eq:definition of velocity transformation} then reduces to
\begin{equation}
    \bar U^+ =  u_\infty^+ \psi(z), \quad \psi(z) = \int_0^z{\mathcal{G}(\zeta) \, \mathrm{d} \zeta},
    \label{eq:the transformed mean velocity profile}
\end{equation}
from which the free-stream velocity ratio follows directly as $F= U_\infty^+/ u_\infty^+=\psi(1)$. Correspondingly, the compressible law of the wall associated with the presumed logarithmic relation $Y^+ = \exp (\kappa_U \bar U^+) / E$ is formulated via the inverse velocity transformation as:
\begin{equation}
\mathrm{d} y^+ = \frac{a}{E} \frac{\mathcal{G}(z)}{\mathcal{F}(z)} \exp [a \psi(z)] \mathrm{d}z.
\label{eq:compressible law of the wall}
\end{equation}
Substituting \eqref{eq:compressible law of the wall} into \eqref{eq:compressible momentum thickness}, and invoking the thermodynamic relation~\eqref{eq:mean rho-T relation}, the compressible momentum thickness admits the following integral expression:
\begin{equation}
    \theta = \frac{\bar{\mu}_w}{{\rho}_{\infty}{u}_{\infty}}\frac{a^2J}{\kappa_U E}, \quad
    J=\int_0^{z_e}{\frac{\mathcal{G} (z)}{\mathcal{F} (z)}\frac{z(1-z)}{1+Bz-A^2z^2}\exp [ a\psi (z) ] \, \mathrm{d}z},
    \label{eq:original expression for theta}
\end{equation}
where $z_e = \bar u_e^+ / u_\infty^+$ corresponds to the boundary-layer edge $y = \delta_e$.

Finally, substitution of \eqref{eq:original expression for theta_i} and \eqref{eq:original expression for theta} into the present mapping-based relations~\eqref{eq:a general formula for Cf transformation} and \eqref{eq:a general formula for Re_theta transformation} yields the skin-friction transformation factors:
\begin{equation}
    F_C=\frac{{\rho}_{\infty}}{\bar{\rho}_w} F^{-2}, \quad F_{\theta}=\frac{\mu _{\infty}}{\bar{\mu}_w}F^2\frac{I}{J}.
    \label{eq:final expression for F_C and F_theta}
\end{equation}
Equations~\eqref{eq:original expression for theta_i}, \eqref{eq:the transformed mean velocity profile}, \eqref{eq:original expression for theta} and \eqref{eq:final expression for F_C and F_theta} constitute the exact-integral formulation of the skin-friction transformation factors for a prescribed velocity mapping and the specified closure ingredients. Here and throughout, `exact-integral' refers to retaining the paired integrals $I$ and $J$ in their finite-Reynolds-number forms, rather than to exactness of the prescribed modelling ingredients. It is worth noting that the final expression for $F_\theta$ depends on the effective slope of the presumed logarithmic law $\kappa_U$ and is independent of the corresponding intercept $C_U$, as the factor $E = \exp(\kappa_U C_U)$ cancels out in the ratio $\theta_i / \theta$.

\subsection{Leading-order asymptotic analysis and unification of van Driest's theory}
\label{sec:subsection 4.2}
To facilitate the derivation and demonstrate that the exact-integral formulation~\eqref{eq:final expression for F_C and F_theta} recovers van Driest's classical transformations under specific approximations, the upper integration limits in \eqref{eq:original expression for theta_i} and \eqref{eq:original expression for theta} are temporarily extended to the free stream within this subsection, i.e., $Z_e = 1$ and $z_e = 1$, respectively. Consequently, the `incompressible' integral $I$ can be evaluated analytically via integrating by parts twice, yielding:
\begin{equation}
    I=\frac{1+\exp (aF)}{a^2F^2}+\frac{2[ 1-\exp (aF) ]}{a^3F^3}.
\end{equation}
According to the CF relation~\eqref{eq:Coles-Fernholz relation}, $U_\infty^+ \gtrsim 22$ for typical Reynolds numbers $Re_{\theta,i} \gtrsim 1000$. For an estimated value of $\kappa_U \approx 0.4$, it follows that $aF=\kappa_U U_\infty^+=\mathcal{O}(10)$ \citep{zhao2025revisiting}. This implies that the exponential term dominates, scaling as $\exp (aF) \sim \mathcal{O}(10^4) \gg 1$, thereby justifying the further approximation:
\begin{equation}
    I \approx \frac{\exp (aF)}{a^2F^2}\left( 1-\frac{2}{aF} \right).
    \label{eq:approximate expression for I}
\end{equation}
Substituting \eqref{eq:approximate expression for I} into \eqref{eq:original expression for theta_i} leads to:
\begin{equation}
    {U}_{\infty}^{+}=\sqrt{\frac{2}{C_{f,i}}}=\frac{1}{\kappa_U}\ln{Re}_{\theta ,i}+C_U+\frac{\ln \kappa_U}{\kappa_U}-\frac{1}{\kappa_U}\ln \left( 1-\frac{2}{aF} \right) .
    \label{eq:derived relation between U_infty^+ and Re_theta_i}
\end{equation}

Equation~\eqref{eq:derived relation between U_infty^+ and Re_theta_i} may be viewed as a finite-$aF$, or finite-Reynolds-number form of the logarithmic relation for the mapped `incompressible' state, but the effective parameters $\kappa_U$ and $C_U$ remain unspecified. In the present formulation, they are closed by requiring compatibility with the CF relation~\eqref{eq:Coles-Fernholz relation}, which is adopted as the model-consistent incompressible reference and, owing to its single-logarithmic form, yields explicit matching conditions for the slope and intercept. Specifically, matching the effective slope and intercept to $\kappa_\theta$ and $C_\theta$ gives
\begin{equation}
    \kappa _{\theta}=\left[ \frac{\mathrm{d} U_{\infty}^{+}}{\mathrm{d} \ln Re_{\theta ,i}} \right]^{-1} = \kappa_U\left[ 1+\frac{2}{aF ( aF-2 )} \right] ,
    \label{eq:relation between kappa_u and kappa_theta}
\end{equation}
\begin{equation}
    C_{\theta}=U_{\infty}^{+}-\frac{1}{\kappa _{\theta}}\ln Re_{\theta,i} = \frac{\kappa_U}{\kappa_\theta}C_U + \frac{\ln \kappa_U}{\kappa_\theta} + \frac{1}{\kappa_\theta}\left[ \frac{2}{aF-2} - \ln \left( 1 - \frac{2}{aF} \right) \right].
    \label{eq:relation between C_U and C_theta}
\end{equation}
Notably, in the infinite-Reynolds-number limit ($aF \rightarrow \infty$), \eqref{eq:derived relation between U_infty^+ and Re_theta_i} becomes asymptotically equivalent to the CF relation, corresponding to the limiting parameter values of $\kappa_U = \kappa_\theta \approx 0.384$ and $C_U = C_\theta - \kappa_\theta^{-1}\ln \kappa_\theta \approx 6.619$. However, due to the logarithmic growth of $U_\infty^+$ with $Re_{\theta,i}$, the approach to this asymptotic state is exceedingly slow. To quantify this finite-Reynolds-number behaviour, \eqref{eq:relation between kappa_u and kappa_theta} and \eqref{eq:relation between C_U and C_theta} are evaluated together with the CF relation over the practical range $10^3 \lesssim Re_{\theta,i} \lesssim 10^5$, yielding $8.17 \lesssim aF \lesssim 12.91$. Over this range, the effective slope varies only weakly, with $0.369 \lesssim \kappa_U \lesssim 0.379$, whereas the corresponding intercept spans $5.35 \lesssim C_U \lesssim 5.82$ and remains appreciably below its infinite-Reynolds-number limit. These estimates show that the parameter $aF$ remains only of order ten under typical flow conditions, indicating that the finite-Reynolds-number correction terms involving $aF$ cannot be neglected. Meanwhile, the weak Reynolds-number dependence of $\kappa_U$ indicates that while the effective logarithmic slope can be evaluated dynamically from the compatibility condition \eqref{eq:relation between kappa_u and kappa_theta}, it may also be approximated by a fixed representative value in simplified practical implementations. For example, adopting a nominal $aF \approx 10$ yields $\kappa_U \approx 0.375$, which lies near the centre of the above finite-Reynolds-number variation.

At this point, it is useful to contrast the present treatment with the nominal-thickness asymptotics of \citet{zhao2025revisiting}. Their asymptotic analysis is applied to the density-weighted nominal thickness $\hat{\theta}$, in a manner formally analogous to the treatment of $\theta_i$ above, but no corresponding semi-analytical closure is developed for the physical compressible momentum thickness $\theta$. Consequently, $(F_\theta)_{\mathrm{ZF}}$ in \eqref{eq:transformation factors of Zhao and Fu} retains the ratio $\hat{\theta}/\theta$ and, in its original implementation, remains dependent on available mean-flow profiles. An analogous integration-by-parts expansion of $\hat{\theta}$ shows that $\hat{\theta}$ and $\theta_i$ share the same formal leading-order asymptotic structure as $aF\rightarrow\infty$. However, for $\hat{\theta}$, this correspondence is obtained only after discarding all higher-order terms containing the density and density-gradient dependence, whereas the present treatment of the constant-property mapped thickness $\theta_i$ retains the corresponding finite-Reynolds-number correction factor, $1-2/(aF)$. Since the preceding estimates show that $aF$ remains only $\mathcal{O}(10)$ over practical Reynolds-number ranges, the neglected density-dependent corrections in the nominal-thickness asymptotics cannot be assumed negligible \emph{a priori}, especially for high-Mach-number flows with strong wall-normal property variations induced by wall cooling or heating. Thus, this leading-order correspondence neither makes $\hat{\theta}$ equivalent to $\theta_i$ nor closes $(F_\theta)_{\mathrm{ZF}}$ for standalone prediction.

We now turn to the corresponding evaluation of the compressible momentum thickness $\theta$ through the integral $J$, which is required to close $F_\theta$ within the present finite-Reynolds-number formulation. For analytical convenience, we introduce an auxiliary function:
\begin{equation}
    \varphi(z) = \frac{\mathcal{G}(z)}{\mathcal{F}(z)} \frac{z(1-z)}{1+Bz-A^2z^2}.
    \label{eq:definition of phi(z)}
\end{equation}
Repeated integration by parts yields the following asymptotic series expansion:
\begin{equation}
    \begin{aligned}
      J & =\int_0^1{\varphi ( z ) \exp [a\psi (z)] \, \mathrm{d}z} 
          = \left[ \sum_{k=0}^{\infty} (-1)^k \frac{w_k(z)}{a^{k+1} \psi^{\prime}(z)} \exp [a\psi (z)] \right]_{0}^{1} \\
        & =\sum_{k=1}^{\infty}{( -1 ) ^k \left[ \frac{w_k(1)}{a^{k+1}\psi ^{\prime}(1)}\exp (aF) -\frac{w_k(0)}{a^{k+1}\psi^{\prime}(0)} \right]} \\
        & =-\frac{w_1(1)}{a^2\psi ^{\prime}(1)}\exp (aF) \sum_{n=0}^{\infty}{(-a) ^{-n}\frac{w_{n+1}(1)}{w_1(1)}}+\frac{w_1(0)}{a^2\psi ^{\prime}(0)}\sum_{n=0}^{\infty}{(-a) ^{-n}\frac{w_{n+1}(0)}{w_1(0)}}.
    \end{aligned}
    \label{eq:asymptotic expansion of J}
\end{equation}
Here, the summation in the second line commences at $k=1$ (corresponding to terms of $\mathcal{O}(a^{-2})$) because the apparent leading-order boundary terms ($k=0$) vanish due to $\varphi(0)=\varphi(1)=0$. The final expression is then obtained by separating the contributions from the boundary-layer edge ($z=1$) and the wall ($z=0$), with the leading-order terms explicitly factored out. The recursive functions are defined by
\begin{equation}
    w_0(z) =\varphi (z), 
    \quad w_{n+1}(z) =\frac{\mathrm{d}}{\mathrm{d}z}\left[ \frac{w_n(z)}{\psi ^{\prime}(z)} \right] 
    \quad \text{for } n \ge 0.
    \label{eq:recursive functions definition}
\end{equation}
Following a similar order-of-magnitude argument, the term associated with the wall ($z=0$) is exponentially small compared to the contribution from the boundary-layer edge ($z=1$) and is therefore neglected. Consequently, the expression for $J$ is further approximated as:
\begin{equation}
    J \approx -\frac{w_1(1)}{a^2\psi ^{\prime}(1)}\exp (aF) [ 1+\mathcal{D} (a) ],
    \label{eq:approximate expression for J} 
\end{equation}
where $w_1(1)$ is derived from \eqref{eq:recursive functions definition} as $w_1(1) = \varphi'(1) / \psi'(1)$ (noting that $\varphi(1) = 0$), and $\mathcal{D}(a)$ denotes the asymptotic series representing higher-order corrections starting at $\mathcal{O}(a^{-1})$:
\begin{equation}
    \mathcal{D} (a) = \sum_{n=1}^{\infty}{(-a) ^{-n}\frac{w_{n+1}(1)}{w_1(1)}}.
    \label{eq:Expression for D(a)}
\end{equation}

Finally, substituting \eqref{eq:approximate expression for I} and \eqref{eq:approximate expression for J} into \eqref{eq:final expression for F_C and F_theta}, while noting that $\psi'(1) = \mathcal{G}(1)$, we obtain the analytical expressions for the transformation factors:
\begin{equation}
    F_C=\frac{{\rho}_{\infty}}{\bar{\rho}_w}[\psi(1)]^{-2}, \quad 
    F_{\theta}=F_{\theta}^{\text{asy}}\mathcal{C}(a)=-\frac{{\mu}_{\infty}}{\bar{\mu}_w}\frac{[\mathcal{G}(1)] ^2}{\varphi ^{\prime}(1)}\frac{1-2/(aF)}{1+\mathcal{D}(a)}.
    \label{eq:analytical expression for F_C and F_theta}
\end{equation}

It is worth noting that the classical analyses of \citet{vanDriest1951turbulent,vanDriest1956problem}, although derived through different routes using momentum-integral arguments, Walz's relation, and specific mixing-length assumptions, can be interpreted in the present notation as particular leading-order reductions of \eqref{eq:analytical expression for F_C and F_theta}. Specifically, they correspond to the assumption that $a$ is sufficiently large for the finite-Reynolds-number correction terms of $\mathcal{O}(a^{-1})$ and higher in $\mathcal{C}(a)$ to be neglected. In the resulting asymptotic limit $a \to \infty$, the finite-Reynolds-number correction factor $\mathcal{C}(a) = [1 - 2 / (aF)] / [1 + \mathcal{D}(a)]$ reduces to unity, and $F_\theta$ converges to its asymptotic form $F_{\theta}^{\text{asy}} = -\mu_\infty / \bar \mu_w \times [\mathcal{G}(1)]^2 / \varphi^{\prime}(1)$. This simplification readily enables the derivation of closed-form skin-friction transformations once a specific velocity mapping is prescribed. Within the present exact-integral formulation, the classical vD~I and vD~II transformations are recovered as particular leading-order reductions when the corresponding vD-type velocity transformations \eqref{eq:vD I velocity transformation} and \eqref{eq:vD II velocity transformation}, and Walz's equation \eqref{eq:Walz's equation} are specified. Furthermore, a corresponding analytical expression induced by the VIPL velocity transformation assessed in \textsection~\ref{sec:section 3} is obtained in an analogous manner (see \eqref{eq:VIPL skin friction transformation, power law} and \eqref{eq:VIPL skin friction transformation, Sutherland law}). Detailed derivations for these three cases are provided in Appendix~\ref{app A}.

\begin{figure}
  \centering
  \includegraphics[width=0.49\textwidth]{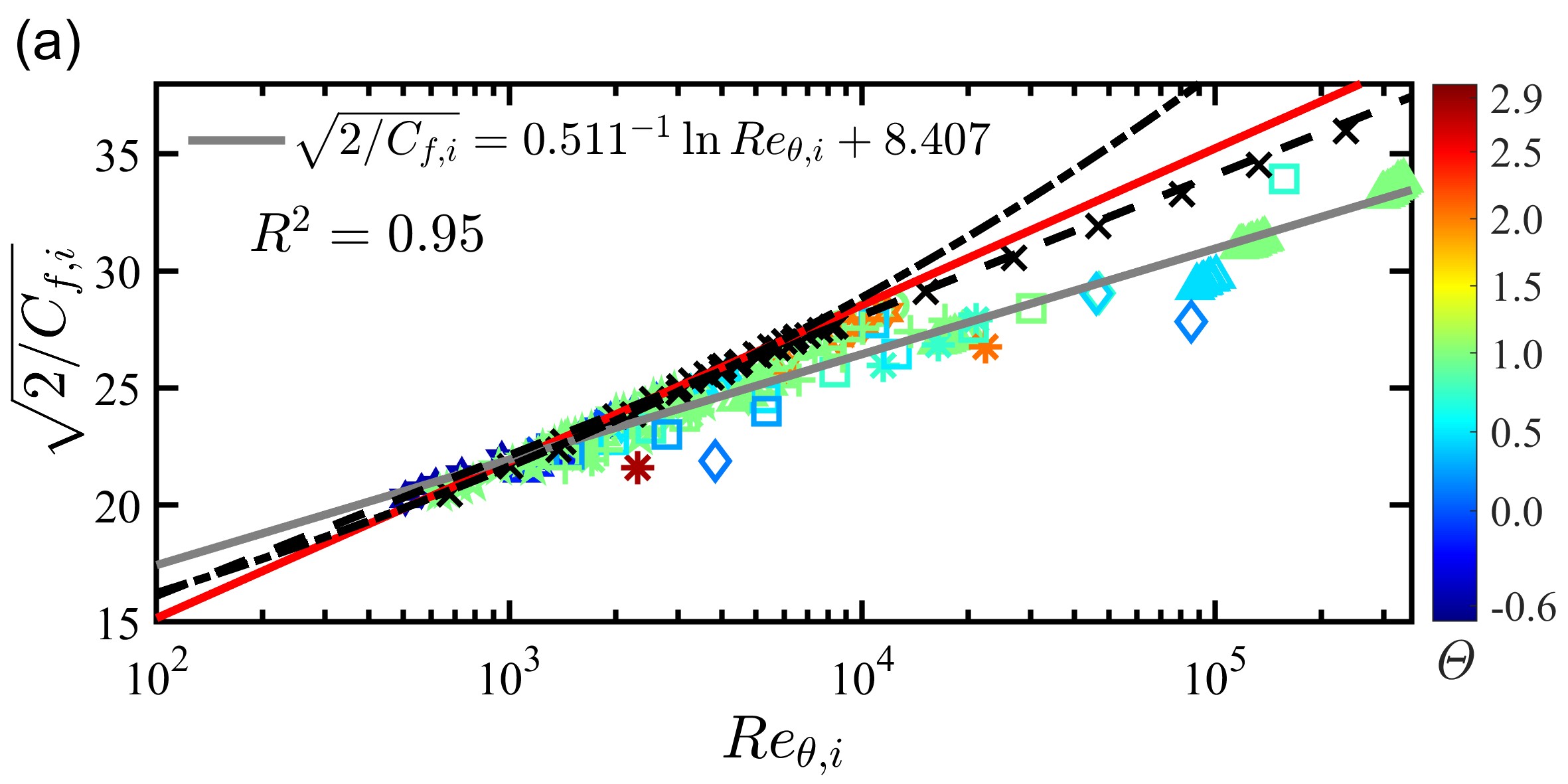}
  \includegraphics[width=0.49\textwidth]{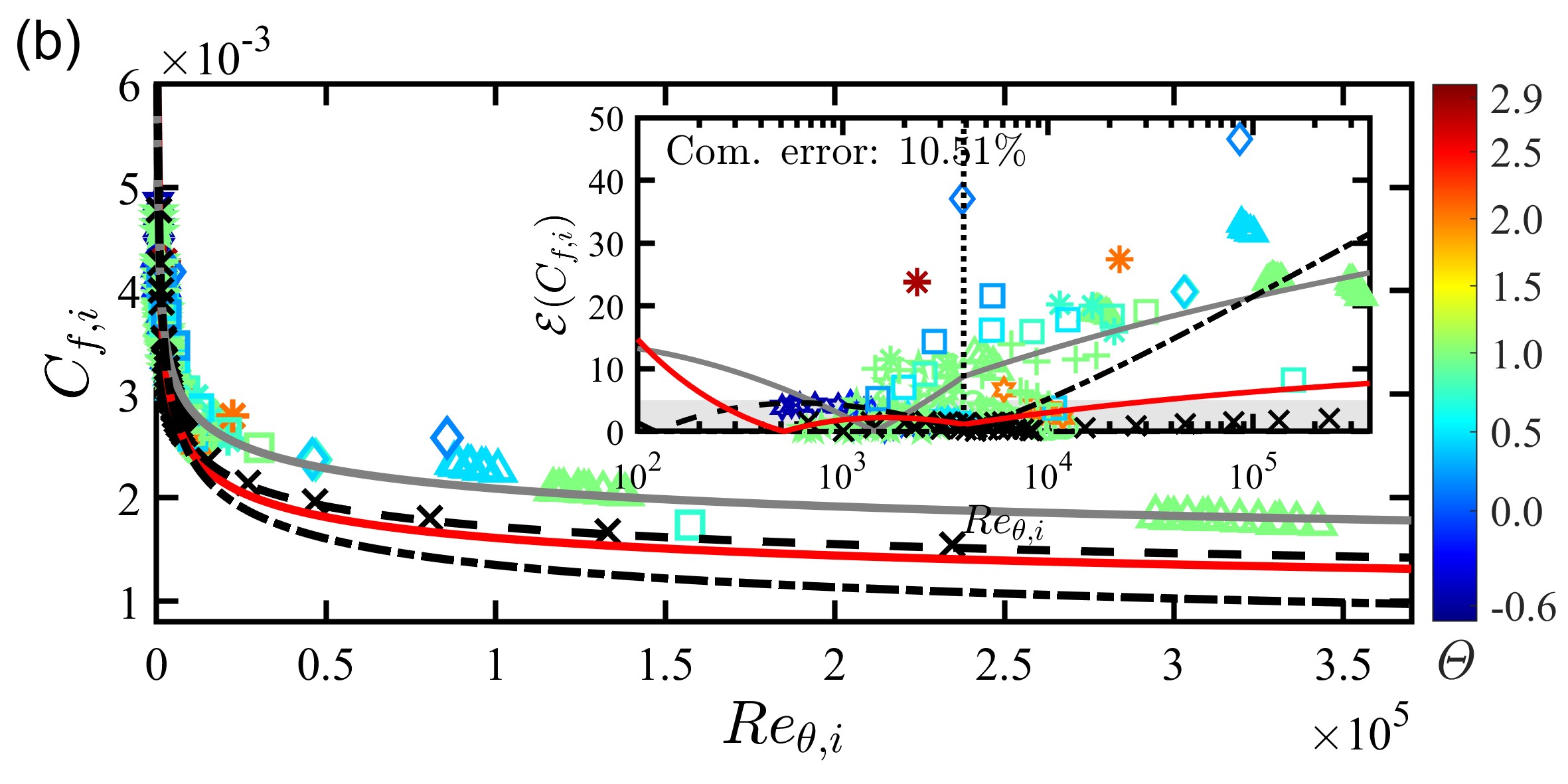}
  \caption{Transformed `incompressible' (a) $\sqrt{2 / C_{f,i}}$ and (b) $C_{f,i}$ versus $Re_{\theta,i}$ using the analytical asymptotic skin-friction transformation induced by the VIPL velocity mapping given by \eqref{eq:VIPL skin friction transformation, power law} and \eqref{eq:VIPL skin friction transformation, Sutherland law}. Insets of panel (b) show the absolute relative errors $\mathcal{E}(C_{f,i})$ defined by \eqref{eq:error against composite incompressible reference}. Plotting, inset, and error-band conventions follow those in \hyperref[fig:GFM and VIPL transformations (ZF, linear)]{figure~\ref{fig:GFM and VIPL transformations (ZF, linear)}}.}
\label{fig:original VIPL transformation}
\end{figure}

However, the validity of such an asymptotic reduction is questionable for practical Reynolds numbers. As evidenced in \hyperref[fig:original vD I, vD II, SC transformations (logarithmic)]{figure~\ref{fig:original vD I, vD II, SC transformations (logarithmic)}} and \hyperref[fig:original vD I, vD II, SC transformations (linear)]{figure~\ref{fig:original vD I, vD II, SC transformations (linear)}}, the classical vD~I transformation, which is derived as the asymptotic limit of \eqref{eq:analytical expression for F_C and F_theta} based on the widely used vD~I mean velocity mapping~\eqref{eq:vD I velocity transformation}, fails to yield a satisfactory data collapse, even for quasi-adiabatic cases. In contrast, the classical vD~II transformation grounded in the less commonly adopted vD~II mean velocity mapping~\eqref{eq:vD II velocity transformation} yields relatively better results. This contrast is not readily explained by the intrinsic quality of the underlying velocity mappings alone, as profile-based assessments of compressible DNS data do not suggest that the vD~I velocity transformation is fundamentally inferior to vD~II. Furthermore, \hyperref[fig:original VIPL transformation]{figure~\ref{fig:original VIPL transformation}} presents the distributions of the transformed `incompressible' $\sqrt{2 / C_{f,i}}$ and $C_{f,i}$ as functions of $Re_{\theta,i}$, obtained using the analytical VIPL skin-friction transformations~\eqref{eq:VIPL skin friction transformation, power law} and \eqref{eq:VIPL skin friction transformation, Sutherland law} derived under the asymptotic approximation $a \rightarrow \infty$. The transformed data exhibit significant scatter and a systematic deviation from the incompressible reference. This performance is markedly inferior to the \emph{a priori} assessment results presented in \textsection~\ref{sec:section 3} (see \hyperref[fig:GFM and VIPL transformations (present without modification, logarithmic)]{figure~\ref{fig:GFM and VIPL transformations (present without modification, logarithmic)}(b)} and \hyperref[fig:GFM and VIPL transformations (present without modification, linear)]{figure~\ref{fig:GFM and VIPL transformations (present without modification, linear)}(b)}), where the VIPL velocity transformation is directly substituted into the exact mapping-based formulation \eqref{eq:a general formula for Cf transformation} and \eqref{eq:a general formula for Re_theta transformation} without any asymptotic truncation.

Indeed, replacing the finite-Reynolds-number correction factor $1 - 2 / (aF)$ in \eqref{eq:approximate expression for I} by unity already overestimates the `incompressible' integral $I$ by approximately 25\% for a representative value of $aF \approx 10$. Concurrently, Appendix~\ref{app B} shows that replacing the correction factor $1+\mathcal{D}(a)$ in \eqref{eq:approximate expression for J} by unity can also lead to significant deviations in the compressible integral $J$ within the range $5 \lesssim a \lesssim 12$ covered by the present DNS datasets. The same appendix also suggests that the asymptotic series $\mathcal{D}(a)$ exhibits poor convergence properties, implying that retaining only a few higher-order terms does not reliably restore accuracy across different velocity mappings and wall-thermal conditions. These observations indicate that the leading-order asymptotic reduction underlying the closed-form transformations in Appendix~\ref{app A} is not uniformly reliable at practical Reynolds numbers. Therefore, the comparatively successful behaviour of the classical vD~II transformation in \hyperref[fig:original vD I, vD II, SC transformations (logarithmic)]{figure~\ref{fig:original vD I, vD II, SC transformations (logarithmic)}(b)} and \hyperref[fig:original vD I, vD II, SC transformations (linear)]{figure~\ref{fig:original vD I, vD II, SC transformations (linear)}(b)} should be interpreted with care, rather than taken as evidence that the leading-order asymptotic reduction is generally accurate.

\subsection{Error cancellation behind the success of the classical vD~II transformation}
\label{sec:subsection 4.3}
The preceding subsection established that the leading-order asymptotic approximations applied to both integrals $I$ and $J$ are individually questionable at practical Reynolds numbers. The comparatively good performance of the classical vD~II transformation, as observed in \textsection~\ref{sec:section 2}, therefore requires further explanation. As shown in Appendix~\ref{app B}, the zeroth-order approximation $J(0)$ for vD~II systematically overpredicts the exact compressible integral $J$; concurrently, approximating the finite-Reynolds-number correction factor $1-2/(aF)$ as unity similarly overestimates the `incompressible' integral $I$. The combined effect of these two truncations is measured by
\begin{equation}
    \mathcal{C}(a)
    = \frac{1-2/(aF)}{1+\mathcal{D}(a)}
    = \left(1-\frac{2}{aF}\right)\frac{J(0)}{J},
    \label{eq:error cancellation factor}
\end{equation}
where the second equality follows from \eqref{eq:partial sum of J}, \eqref{eq:approximate expression for J}, and \eqref{eq:Expression for D(a)}. For vD~II, the two factors entering $\mathcal{C}(a)$ tend to counterbalance each other: the factor $1-2/(aF)$ is inherently less than unity, whereas the ratio $J(0)/J$ exceeds unity. Consequently, their product may remain close to unity even though neither truncation is individually accurate.

\begin{figure}
  \centering
  \includegraphics[width=0.45\textwidth]{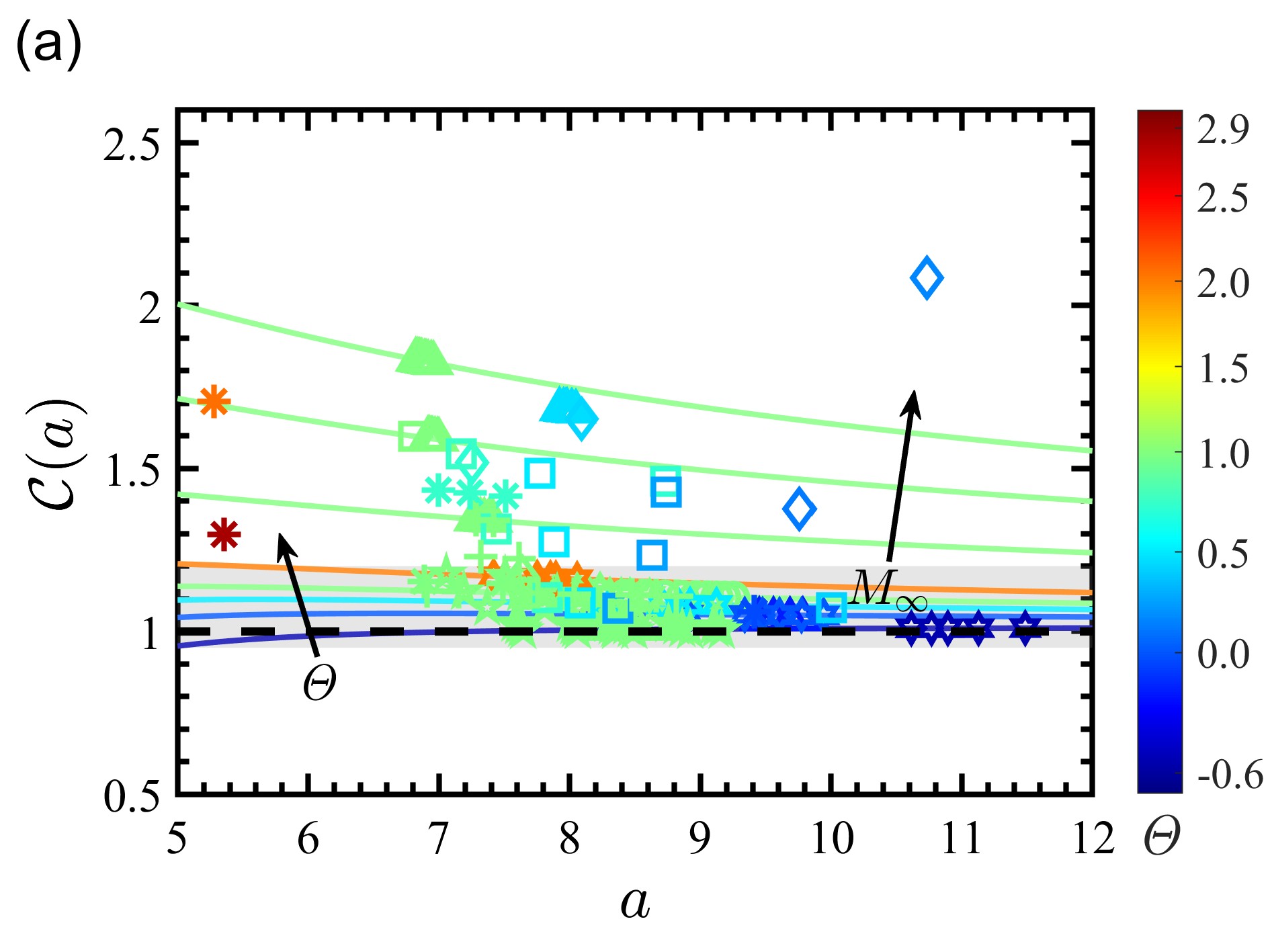}
  \includegraphics[width=0.45\textwidth]{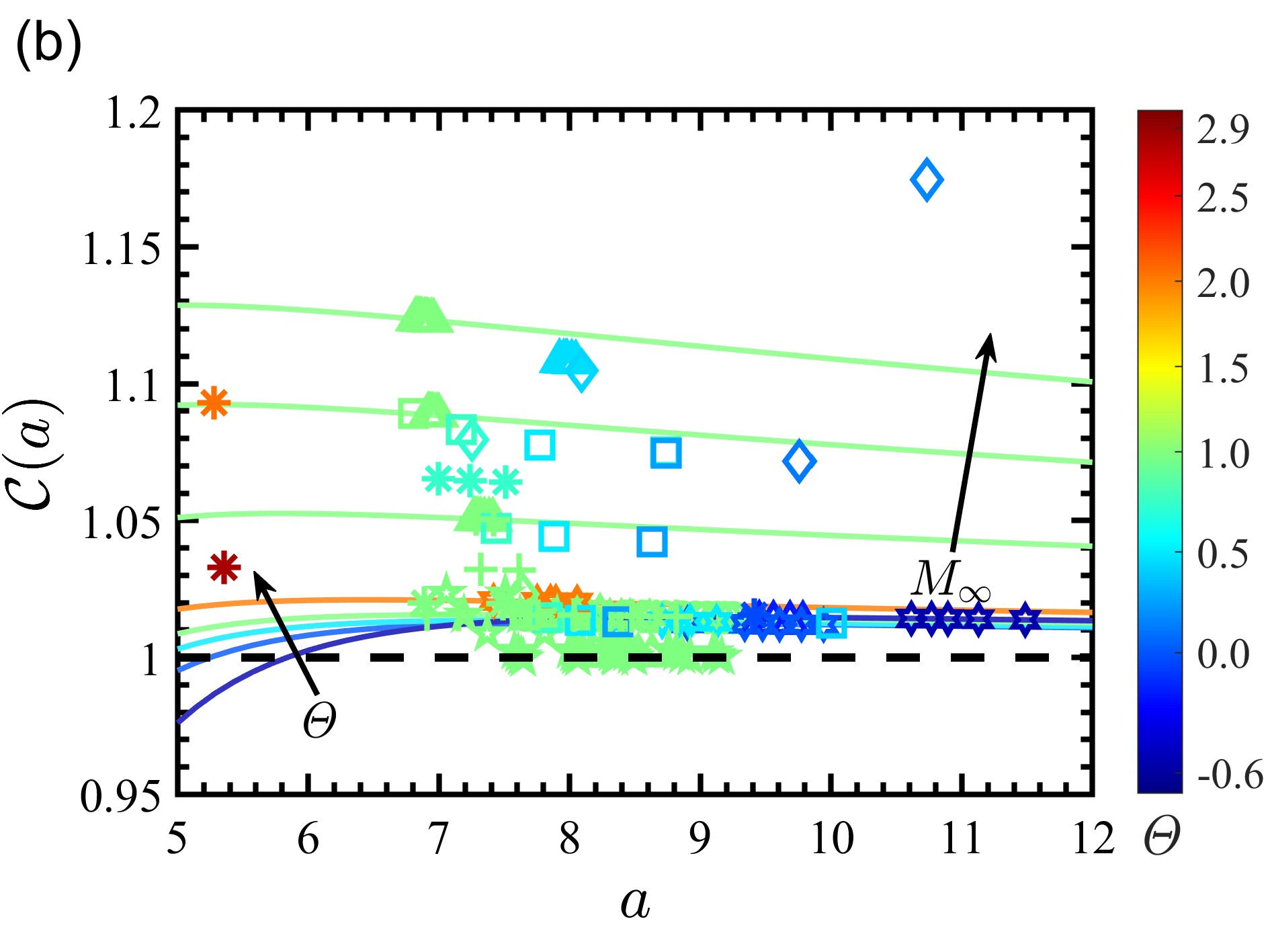}
  \caption{Finite-Reynolds-number correction factor $\mathcal{C}(a)$ defined by \eqref{eq:error cancellation factor} as a function of $a=\kappa_U u_\infty^+$ for the (a) vD~I and (b) vD~II velocity mappings, \eqref{eq:vD I velocity transformation} and \eqref{eq:vD II velocity transformation}, respectively. Coloured solid lines denote representative cases at fixed free-stream Mach numbers and wall-thermal conditions, with $a$ varied continuously to represent the Reynolds-number effect, while symbols indicate individual DNS cases from \hyperref[tab:the ZPG CTBLs database 1]{tables~\ref{tab:the ZPG CTBLs database 1}--\ref{tab:the ZPG CTBLs database 2}}, with colour representing the diabatic parameter $\varTheta$. The selected curves include four quasi-adiabatic cases with $M_\infty=2,4,6$ and $8$ to illustrate Mach-number effects, and five $M_\infty=2$ cases with $\varTheta=-0.55,0.00,0.50,1.00$ and $2.02$ to illustrate wall-thermal effects; the $M_\infty=2$, $\varTheta=1.00$ case is shared by the two groups. Arrows indicate the direction of increasing $M_\infty$ and $\varTheta$. The horizontal black dashed line marks the infinite-Reynolds-number asymptotic limit, $\mathcal{C}(a) = 1$. The shaded band in panel (a) indicates the vertical range displayed in panel (b).}
  \label{fig:error cancellation factor}
\end{figure}

\hyperref[fig:error cancellation factor]{Figure~\ref{fig:error cancellation factor}} quantifies the behaviour of $\mathcal{C}(a)$ for the vD~I and vD~II transformations. Along each coloured curve, the free-stream Mach number and wall-thermal condition are fixed, so that variation along the abscissa isolates the Reynolds-number dependence through the parameter $a=\kappa_U u_\infty^+$. The discrete symbols represent the individual DNS cases from \hyperref[tab:the ZPG CTBLs database 1]{tables~\ref{tab:the ZPG CTBLs database 1}--\ref{tab:the ZPG CTBLs database 2}}, each corresponding to a specific physical flow state defined by $M_\infty$, $\varTheta$ and $Re_{\theta}$. As expected from the asymptotic construction, $\mathcal{C}(a)$ approaches the formal limit $\mathcal{C}(a)=1$ as $a\to\infty$, corresponding to the infinite-Reynolds-number limit. However, over the practical Reynolds-number range represented here by $5 \lesssim a \lesssim 12$, this convergence is evidently incomplete, particularly for high-Mach-number flows.

The contrast between panels (a) and (b) is pronounced. For the vD~I transformation, $\mathcal{C}(a)$ exhibits a broad distribution and often departs substantially from unity. In particular, the quasi-adiabatic high-Mach-number curves lie well above the asymptotic baseline $\mathcal{C}=1$, with the deviation becoming more pronounced as $M_\infty$ increases. This indicates that the overestimation associated with $J(0)/J>1$ is not sufficiently compensated by the finite-Reynolds-number factor $1-2/(aF)<1$. Since $F_\theta^{\text{asy}} = F_\theta / \mathcal{C}(a)$, the leading-order asymptotic transformation underestimates the true mapped $Re_{\theta,i}$, consistent with the tendency of the transformed data to be shifted towards lower $Re_{\theta,i}$ in \hyperref[fig:original vD I, vD II, SC transformations (logarithmic)]{figure~\ref{fig:original vD I, vD II, SC transformations (logarithmic)}(a)} and \hyperref[fig:original vD I, vD II, SC transformations (linear)]{figure~\ref{fig:original vD I, vD II, SC transformations (linear)}(a)}. The lower group of curves at fixed $M_\infty=2$ shows that wall heat transfer also affects $\mathcal{C}(a)$; within the representative cases shown, increasing $\varTheta$ generally shifts the curves upward, although this wall-thermal effect is milder than the Mach-number effect. Overall, the large spread in panel (a) demonstrates that the leading-order asymptotic reduction is not a reliable approximation for the vD~I transformation at practical Reynolds numbers.

By contrast, the vD~II transformation displays a much narrower distribution of $\mathcal{C}(a)$, as shown in panel (b). For the selected curves and DNS cases considered here, $\mathcal{C}(a)$ remains approximately within the range $0.97\lesssim \mathcal{C}(a)\lesssim 1.18$, in sharp contrast to the much broader variation observed for vD~I. The sensitivity to both Mach number and wall-thermal condition is markedly weaker than in the vD~I case. Although several high-Mach-number cases still yield values above unity, their deviations remain modest compared with those observed in panel (a). The majority of DNS points fall close to the asymptotic baseline $\mathcal{C}=1$, demonstrating that the two individually inaccurate asymptotic approximations compensate each other rather effectively over a broad portion of the database. The shaded band in panel (a), corresponding to the ordinate range of panel (b), further highlights the exceptional narrowing of $\mathcal{C}(a)$ for the vD~II transformation.

These results provide a quantitative explanation for the historical success of the classical vD~II transformation. Its comparatively good performance does not imply that the leading-order asymptotic approximations accurately represent the exact integrals $I$ or $J$ separately. Rather, this performance arises because the finite-Reynolds-number correction omitted from $I$ and the leading-order truncation error in $J$ counterbalance each other over much of the practical parameter space considered here. In this sense, the apparent superiority of the classical vD~II skin-friction transformation over vD~I, as evidenced in \hyperref[fig:original vD I, vD II, SC transformations (logarithmic)]{figure~\ref{fig:original vD I, vD II, SC transformations (logarithmic)}} and \hyperref[fig:original vD I, vD II, SC transformations (linear)]{figure~\ref{fig:original vD I, vD II, SC transformations (linear)}}, can be attributed to a fortuitous cancellation of errors, rather than to a uniformly valid leading-order asymptotic reduction. This behaviour, however, should be regarded as an exception rather than a general rule. When such cancellation is absent, the same type of leading-order asymptotic reduction can lead to pronounced performance degradation, as demonstrated by the vD~I case in \hyperref[fig:original vD I, vD II, SC transformations (logarithmic)]{figure~\ref{fig:original vD I, vD II, SC transformations (logarithmic)}(a)} and \hyperref[fig:original vD I, vD II, SC transformations (linear)]{figure~\ref{fig:original vD I, vD II, SC transformations (linear)}(a)}, as well as by the VIPL case in \hyperref[fig:original VIPL transformation]{figure~\ref{fig:original VIPL transformation}}.

\section{\emph{A priori} scaling assessment and \emph{a posteriori} skin-friction prediction}
\label{sec:section 5}
As established by the asymptotic analysis in \textsection~\ref{sec:subsection 4.2} and the error-cancellation assessment in \textsection~\ref{sec:subsection 4.3}, the asymptotic representation of $F_\theta$ in \eqref{eq:analytical expression for F_C and F_theta} exhibits significant limitations for practical implementation. Because the Reynolds numbers of interest are generally insufficient to reach the formal asymptotic regime, the classical leading-order asymptotic reduction does not provide a generally reliable approximation unless a fortuitous error cancellation occurs. Moreover, as detailed in Appendix~\ref{app B}, retaining a finite number of higher-order terms does not provide a robust remedy, as the underlying asymptotic series exhibits poor convergence properties. Consequently, we return to the exact-integral formulation~\eqref{eq:final expression for F_C and F_theta} for the subsequent implementation and assessment. For clarity, skin-friction transformations constructed using this formulation are hereafter referred to as `modified' transformations, to distinguish them from their classical counterparts based on leading-order asymptotic reductions.

To maintain consistency with the DNS evaluation of the momentum thickness \eqref{eq:compressible momentum thickness}, the upper limit of the compressible integral $J$ is set by the physical boundary-layer edge $y=\delta_e=\delta_{99}$, which corresponds to $z_e=\bar u_e/u_\infty=0.99$. Consistently, based on the mapped momentum-thickness definition \eqref{eq:incompressible momentum thickness}, the upper limit of the `incompressible' integral $I$ is taken as the same edge expressed in the transformed state, namely $Y=\Delta_e$. Applying the velocity mapping \eqref{eq:the transformed mean velocity profile} gives $Z_e = \psi(z_e) / \psi(1)$. With this choice, the `incompressible' integral $I$ admits the closed-form expression
\begin{equation}
    \begin{aligned}
    I & =\int_0^{Z_e}{Z(1-Z) \exp(aFZ) \, \mathrm{d}Z} \\
      & =\exp(aFZ_e) \left[ \frac{Z_e(1-Z_e)}{aF}+\frac{2Z_e-1}{a^2F^2}-\frac{2}{a^3F^3} \right] +\frac{1}{a^2F^2}+\frac{2}{a^3F^3}.
    \end{aligned}
    \label{eq:analytical expression for I}
\end{equation}
In contrast, the compressible integral $J$ is evaluated numerically.

Once a specific velocity transformation is prescribed, $F_C$ is explicitly determined by the governing flow parameters ($M_\infty, Pr, T_\infty$), the wall-thermal condition ($\varTheta$ or $\bar T_w / T_r$), and the adopted viscosity model (characterized by $n$ or $T_s$). In contrast, the expression for $F_\theta$ involves the parameter $a = \kappa_U u_\infty^+$, which is \emph{a priori} unknown. Following the compatibility analysis in \textsection~\ref{sec:subsection 4.2}, the effective logarithmic slope $\kappa_U$ is determined dynamically by solving \eqref{eq:relation between kappa_u and kappa_theta} together with the CF relation~\eqref{eq:Coles-Fernholz relation} in this section. This treatment establishes the CF relation as the model-consistent incompressible reference for the present closure. Because $\kappa_U$ varies only weakly over the practical range of $Re_{\theta,i}$, the fixed representative value $\kappa_U=0.375$ may also be used in simplified implementations; as shown in \textsection~\ref{sec:subsection 4.2}, it deviates by no more than approximately $1.54\%$ for $10^3 \lesssim Re_{\theta,i} \lesssim 10^5$ and differs by only about $2.34\%$ from its infinite-Reynolds-number asymptote. However, since $a=\kappa_U u_\infty^+$ still involves the unknown free-stream velocity $u_\infty^+$, and since $\kappa_U$ itself depends implicitly on $Re_{\theta,i}$, the transformation remains formally unclosed. This necessitates a two-stage validation strategy: first, an \emph{a priori} assessment using DNS-extracted $u_\infty^+$ to evaluate the intrinsic performance of the modified transformations as skin-friction scalings, followed by an \emph{a posteriori} assessment to evaluate their capability to directly predict the compressible skin-friction coefficient $C_f$.

\begin{figure}
  \centering
  \includegraphics[width=0.49\textwidth]{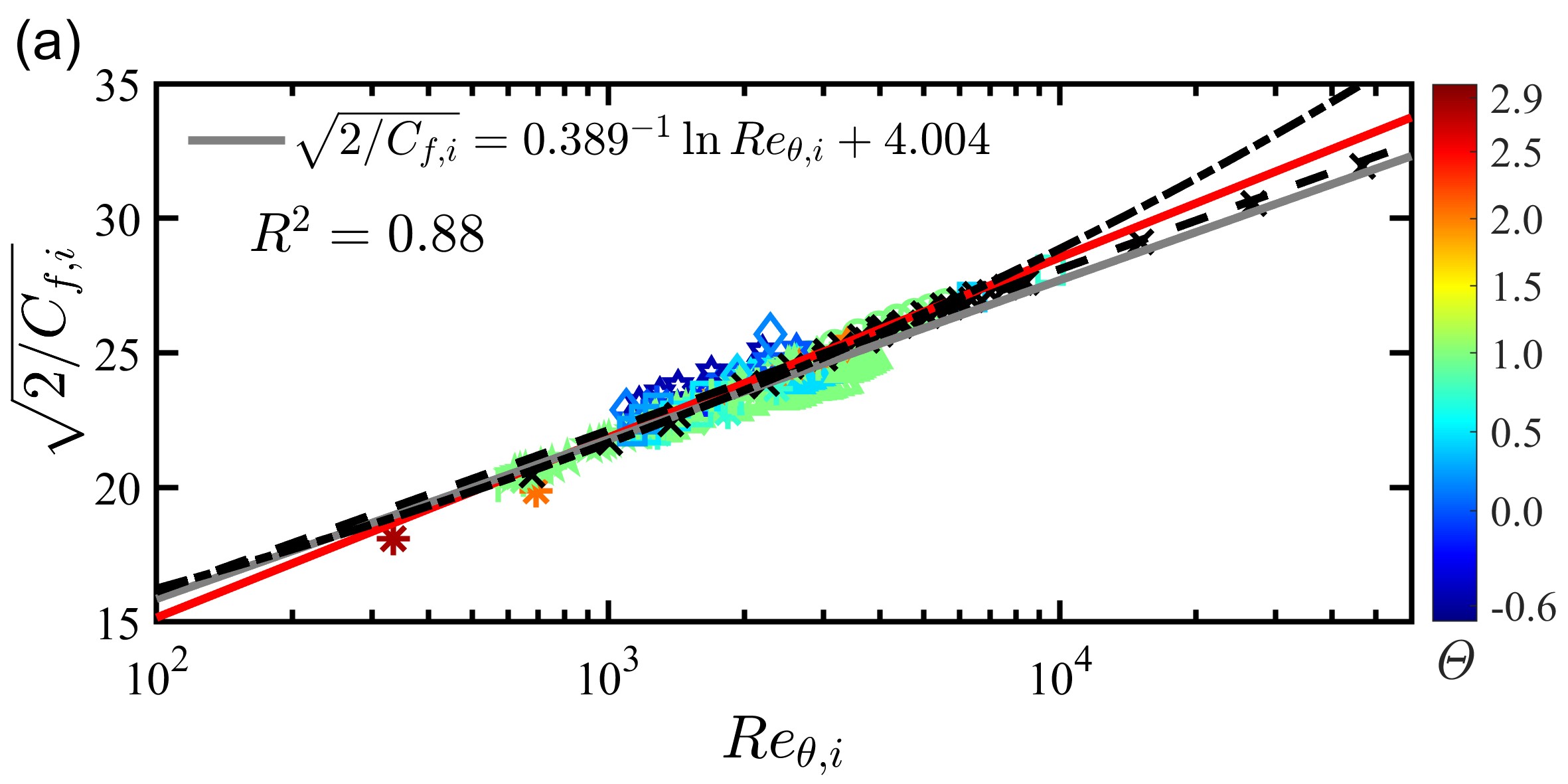}
  \includegraphics[width=0.49\textwidth]{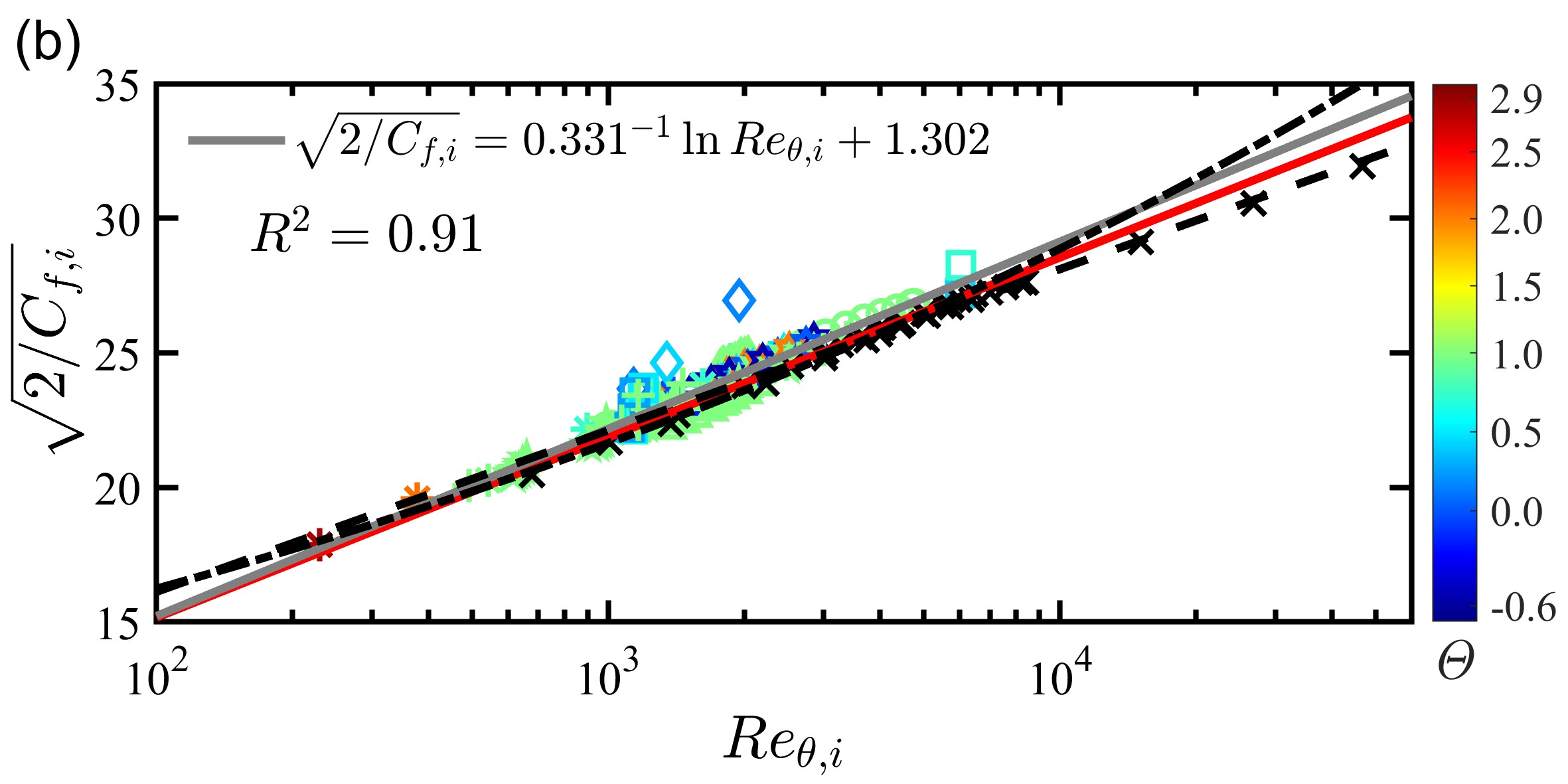}
  \includegraphics[width=0.49\textwidth]{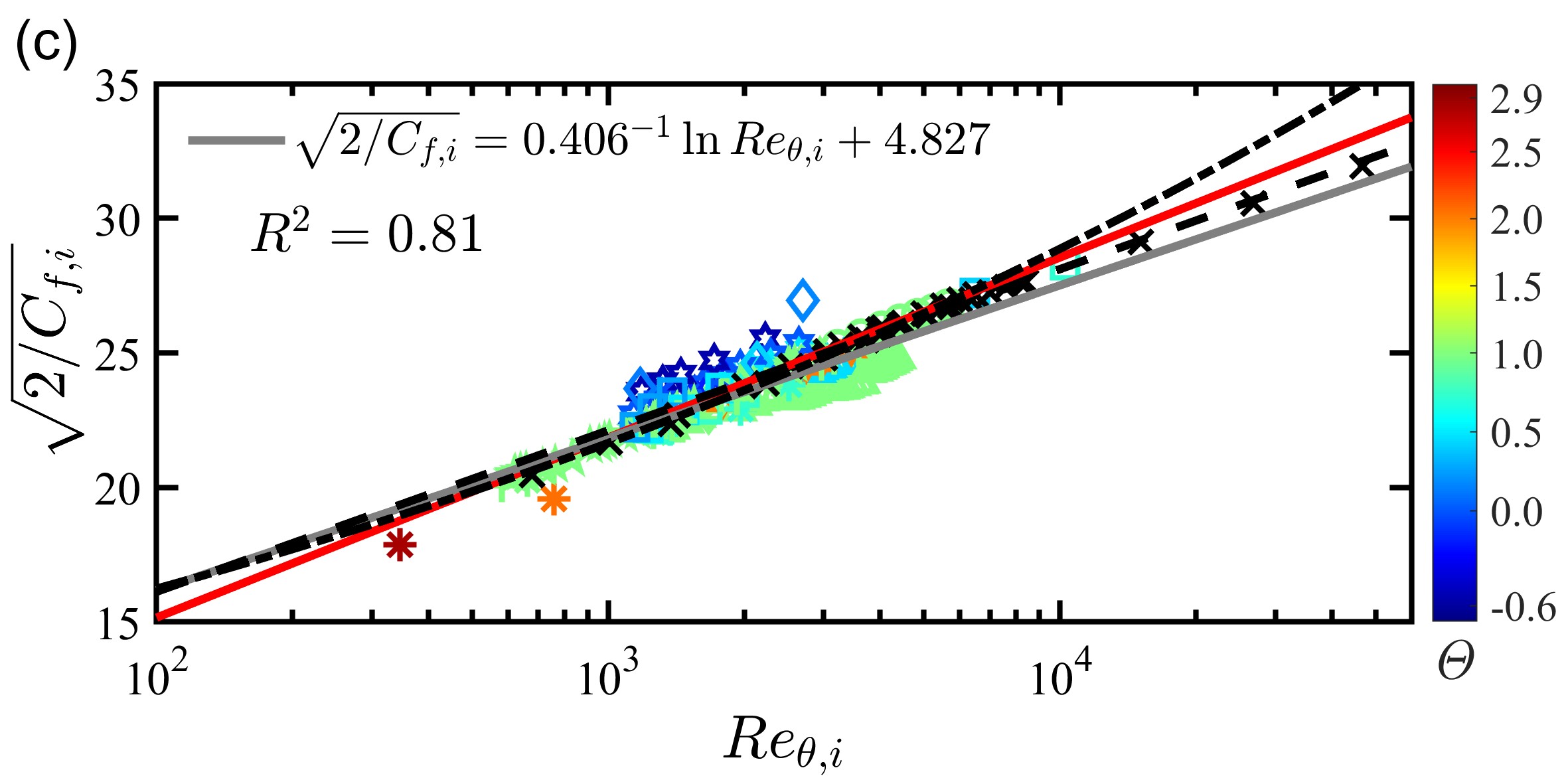}
  \includegraphics[width=0.49\textwidth]{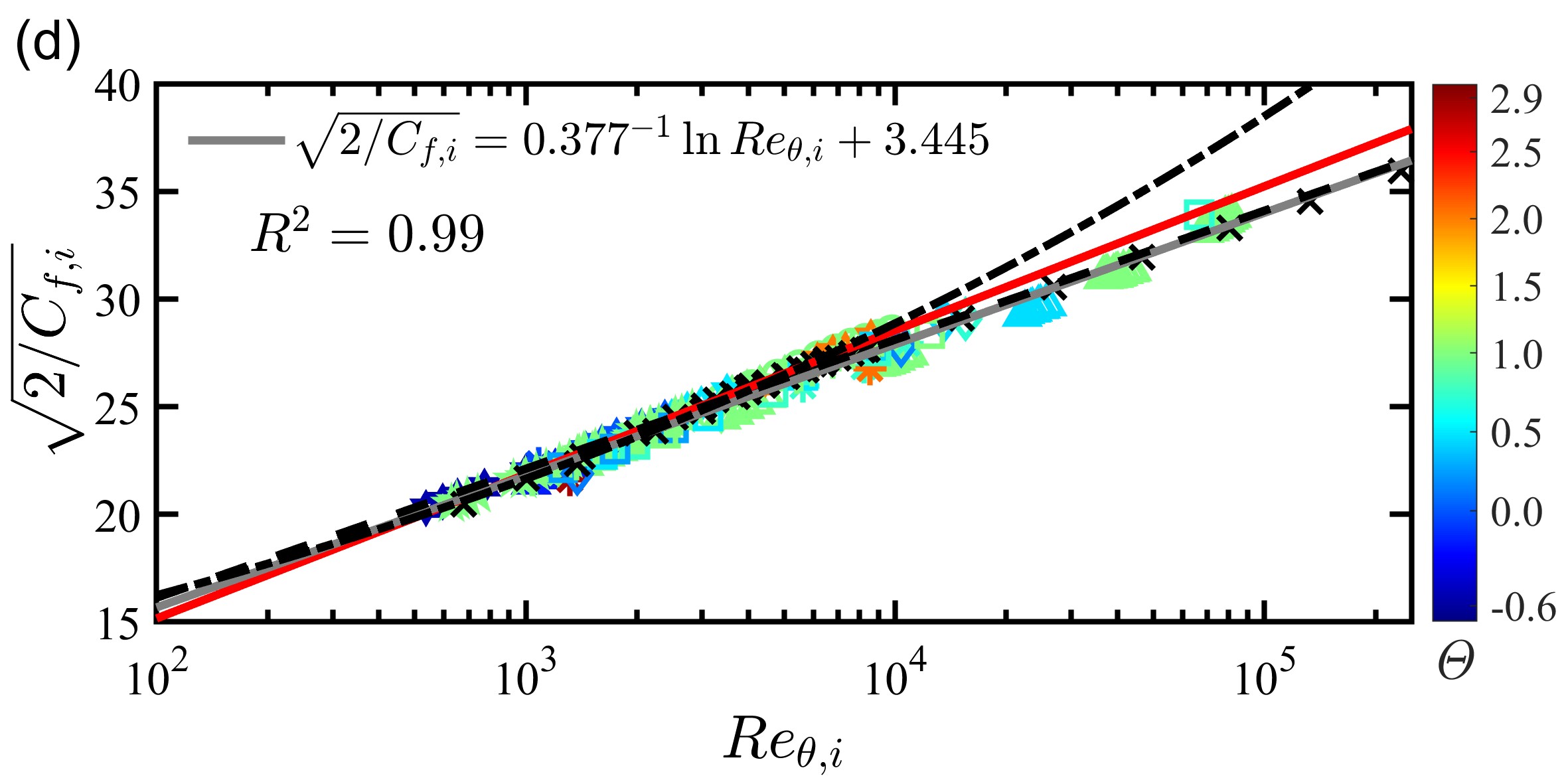}
  \caption{Transformed `incompressible' $\sqrt{2 / C_{f,i}}$ versus $Re_{\theta,i}$ in semi-logarithmic coordinates. (a) Baseline results using the classical vD~II transformation~\eqref{eq:vD II skin friction transformation}. (b--d) Results using the modified (b) vD~I, (c) vD~II, and (d) VIPL transformations based on the exact-integral formulation~\eqref{eq:final expression for F_C and F_theta}, with $\kappa_U$ dynamically determined from the compatibility condition \eqref{eq:relation between kappa_u and kappa_theta} together with the CF relation~\eqref{eq:Coles-Fernholz relation}. Symbols, colours, and line styles follow those in \hyperref[fig:original vD I, vD II, SC transformations (logarithmic)]{figure~\ref{fig:original vD I, vD II, SC transformations (logarithmic)}}.}
\label{fig:modified vD I, vD II, SC transformations (logarithmic)}
\end{figure}

\begin{figure}
  \centering
  \includegraphics[width=0.49\textwidth]{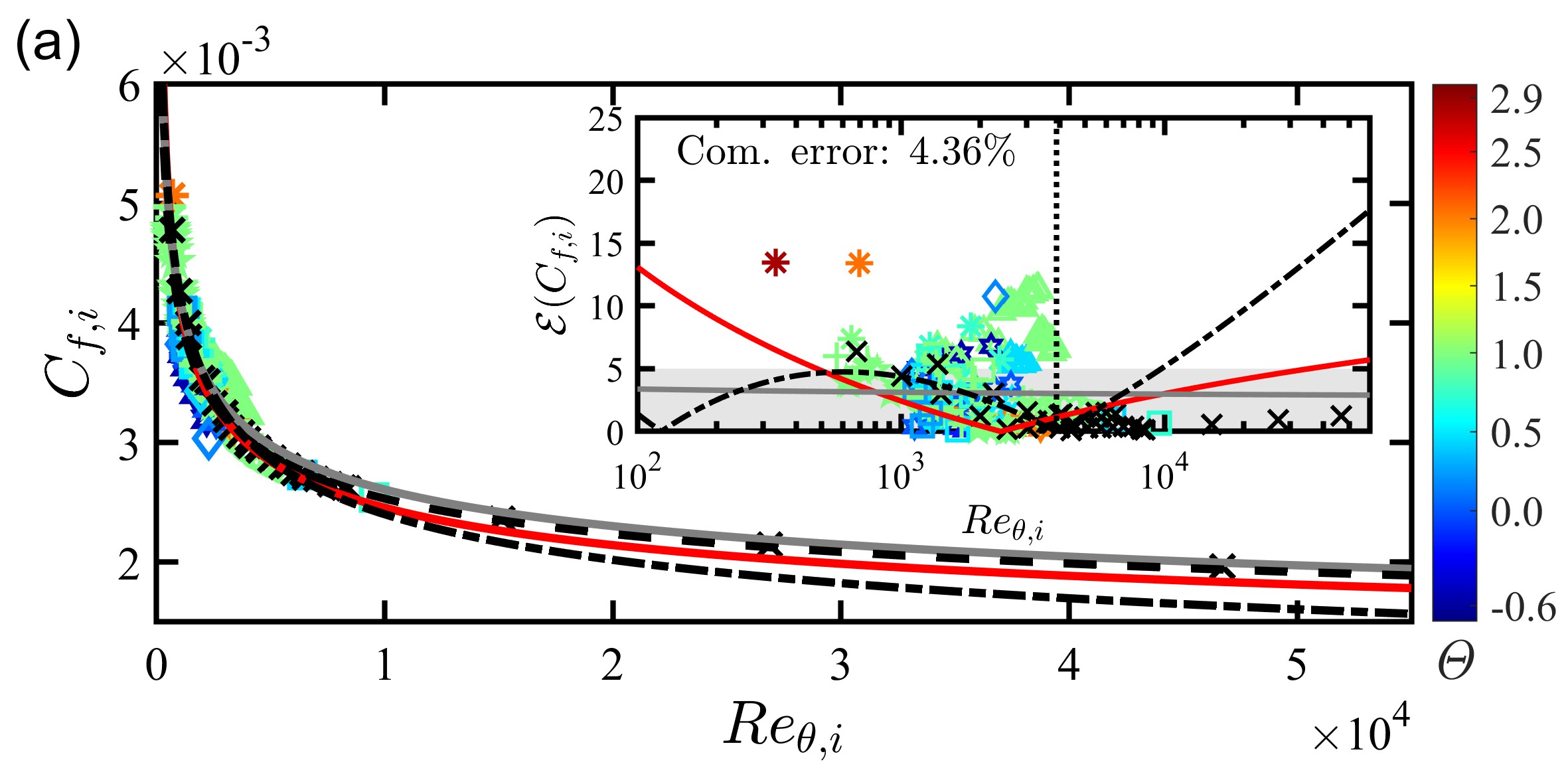}
  \includegraphics[width=0.49\textwidth]{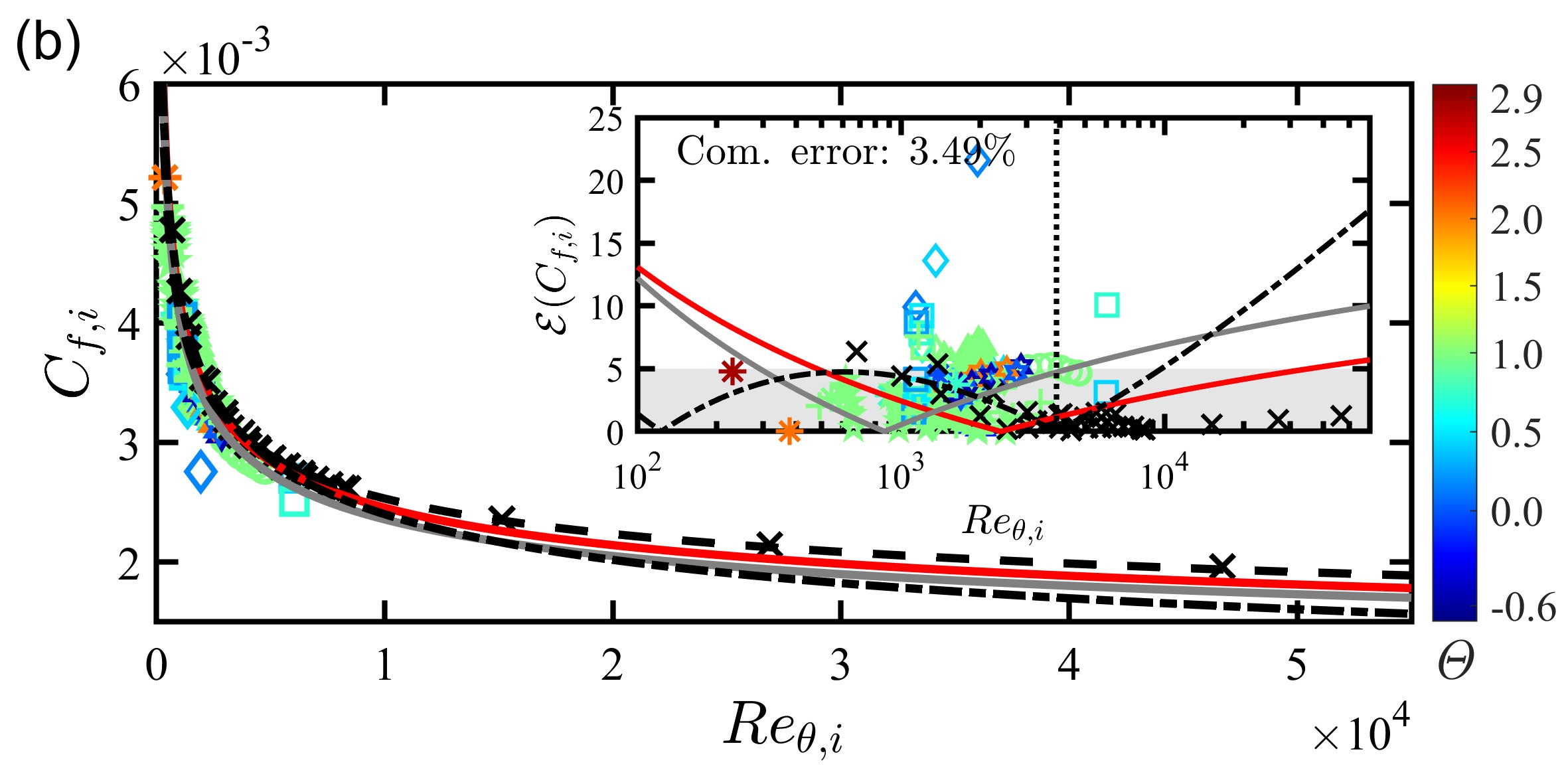}
  \includegraphics[width=0.49\textwidth]{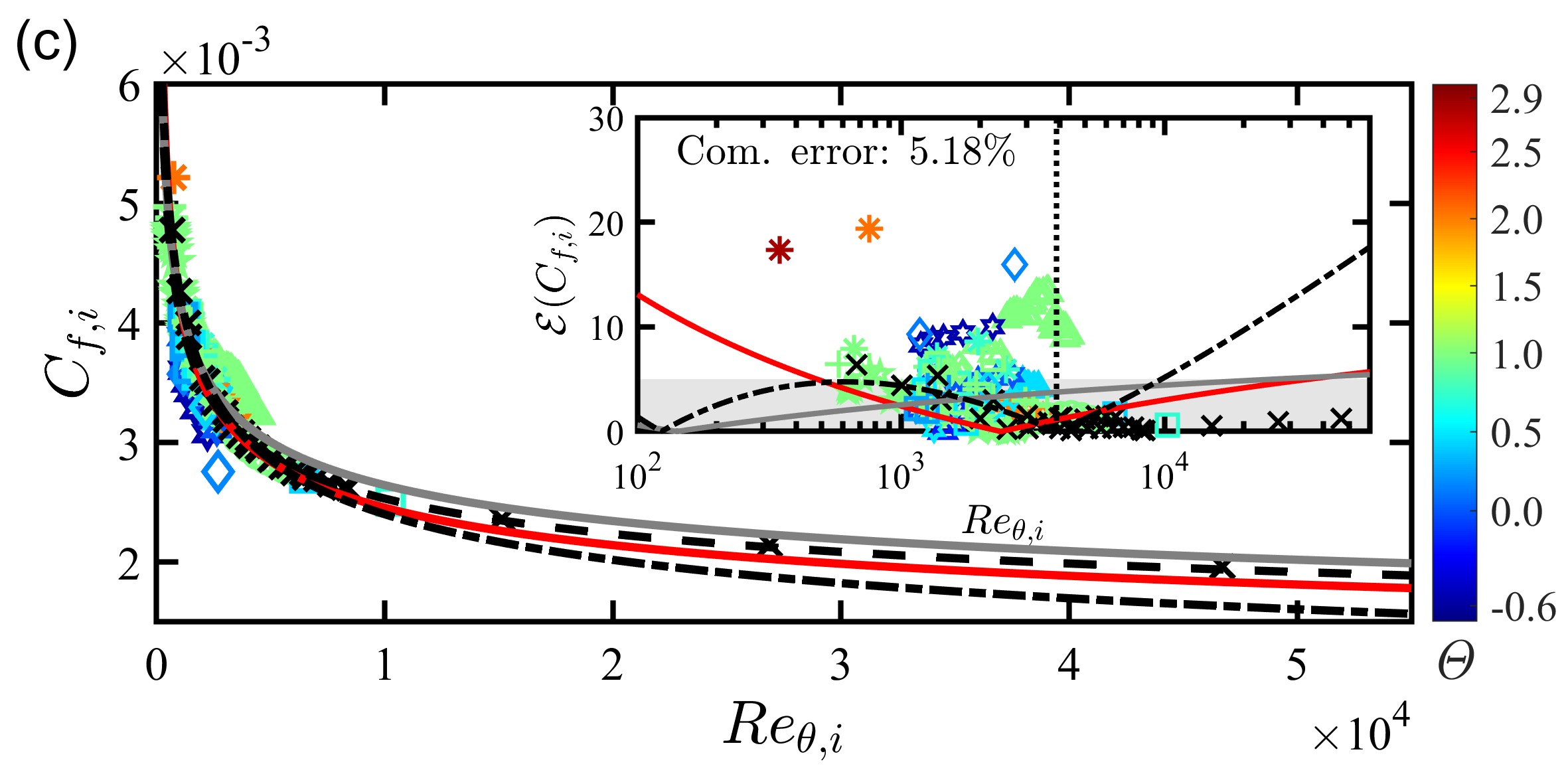}
  \includegraphics[width=0.49\textwidth]{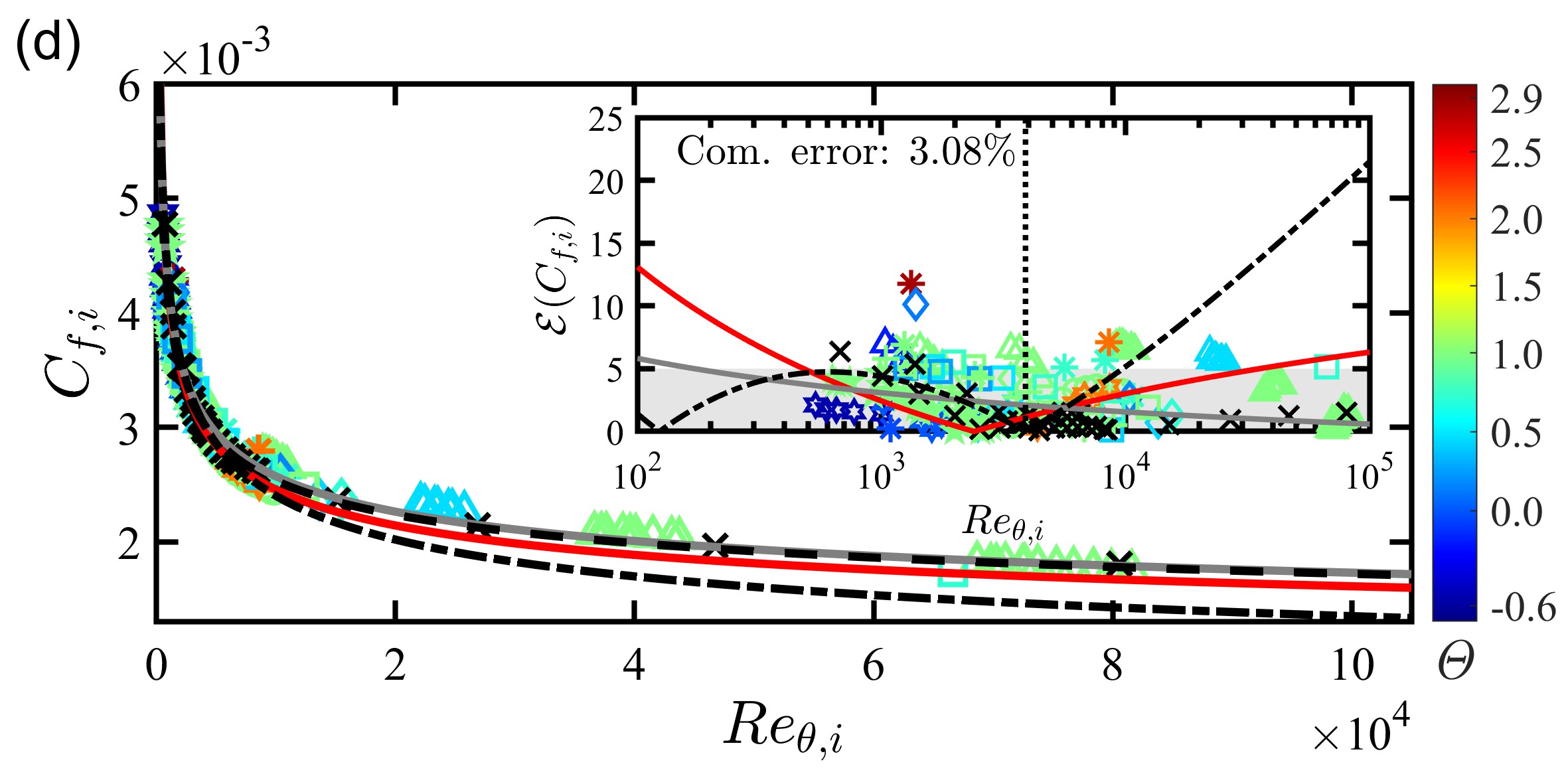}
  \caption{Transformed `incompressible' $C_{f,i}$ versus $Re_{\theta,i}$. (a) Baseline results using the classical vD~II transformation~\eqref{eq:vD II skin friction transformation}. (b--d) Results using the modified (b) vD~I, (c) vD~II, and (d) VIPL transformations based on the exact-integral formulation~\eqref{eq:final expression for F_C and F_theta}, with $\kappa_U$ dynamically determined from the compatibility condition \eqref{eq:relation between kappa_u and kappa_theta} together with the CF relation~\eqref{eq:Coles-Fernholz relation}. Insets show the absolute relative errors $\mathcal{E}(C_{f,i})$ of the transformed data and the plotted correlations, evaluated with respect to the CF relation~\eqref{eq:Coles-Fernholz relation}, which serves as the model-consistent incompressible reference for the present closure, i.e. $\mathcal{E}(C_{f,i}) = \left| C_{f,i} - (C_{f,i})_{\mathrm{CF}} \right| / (C_{f,i})_{\mathrm{CF}} \times 100\%$. The vertical dotted line marks the transition Reynolds number $Re_{\theta,i}^{\mathrm{cr}} \approx 3887$. The mean error over all compressible cases is reported in the upper-left corner of each panel, with the grey shaded region highlighting a 5\% error bound. Symbols, colours and line styles follow those in \hyperref[fig:original vD I, vD II, SC transformations (logarithmic)]{figure~\ref{fig:original vD I, vD II, SC transformations (logarithmic)}}.}
\label{fig:modified vD I, vD II, SC transformations (linear)}
\end{figure}

The results of the \emph{a priori} assessment for the modified vD~I, vD~II, and VIPL transformations are presented in \hyperref[fig:modified vD I, vD II, SC transformations (logarithmic)]{figure~\ref{fig:modified vD I, vD II, SC transformations (logarithmic)}} and \hyperref[fig:modified vD I, vD II, SC transformations (linear)]{figure~\ref{fig:modified vD I, vD II, SC transformations (linear)}}. A key distinction from the assessments in \textsection~\ref{sec:section 2} and \textsection~\ref{sec:section 3} is that the inset errors $\mathcal{E}(C_{f,i})$ in \hyperref[fig:modified vD I, vD II, SC transformations (linear)]{figure~\ref{fig:modified vD I, vD II, SC transformations (linear)}} are now evaluated with respect to the CF relation~\eqref{eq:Coles-Fernholz relation}, which serves as the model-consistent incompressible reference for the present closure. The classical vD~II transformation \eqref{eq:vD II skin friction transformation} is included in panel (a) as a baseline, since it was found in \textsection~\ref{sec:subsection 2.1} to be the most accurate of the classical skin-friction transformations assessed there. It yields a reasonably organized collapse and a mean relative error of 4.36\%.

The modified vD~I transformation in panel (b) demonstrates a marked improvement over its classical counterpart~\eqref{eq:vD I skin friction transformation}, as evidenced by comparisons with \hyperref[fig:original vD I, vD II, SC transformations (logarithmic)]{figure~\ref{fig:original vD I, vD II, SC transformations (logarithmic)}(a)} and \hyperref[fig:original vD I, vD II, SC transformations (linear)]{figure~\ref{fig:original vD I, vD II, SC transformations (linear)}(a)}. The transformed $\sqrt{2/C_{f,i}}$ data exhibit noticeably reduced scatter and are much better organized by a logarithmic trend against $Re_{\theta,i}$. This improvement is especially clear for quasi-adiabatic and heated-wall cases, consistent with the well-established effectiveness of the vD~I velocity transformation~\eqref{eq:vD I velocity transformation} in these regimes. Although residual deviations persist for high-Mach-number flows with moderate-to-strong wall cooling, the systematic underprediction of $C_{f,i}$ produced by the classical vD~I skin-friction transformation~\eqref{eq:vD I skin friction transformation} is substantially reduced. The relative errors with respect to the CF relation remain predominantly below 10\%, with a mean error of 3.49\%.

By contrast, the modified vD~II transformation, shown in panel (c), does not improve upon its classical baseline. Its mean error relative to the CF relation increases from 4.36\% to 5.18\%, and the transformed data exhibit slightly more scatter. This behaviour is consistent with the error-cancellation analysis in \textsection~\ref{sec:subsection 4.3}. For the vD~II velocity transformation~\eqref{eq:vD II velocity transformation}, the ratio between the exact-integral and asymptotic transformation factors, $\mathcal{C}(a)=F_\theta/F_{\theta}^{\text{asy}}$, remains confined to a relatively narrow range, approximately $0.97\lesssim \mathcal{C}(a)\lesssim 1.18$. Thus, the exact-integral formulation modifies the transformed data only modestly, while disrupting part of the fortuitous error cancellation that underlies the apparent success of the classical vD~II transformation. As a result, once the leading-order asymptotic reduction is removed, the modified vD~I transformation yields a lower error than its vD~II counterpart when assessed against the model-consistent CF relation, especially for quasi-adiabatic cases. This reversal aligns with the historical preference for the vD~I velocity mapping~\eqref{eq:vD I velocity transformation} over the vD~II mapping~\eqref{eq:vD II velocity transformation}, supporting the interpretation that the poor performance of the classical vD~I skin-friction scaling mainly reflects its leading-order asymptotic reduction rather than an intrinsic deficiency of the underlying velocity mapping.

The modified VIPL transformation in panel (d) exhibits the best overall performance among the modified transformations assessed here. It demonstrates broad consistency with the direct mapping-based assessment in \textsection~\ref{sec:subsection 3.2}, where the VIPL velocity transformation was substituted directly into \eqref{eq:a general formula for Cf transformation} and \eqref{eq:a general formula for Re_theta transformation} (see \hyperref[fig:GFM and VIPL transformations (present without modification, logarithmic)]{figure~\ref{fig:GFM and VIPL transformations (present without modification, logarithmic)}(b)} and \hyperref[fig:GFM and VIPL transformations (present without modification, linear)]{figure~\ref{fig:GFM and VIPL transformations (present without modification, linear)}(b)}). It also provides a substantial improvement over the leading-order asymptotic VIPL skin-friction transformations~\eqref{eq:VIPL skin friction transformation, power law} and \eqref{eq:VIPL skin friction transformation, Sutherland law}, as shown by comparison with \hyperref[fig:original VIPL transformation]{figure~\ref{fig:original VIPL transformation}}. In the present semi-analytical implementation, the transformed $\sqrt{2/C_{f,i}}$ data exhibit substantially reduced scatter compared with the leading-order asymptotic result and are well organized by a logarithmic trend against $Re_{\theta,i}$, with $R^2=0.99$. The relative errors of the transformed $C_{f,i}$ remain mostly within the 5\% error band over a large portion of the database and are generally below 7\% for $Re_{\theta,i}\gtrsim 2000$, yielding a mean error of 3.08\% relative to the CF relation. It is noteworthy that, although the CF relation serves as the model-consistent incompressible reference for the inset errors, the modified VIPL-transformed data also reproduce the low-Reynolds-number curvature represented by the power-law correlation~\eqref{eq:Smits relation} for $Re_{\theta,i}\leq Re_{\theta,i}^{\mathrm{cr}}$, before approaching the CF relation~\eqref{eq:Coles-Fernholz relation}. Indeed, if the composite incompressible reference~\eqref{eq:incompressible skin-friction correlation} adopted in \textsection~\ref{sec:section 2}--\ref{sec:section 3} were used instead to define the error, the low-Reynolds-number deviations would be further reduced and the overall mean error would decrease to 2.54\% (not shown). This favourable finite-Reynolds-number behaviour is not explicitly imposed by the closure, since the dynamic determination of $\kappa_U$ is tied to the CF relation rather than to the composite reference. Rather, it indicates that the exact-integral formulation, when combined with the VIPL mapping, retains the main finite-Reynolds-number behaviour expected from the mapping-based formulation while introducing a practically useful semi-analytical closure. Despite these improvements, residual deviations persist primarily in high-Mach-number diabatic cases, reflecting the inherited outer-layer limitations of the VIPL velocity transformation identified in \textsection~\ref{sec:subsection 3.2}.

Overall, these results indicate that the semi-analytical exact-integral closure developed in \textsection~\ref{sec:section 4} largely preserves the skin-friction behaviour implied by the underlying velocity mappings, while avoiding the unreliable leading-order asymptotic truncation. This structural consistency with the direct mapping-based formulation of \textsection~\ref{sec:subsection 3.2} is explicit for the VIPL case through the preceding comparisons; analogous diagnostic checks for the vD~I and vD~II mappings, not shown here for brevity, yield the same qualitative alignment.

Having established the intrinsic \emph{a priori} scaling performance of the modified transformations, we now assess their capability for the conditional standalone prediction of the local compressible skin-friction coefficient $C_f$ at a prescribed $Re_\theta$, which serves as a scalar measure of the boundary-layer development state. Here, `standalone' signifies that the prediction requires neither DNS-extracted wall quantities nor wall-normal mean-flow profiles once the relevant macroscopic inputs are prescribed; it does not imply simultaneous prediction of $Re_{\theta} (x)$ or of the complete streamwise evolution of the boundary layer. The required inputs are $M_\infty$, $Re_\theta$, $\bar T_w/T_r$ (or $\varTheta$), $Pr$ and $T_\infty$, together with the viscosity-law parameters, namely $n$ for the power-law model or $T_s$ for Sutherland's law. The CF relation~\eqref{eq:Coles-Fernholz relation} is adopted as the incompressible closure not merely as a high-Reynolds-number correlation, but because it serves as the model-consistent incompressible reference used to dynamically determine $\kappa_U$ in the preceding exact-integral formulation. No parameter in this primary predictive closure is fitted to the compressible cases evaluated below. Consequently, the definition of the skin-friction transformation \eqref{eq:skin friction transformation} implies that $C_f$ follows the compressible extension given by \eqref{eq:Coles-Fernholz relation (compressible extension)}. For the classical transformations \eqref{eq:vD I skin friction transformation}, \eqref{eq:vD II skin friction transformation} and \eqref{eq:SC skin friction transformation}, the factors $F_C$ and $F_\theta$ are explicit functions of the prescribed inputs. Hence, \eqref{eq:Coles-Fernholz relation (compressible extension)} constitutes an explicit algebraic equation for $C_f$ that can be evaluated directly. By contrast, for the modified transformations based on the exact-integral formulation, $F_\theta$ depends on the unknown parameter $a=\kappa_U u_\infty^+$ through the integrals $I(a)$ and $J(a)$, while $\kappa_U$ is dynamically coupled to $a$ through the compatibility condition~\eqref{eq:relation between kappa_u and kappa_theta}. This coupling renders \eqref{eq:Coles-Fernholz relation (compressible extension)} an implicit nonlinear equation, which is solved here using a Newton--Raphson iteration with $a$ as the iteration variable. Once convergence is reached, $C_f$ is directly obtained from the corresponding values of $a$ and $\kappa_U$. For all cases considered, the iteration converges rapidly to a unique physically admissible solution.

\begin{figure}
  \centering
  \includegraphics[width=0.49\textwidth]{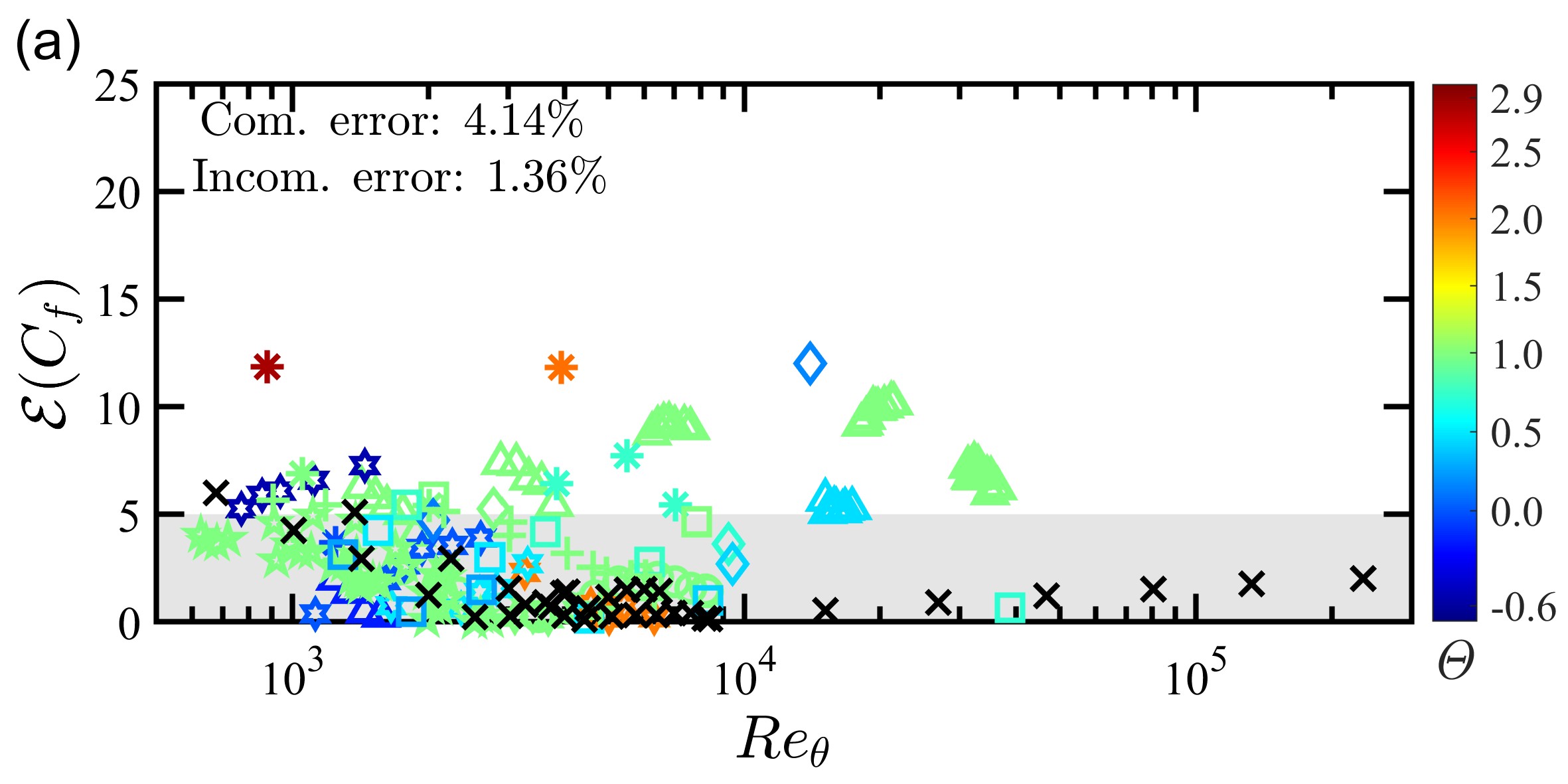}
  \includegraphics[width=0.49\textwidth]{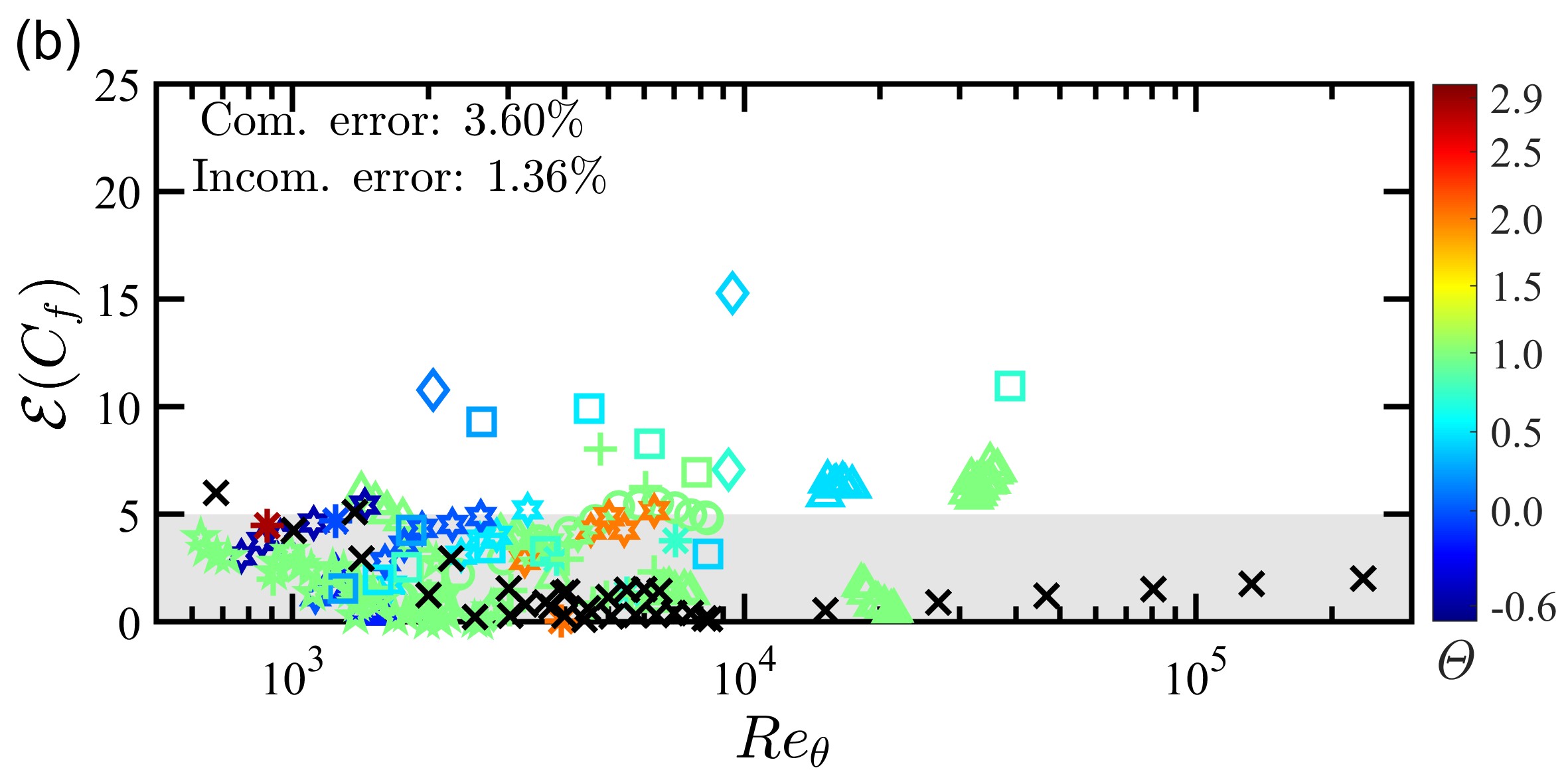}
  \includegraphics[width=0.49\textwidth]{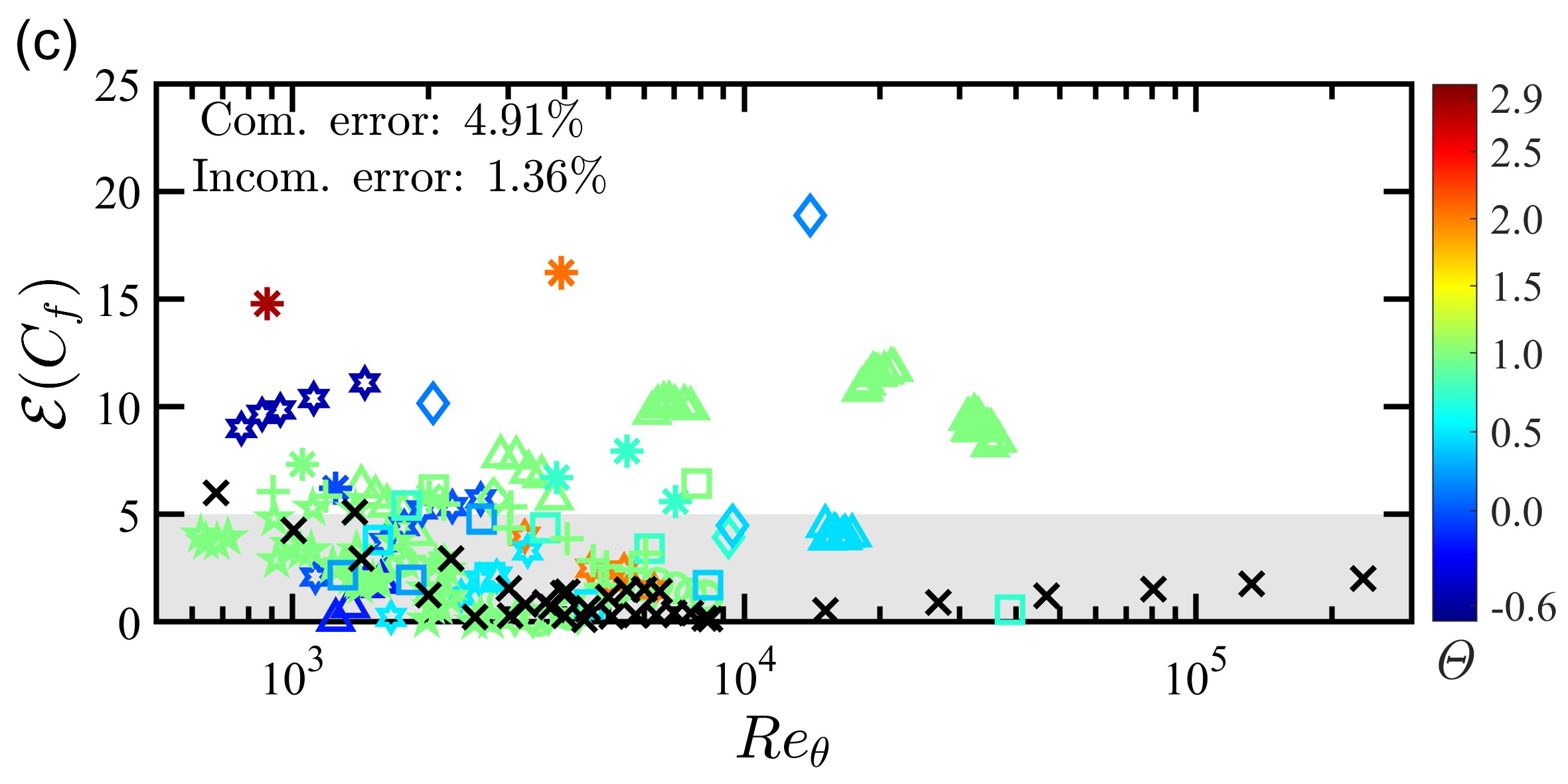}
  \includegraphics[width=0.49\textwidth]{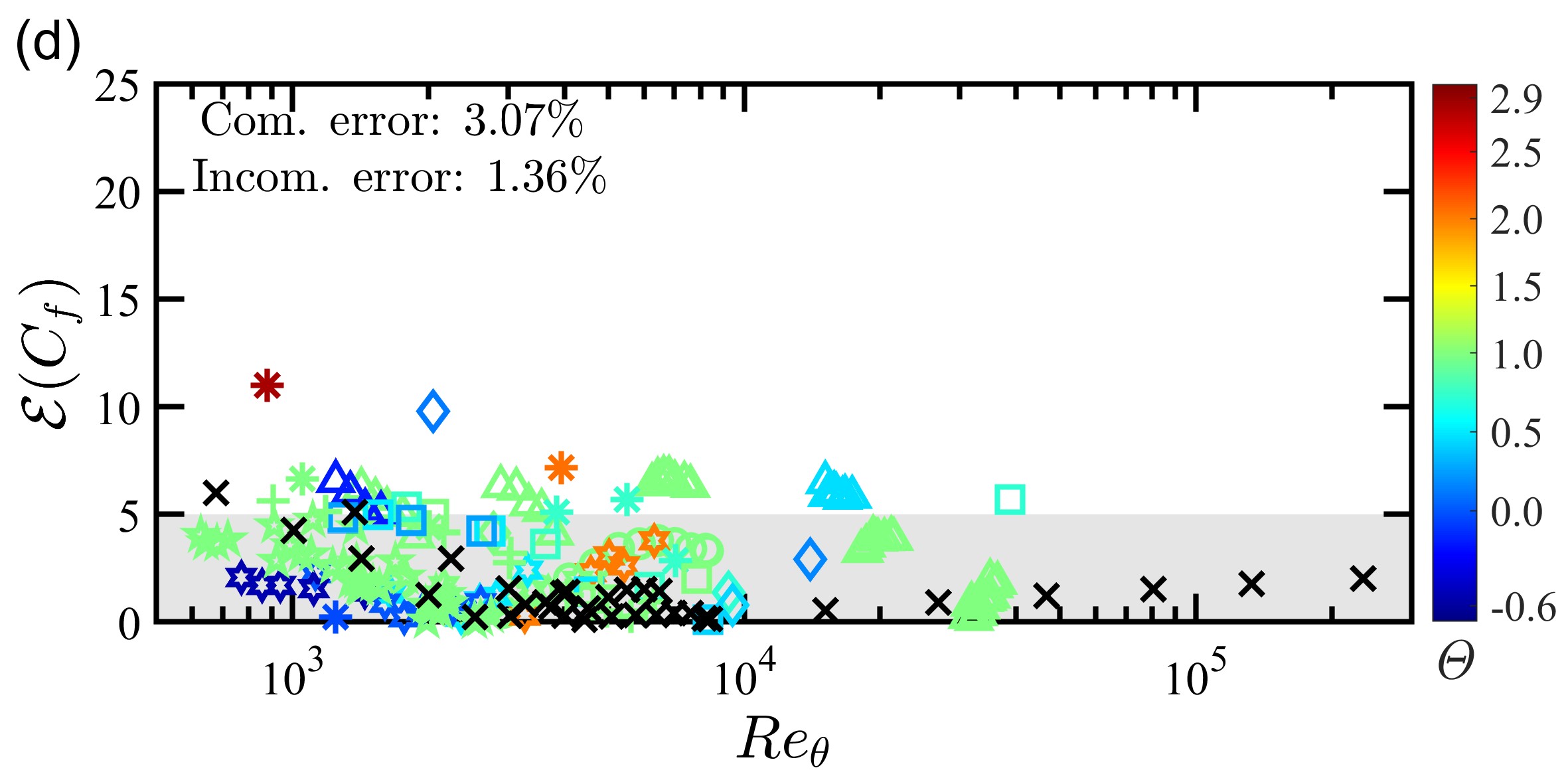}
  \caption{Absolute relative errors of the predicted skin-friction coefficient $C_f$, defined as $\mathcal{E}(C_f) = | C_{f} - C_{f,\mathrm{ref}} | / C_{f,\mathrm{ref}} \times 100\%$, where $C_{f,\mathrm{ref}}$ denotes the DNS value for ZPG CTBL cases and the experimental or DNS value for ZPG ITBL cases. Errors are plotted as a function of $Re_\theta$. (a) Baseline results using the classical vD~II transformation~\eqref{eq:vD II skin friction transformation}. (b--d) Results obtained using the modified (b) vD~I, (c) vD~II, and (d) VIPL transformations based on the exact-integral formulation~\eqref{eq:final expression for F_C and F_theta}, with $\kappa_U$ dynamically determined from the compatibility condition \eqref{eq:relation between kappa_u and kappa_theta} together with the CF relation~\eqref{eq:Coles-Fernholz relation}. Predictions for the incompressible reference cases are denoted by black $\times$ symbols. The mean errors over the compressible and incompressible datasets are reported in the top-left corner of each panel, with the grey shaded region highlighting a 5\% error bound. All predictions employ the CF relation~\eqref{eq:Coles-Fernholz relation} as the incompressible closure. Symbols and colours follow those in \hyperref[fig:original vD I, vD II, SC transformations (logarithmic)]{figure~\ref{fig:original vD I, vD II, SC transformations (logarithmic)}}.}
    \label{fig:error of predicted Cf (CF relation)}
\end{figure}
   
\hyperref[fig:error of predicted Cf (CF relation)]{Figure~\ref{fig:error of predicted Cf (CF relation)}(b--d)} presents the absolute relative error distributions for the \emph{a posteriori} $C_f$ predictions obtained using the modified vD~I, vD~II, and VIPL transformations, respectively. For comparison, the results obtained using the classical vD~II transformation are included in panel (a) as a baseline, yielding a mean error of 4.14\% over the compressible cases. Overall, the observed predictive performance is broadly consistent with the \emph{a priori} assessment presented in \hyperref[fig:modified vD I, vD II, SC transformations (logarithmic)]{figure~\ref{fig:modified vD I, vD II, SC transformations (logarithmic)}} and \hyperref[fig:modified vD I, vD II, SC transformations (linear)]{figure~\ref{fig:modified vD I, vD II, SC transformations (linear)}}. As anticipated, the modified vD~II transformation performs worse than its classical counterpart \eqref{eq:vD II skin friction transformation}, with the mean error increasing to 4.91\% and the deviations for several quasi-adiabatic cases approaching 10\%. In contrast, the modified vD~I transformation yields reliable predictions under quasi-adiabatic and heated-wall conditions, with the corresponding relative errors confined within 8\%. Although its performance degrades in the presence of wall cooling, with maximum errors reaching approximately 15\%, it nevertheless outperforms the classical vD~II baseline, reducing the overall mean error to 3.60\% over all compressible cases. The modified VIPL transformation demonstrates the best overall predictive performance among the approaches assessed here, with relative errors for the entire compressible database remaining below 11\% and a further reduced mean error of 3.07\%. Notably, the errors remain below 7\% for all quasi-adiabatic cases. However, its predictive accuracy for cases combining high Mach numbers with moderate-to-strong wall heat transfer remains to be improved. Finally, since all methods employ the CF relation~\eqref{eq:Coles-Fernholz relation} as the incompressible closure, their predictions for the incompressible reference cases yield identical relative errors, with a mean value of 1.36\%.

Finally, it is worth emphasizing that the present \emph{a posteriori} assessment adopts the CF relation~\eqref{eq:Coles-Fernholz relation} as a prescribed, model-consistent incompressible closure. Accordingly, the errors reported in \hyperref[fig:error of predicted Cf (CF relation)]{figure~\ref{fig:error of predicted Cf (CF relation)}} quantify the performance of a prescribed, profile-free predictive model rather than in-sample residuals obtained by fitting a closure to the same compressible database. Because the same DNS collection also informs the preceding \emph{a priori} assessments, these results are presented as an \emph{a posteriori} assessment on the present database rather than as validation against an independent dataset. A diagnostic variant based on method-specific logarithmic closures fitted to the transformed compressible data is examined in Appendix~\ref{app C}. Although such calibration can reduce residuals over the fitted compressible database, the resulting errors are in-sample diagnostics and must additionally be assessed for consistency with the expected ZPG ITBL behaviour. This issue is particularly relevant for transformations that organize mapped compressible data around an empirical logarithmic trend that nevertheless deviates from the incompressible reference behaviour. The results based on the CF relation in \hyperref[fig:error of predicted Cf (CF relation)]{figure~\ref{fig:error of predicted Cf (CF relation)}} are therefore retained as the primary \emph{a posteriori} prediction assessment, whereas Appendix~\ref{app C} is used only to evaluate the potential practical utility and incompressible-limit consistency of the corresponding calibrated variants.

\section{Summary and conclusions}
\label{sec:conclusions}
This study recasts classical skin-friction transformations for canonical smooth-wall ZPG CTBLs into a mapping-based, exact-integral formulation, thereby constructing \emph{a priori} scalings and standalone \emph{a posteriori} predictions of $C_f$ that remain consistent with the compressible law of the wall implied by the prescribed velocity transformation over practical finite-Reynolds-number ranges.

As a starting point, the classical vD~I~\eqref{eq:vD I skin friction transformation}, vD~II~\eqref{eq:vD II skin friction transformation} and SC~\eqref{eq:SC skin friction transformation} transformations are systematically reassessed using an extensive published DNS database. The assessment is not restricted to the empirical collapse of the transformed compressible data, but asks whether the transformed $(C_{f,i},Re_{\theta,i})$ pairs recover the reference skin-friction behaviour of ZPG ITBLs. None of these classical transformations yields a uniformly satisfactory agreement with the incompressible baseline \eqref{eq:incompressible skin-friction correlation} across the broad range of Mach numbers and wall-thermal conditions considered, although the vD~II transformation exhibits the best overall performance. Their relative performance is nevertheless not readily explained by their stated rationales: the semi-analytical vD~I and vD~II transformations do not behave as expected from their respective mixing-length arguments, while the empirical wall-temperature correction in the SC transformation does not systematically improve diabatic cases. Furthermore, while the recent nominal-thickness scaling~\eqref{eq:transformation factors of Zhao and Fu} of \citet{zhao2025revisiting} improves the profile-based organization of the transformed data when combined with the GFM velocity transformation~\eqref{eq:GFM transformation}, this improvement does not constitute a genuine recovery of the mapped incompressible state. The retained density weighting in the transformed $Re_{\theta,i}$ \eqref{eq:implied transformed variables (Zhao and Fu)} leaves residual Mach-number and wall-temperature effects, and this limitation becomes more evident when the same construction is combined with the VIPL transformation~\eqref{eq:VIPL transformation}. More critically for predictive applications, the formulation requires available mean-flow profiles and therefore does not provide a standalone \emph{a posteriori} procedure for predicting $C_f$ based solely on prescribed macroscopic flow parameters.

To address these limitations, the mapped target state is specified as a constant-property ZPG ITBL associated with the physical ZPG CTBL. Standard incompressible definitions in this state directly give the transformed $C_{f,i}$ and $Re_{\theta,i}$, so that the transformation factors $F_C$~\eqref{eq:a general formula for Cf transformation} and $F_\theta$~\eqref{eq:a general formula for Re_theta transformation} emerge as exact consequences of the prescribed mean-velocity and wall-normal-coordinate mappings~\eqref{eq:definition of velocity transformation}, rather than as additional empirical corrections. This definition-first approach also clarifies the conceptual distinction from the nominal-thickness construction of \citet{zhao2025revisiting}: $\theta_i$~\eqref{eq:incompressible momentum thickness} is evaluated entirely within the mapped constant-property state, whereas $\hat \theta$~\eqref{eq:definition of theta_hat} retains the compressible density weighting. The present formulation therefore separates the transformed kinematics from residual density-weighting effects and exposes the full-layer performance of the prescribed velocity mapping. Direct applications to the GFM and VIPL velocity transformations confirm this diagnostic role, showing that the remaining deviations from the incompressible baseline arise mainly from outer-layer errors in the mapped velocity, especially the mapped free-stream value $U_\infty^+$, which enters both $C_{f,i}=2/U_\infty^{+2}$ and $Re_{\theta,i}=U_\infty^+\theta_i^+$. Thus, the practical success of these mapping-based formulations is constrained by the outer-layer accuracy of existing velocity transformations, especially at high Mach numbers with appreciable wall heat transfer.

To facilitate practical applications, the mapping-based relations~\eqref{eq:a general formula for Cf transformation} and \eqref{eq:a general formula for Re_theta transformation} are closed semi-analytically by replacing DNS-extracted profiles with prescribed modelling ingredients. Specifically, exact-integral expressions for the transformation factors $F_C$ and $F_\theta$ are derived in \eqref{eq:final expression for F_C and F_theta} by incorporating an effective logarithmic representation of the mapped `incompressible' velocity profile, the inverse form of the prescribed velocity transformation, and the mean thermodynamic closures~\eqref{eq:mean T-u relation}--\eqref{eq:mean mu-T relation (Sutherland law)}. A leading-order asymptotic analysis of \eqref{eq:analytical expression for F_C and F_theta} shows that, in the formal limit $a\to\infty$, the classical vD~I and vD~II transformations are recovered when their respective vD-type velocity mappings~\eqref{eq:vD I velocity transformation} and \eqref{eq:vD II velocity transformation} are combined with Walz's equation~\eqref{eq:Walz's equation}. Thus, van Driest's theory is recast as a set of leading-order reductions of the present mapping-based exact-integral formulation. The finite-Reynolds-number analysis shows, however, that these reductions are not uniformly reliable over practical parameter ranges: the correction $1-2/(aF)$ associated with the `incompressible' integral $I$ remains non-negligible, while the asymptotic series $\mathcal{D}(a)$ associated with the compressible integral $J$ exhibits poor convergence. The relative success of the classical vD~II transformation is therefore explained by a fortuitous cancellation of these two finite-Reynolds-number truncation effects through the correction factor $\mathcal{C}(a)$ in \eqref{eq:error cancellation factor}, rather than by a generally valid leading-order asymptotic approximation or a superior physical description of ZPG CTBLs.

Guided by this finite-Reynolds-number analysis, the exact-integral formulation~\eqref{eq:final expression for F_C and F_theta} is retained in \textsection~\ref{sec:section 5} to construct modified skin-friction transformations. In this implementation, the effective logarithmic slope $\kappa_U$ is determined through compatibility with the CF relation~\eqref{eq:Coles-Fernholz relation}, which therefore serves as the model-consistent incompressible closure for the modified transformations. The \emph{a priori} assessment shows that these modified transformations largely preserve the skin-friction behaviour implied by their underlying velocity mappings. Accordingly, the modified vD~I transformation substantially outperforms its classical counterpart, whereas the modified vD~II transformation loses part of the fortuitous error cancellation supporting the classical vD~II formula. Among the formulations considered, the VIPL-based modified skin-friction transformation proposed here gives the best overall \emph{a priori} scaling. The same exact-integral formulation further enables standalone \emph{a posteriori} iterative prediction of $C_f$ from prescribed macroscopic inputs, namely $M_\infty$, $Re_\theta$, $\bar T_w/T_r$ or $\varTheta$, $Pr$, $T_\infty$ and the viscosity-law constants, without requiring DNS-extracted wall quantities or mean-flow profiles. The predictive trends are broadly consistent with the preceding \emph{a priori} assessment. Among the formulations considered and across the investigated flow regimes, $0.30 \leq M_\infty \leq 13.64$ and $-0.55 \leq \varTheta \leq 2.85$, the VIPL-based modified transformation gives the best overall performance, with relative errors below 11\% for all compressible cases and a mean error of 3.07\%, compared with 4.14\% for the classical vD~II baseline. The modified vD~I transformation also improves the mean prediction error to 3.60\%, despite some deterioration under high-Mach-number cooled-wall conditions, whereas the modified vD~II transformation remains less effective after the removal of its beneficial asymptotic error cancellation, with a mean error of 4.91\%. Overall, the exact-integral formulation removes the unreliable leading-order asymptotic truncation while maintaining consistency with the compressible law of the wall implied by the underlying velocity mapping. The remaining errors are primarily linked to the outer-layer limitations of existing velocity transformations, especially for high-Mach-number flows with appreciable wall heat transfer.

Taken together, the present study establishes a mapping-based, exact-integral basis for skin-friction transformations. The definition-first mapped-state construction determines the transformation factors $F_C$ and $F_\theta$ from their standard definitions once the mapped constant-property state and the associated mean-velocity and wall-normal-coordinate mappings are prescribed. The finite-Reynolds-number exact-integral reformulation of van Driest's theory further shows that the classical vD~I and vD~II formulae emerge as leading-order asymptotic reductions that are not uniformly reliable over practical Reynolds-number ranges, with the relative success of vD~II traced to a fortuitous cancellation of finite-Reynolds-number truncation errors. Together, these two elements address two longstanding criticisms of classical approaches: the transformation factors are no longer introduced as arbitrary empirical adjustments, and the resulting finite-Reynolds-number scalings remain consistent with the compressible law of the wall implied by the prescribed velocity transformation.

From a practical standpoint, the modified skin-friction transformations provide a closed, standalone procedure for rapidly predicting $C_f$ and generating model-consistent $C_f$--$Re_\theta$ relations for canonical ZPG CTBLs over the Mach-number, Reynolds-number and wall-thermal ranges considered here, thereby supporting data assessment and parametric exploration of compressibility and wall-temperature effects. Beyond this use, the resulting $C_f$ may in principle be combined with a prescribed Reynolds-analogy factor to provide a secondary estimate of wall heat transfer \citep{vanDriest1951turbulent,hopkins1971evaluation}; however, its accuracy is necessarily limited by the empirical closure adopted for $s$. The modified transformations may also be incorporated into numerical mean-flow prediction approaches, such as that of \citet{ying2025general}, as compatible skin-friction closures for the simultaneous reconstruction of mean velocity and temperature profiles. These possible extensions are not assessed here and are left for future work.

Several limitations should nevertheless be made explicit. The present analysis is restricted to canonical smooth-wall ZPG CTBLs of calorically perfect, air-like gases, consistent with the DNS database summarized in \hyperref[tab:the ZPG CTBLs database 1]{tables~\ref{tab:the ZPG CTBLs database 1}--\ref{tab:the ZPG CTBLs database 2}}. Accordingly, the exact-integral implementation is closed with the quadratic TV relation~\eqref{eq:mean T-u relation}, the calorically perfect-gas equation of state together with the boundary-layer pressure approximation~\eqref{eq:mean rho-T relation}, and either the power-law~\eqref{eq:mean mu-T relation (power law)} or Sutherland viscosity law~\eqref{eq:mean mu-T relation (Sutherland law)}. Applications to flows with non-air-like viscosity behaviour, supercritical thermophysical properties, or strong real-gas effects would require replacing these thermodynamic and transport-property closures. Because the formulation remains tied to a prescribed velocity transformation, its accuracy also inherits the full-layer performance of that mapping. The residual errors observed for high-Mach-number diabatic cases therefore highlight the need for more robust full-layer or composite velocity transformations, especially in the outer region. Extensions to pressure-gradient boundary layers, rough walls, shock/boundary-layer interactions, high-enthalpy reacting flows and other complex configurations remain to be assessed in future work.



\backsection[Acknowledgements]{The authors gratefully acknowledge the researchers listed in \hyperref[tab:the ZPG CTBLs database 1]{tables~\ref{tab:the ZPG CTBLs database 1}--\ref{tab:the ZPG CTBLs database 2}} for making their direct numerical simulation data available, either through public repositories or private communications. The authors also thank the reviewers for their constructive comments and suggestions, which helped improve the quality of the paper. The OpenAI GPT-5 language model (ChatGPT, \href{https://chat.openai.com}{https://chat.openai.com}) was used solely to improve the grammar and readability of the manuscript and did not contribute to the research content, data analysis or physical interpretations.}

\backsection[Funding]{This work was supported by the National Natural Science Foundation of China (NSFC grant no. U25A6006).}

\backsection[Declaration of interests]{The authors report no conflict of interest.}





\appendix

\section{$F_C$ and $F_{\theta}^{\mathrm{asy}}$ for specific mean velocity scalings}\label{app A}

\subsection{The vD~I-type transformations}\label{app A.1}
\citet{vanDriest1951turbulent} postulated that Prandtl's mixing length hypothesis, $\ell = \kappa y$, remains valid within the logarithmic region of ZPG CTBLs. Invoking the constant-stress-layer approximation and neglecting the viscous contribution, the shear-stress balance $\bar{\tau}_w \approx \bar{\rho} \ell^{2} (\partial \bar{u}/ \partial y)^{2}$ yields
\begin{equation}
    \sqrt{\bar{\rho}^+}\frac{\partial \bar{u}^+}{\partial y^+}=\frac{1}{\kappa y^+}.
    \label{eq:vD I shear stress equation}
\end{equation}
Introducing the transformations
\begin{equation}
    \mathrm{d}Y_{\mathrm{vD\,I}}^+ = \mathrm{d} y^+, \quad \mathrm{d} \bar U_{\mathrm{vD\,I}}^+ = \sqrt{\bar \rho^{+}} \mathrm{d}\bar u^{+},
    \label{eq:vD I velocity transformation}
\end{equation}
the transformed velocity profile $\bar U_{\mathrm{vD\,I}}^+(Y_{\mathrm{vD\,I}}^+)$ is readily shown to satisfy the `incompressible' counterpart of \eqref{eq:vD I shear stress equation}. Its solution follows the classical log-law, $\bar U_{\mathrm{vD\,I}}^+ = \kappa^{-1} \ln Y_{\mathrm{vD\,I}}^+ + C$. Consequently, the physical hypothesis $\ell = \kappa y$ is mathematically equivalent to prescribing the vD~I velocity transformation~\eqref{eq:vD I velocity transformation} with $f_I = 1$ and $g_I = \sqrt{\bar{\rho}^+}$ \citep{zhang2012mach}.

Extending this transformation to the entire boundary layer and incorporating the thermodynamic relation~\eqref{eq:mean rho-T relation}, the corresponding kernel functions become
\begin{equation}
    \mathcal{F}(z) = 1, \quad
    \mathcal{G}(z) = ( 1 + B z - A^2 z^2 )^{-1/2},
    \label{eq:F(z) and G(z), vD I}
\end{equation}
which further yield
\begin{equation}
    \varphi(z) = \frac{z(1-z)}{( 1+Bz-A^2z^2 )^{3/2}}, \quad
    \psi(z) = \frac{1}{A}\left[ \sin ^{-1} \left(\frac{2A^2z-B}{\sqrt{B^2 + 4A^2}} \right) + \sin ^{-1}\beta \right].
\label{eq:phi(z) and psi(z), vD I}
\end{equation}
Accordingly, the free-stream velocity ratio is obtained as
\begin{equation}
    F = \psi(1) = \frac{1}{A}( \sin ^{-1}\alpha+\sin ^{-1}\beta),
    \label{eq:F, vD I}
\end{equation}
where $\alpha = (2A^2 - B) / (B^2 + 4A^2)^{1/2}$, $\beta = B / (B^2 + 4A^2)^{1/2}$, with $A$ and $B$ defined in \textsection~\ref{sec:subsection 4.2}.

Substituting \eqref{eq:F(z) and G(z), vD I}-\eqref{eq:F, vD I} into the general formulation~\eqref{eq:analytical expression for F_C and F_theta} and taking the asymptotic limit $a \rightarrow \infty$ yields the vD~I skin-friction transformation,
\begin{equation}
    ( F_C )_{\mathrm{vD\,I}}=\frac{ (T_r/T_{\infty}-1) + (1 - sPr)(\bar T_w - T_r) / T_\infty}{( \sin ^{-1}\alpha +\sin ^{-1}\beta ) ^2}, \quad
    ( F_{\theta}^{\mathrm{asy}} ) _{\mathrm{vD\,I}}=\sqrt{\frac{T_{\infty}}{\bar{T}_w}}\frac{\mu _{\infty}}{\bar{\mu}_w}.
    \label{eq:vD I skin friction transformation (asymptotic solution)}
\end{equation}
It is noteworthy that if Walz's equation~\eqref{eq:Walz's equation} is adopted instead of \eqref{eq:mean T-u relation} (i.e., replacing the coefficients with $\alpha_0, \beta_0, A_0, B_0$ as defined in \textsection~\ref{sec:section 2}), \eqref{eq:vD I skin friction transformation (asymptotic solution)} fully recovers the classical form~\eqref{eq:vD I skin friction transformation}.

\subsection{The vD~II-type transformations}\label{app A.2}
In contrast, \citet{vanDriest1956problem} adopted von Kármán's mixing length, $\ell=-\kappa(\partial \bar u/\partial y)/(\partial ^2\bar u/\partial y^2)$, within the logarithmic region. Accordingly, the shear-stress balance yields
\begin{equation}
    \frac{\partial ^2\bar u^+}{\partial y^{+2}} = -\sqrt{\bar \rho^+} \kappa \left( \frac{\partial \bar u^+}{\partial y^{+}} \right)^2.
    \label{eq:vD II shear stress equation}
\end{equation}
Introducing the transformations
\begin{equation}
    \mathrm{d}Y_{\mathrm{vD\,II}}^+ = \sqrt{\bar \rho^{+}} \mathrm{d} y^+, \quad \mathrm{d} \bar U_{\mathrm{vD\,II}}^+ = \sqrt{\bar \rho^{+}} \mathrm{d}\bar u^{+},
    \label{eq:vD II velocity transformation}
\end{equation}
the transformed velocity profile $\bar U_{\mathrm{vD\,II}}^+(Y_{\mathrm{vD\,II}}^+)$ satisfies the `incompressible' counterpart of \eqref{eq:vD II shear stress equation}. Its solution, $\bar U_{\mathrm{vD\,II}}^+ = \kappa^{-1} \ln(Y_{\mathrm{vD\,II}}^+ + \kappa^{-1}) + C$, recovers the classical logarithmic law for $Y^+ \gg \kappa^{-1}$. Consequently, adopting von Kármán's mixing length is mathematically equivalent to prescribing the vD~II velocity transformation~\eqref{eq:vD II velocity transformation} with $f_I = g_I = \sqrt{\bar{\rho}^+}$. To the authors’ knowledge, this equivalence has rarely been stated explicitly in the existing literature, with the notable exception of \citet{huang1993skin}.

Extending this transformation to the entire boundary layer and incorporating the thermodynamic relation~\eqref{eq:mean rho-T relation}, the kernel functions become
\begin{equation}
    \mathcal{F}(z) = \mathcal{G}(z) = ( 1 + Bz - A^2 z^2 )^{-1/2}.
    \label{eq:F(z) and G(z), vD II}
\end{equation}
Since $\mathcal{G}(z)$ is identical to that in the vD~I case, the velocity mapping function $\psi(z)$ and the free-stream velocity ratio $F$ remain unchanged, as given by \eqref{eq:phi(z) and psi(z), vD I} and \eqref{eq:F, vD I}, respectively. Note, however, that the distinct form of $\mathcal{F}(z)$ leads to a different auxiliary function $\varphi(z)$. Consequently, substituting these into \eqref{eq:analytical expression for F_C and F_theta} and taking the asymptotic limit $a \rightarrow \infty$ yields the vD~II skin-friction transformation,
\begin{equation}
    ( F_C )_{\mathrm{vD\,II}}=\frac{ (T_r/T_{\infty}-1) + (1 - sPr)(\bar T_w - T_r) / T_\infty}{( \sin ^{-1}\alpha +\sin ^{-1}\beta ) ^2}, \quad
    ( F_{\theta}^{\mathrm{asy}} ) _{\mathrm{vD\,II}}=\frac{\mu _{\infty}}{\bar{\mu}_w}.
    \label{eq:vD II skin friction transformation (asymptotic solution)}
\end{equation}
Likewise, adopting Walz's equation~\eqref{eq:Walz's equation} instead recovers the classical form~\eqref{eq:vD II skin friction transformation} exactly.

\subsection{The VIPL-type transformations}\label{app A.3}
We now consider the VIPL velocity transformation proposed by \citet{volpiani2020data}, which is defined by the kernel functions $f_I = (\bar \rho^+)^{1/2} (\bar \mu^+)^{-3/2}$ and $g_I = (\bar \rho^+)^{1/2} (\bar \mu^+)^{-1/2}$, as given in \eqref{eq:VIPL transformation}. To ensure consistency with the mixing length perspectives discussed in \textsection~\ref{app A.1} and \textsection~\ref{app A.2}, this transformation is shown to be equivalent to prescribing a compressible mixing length of $\ell^* = \kappa y_{\mathrm{V}}^+$ within the logarithmic region \citep{zhu2024velocity}.

We first examine the case where the viscosity-temperature dependence is modelled by the power law. Combining the thermodynamic relations~\eqref{eq:mean rho-T relation} and \eqref{eq:mean mu-T relation (power law)}, the kernel functions are
\begin{equation}
    \mathcal{F}(z) = ( 1 + Bz - A^2 z^2 )^{-(1+3n)/2}, \quad \mathcal{G}(z) = ( 1 + Bz - A^2 z^2 )^{-(1+n)/2},
    \label{eq:F(z) and G(z), VIPL, power law}
\end{equation}
which implies:
\begin{equation}
    \varphi(z) = \frac{z(1-z)}{( 1+Bz-A^2z^2)^{1-n}}, \quad \psi(z) = \int_0^{z}{\left( 1 + B\zeta - A^2\zeta^2 \right)^{-(1+n)/2} \, \mathrm{d}\zeta}.
    \label{eq:phi(z) and psi(z), VIPL, power law}
\end{equation}
Accordingly, the free-stream velocity ratio is expressed as:
\begin{equation}
    F = \psi(1) = \int_0^{1}{\left( 1 + B\zeta - A^2\zeta^2 \right)^{-(1+n)/2} \, \mathrm{d}\zeta}.
    \label{eq:F, VIPL, power law}
\end{equation}
Notably, $\psi(z)$ and $F$ admit analytical expressions. Let $m = (1+n)/2$, and let $z_1$ and $z_2$ denote the two real roots of the quadratic equation $1 + B z - A^2 z^2 = 0$. By introducing the affine transformation $t = (z-z_1)/(z_2-z_1)$, the resulting function $\psi(z)$ can be expressed as
\begin{equation}
    \psi(z) = \chi(t) = A^{-2m} (z_2 - z_1)^{1-2m} [ \beta_{\mathrm{inc}}(t;1-m,1-m) - \beta_{\mathrm{inc}}(t_0;1-m,1-m) ],
\label{eq:final expression for psi, VIPL, power law}
\end{equation}
where $t_0 = -z_1/(z_2-z_1)$, $t_1 = (1-z_1)/(z_2-z_1)$, and $F = \chi(t_1)$. The incomplete beta function $\beta_{\mathrm{inc}} (t;a,b)$ is defined by:
\begin{equation}
    \beta_{\mathrm{inc}} (t;a,b) = \int_0^t{x^{a-1}(1-x)^{b-1}\, \mathrm{d}x}.
    \label{eq:definition of the incomplete beta function}
\end{equation}

Substituting \eqref{eq:F(z) and G(z), VIPL, power law}-\eqref{eq:final expression for psi, VIPL, power law} into the general formulation~\eqref{eq:analytical expression for F_C and F_theta} and taking the asymptotic limit $a \rightarrow \infty$ yields the VIPL skin-friction transformation:
\begin{equation}
    ( F_C ) _{\mathrm{VIPL}}=\frac{\rho_\infty}{\bar \rho_w} [\chi(t_1)]^{-2}, \quad
    ( F_{\theta}^{\mathrm{asy}} )_{\mathrm{VIPL}}=\frac{\bar{\mu}_w}{\mu_{\infty}}.
    \label{eq:VIPL skin friction transformation, power law}
\end{equation}

If Sutherland's law~\eqref{eq:mean mu-T relation (Sutherland law)} is adopted instead, a similar procedure leads to
\begin{equation}
    ( F_C ) _{\mathrm{VIPL}}=\frac{\rho_\infty}{\bar \rho_w} F^{-2}, \quad
    ( F_{\theta}^{\mathrm{asy}} )_{\mathrm{VIPL}}=\frac{\bar{\mu}_w}{\mu_{\infty}},
    \label{eq:VIPL skin friction transformation, Sutherland law}
\end{equation}
where the free-stream velocity ratio is expressed as:
\begin{equation}
    F = \psi (1) = \int_0^1{( 1+B\zeta-A^2\zeta^2 )^{-5/4} \left( \frac{1+{T}_s/\bar{T}_w}{1+B\zeta-A^2\zeta^2+{T}_s/\bar{T}_w} \right) ^{-1/2} \, \mathrm{d}\zeta.}
    \label{eq:final expression for F, VIPL, Sutherland's law}
\end{equation}
It is worth noting that the expression for $F_\theta^{\mathrm{asy}}$ remains identical to that in \eqref{eq:VIPL skin friction transformation, power law}, whereas the complex expression for $F$ in \eqref{eq:final expression for F, VIPL, Sutherland's law} does not admit a closed-form solution and must therefore be evaluated numerically.

\section{Assessment of finite-order asymptotic approximations for $J$}\label{app B}
To investigate whether the asymptotic series $1+\mathcal{D}(a)$ in \eqref{eq:approximate expression for J} yields a reliable approximation when truncated at a finite order, we introduce the partial sum $J(N)$, obtained by retaining terms up to $n = N$:
\begin{equation}
    J(N) = -\frac{w_1(1)}{a^2\psi^{\prime}(1)} \exp(aF) \sum_{n=0}^N{(-a)^{-n}\frac{w_{n+1}(1)}{w_1(1)}}.
    \label{eq:partial sum of J}
\end{equation}
Given the recursive definition in \eqref{eq:recursive functions definition}, the algebraic complexity of the coefficients $w_n(z)$ increases rapidly with $n$, making analytical derivations beyond the first few orders impractical for predictive applications. Consequently, the present analysis is restricted to truncation orders $N \le 3$. By definition, the partial sum $J(N)$ approaches the exact integral $J$ as $a \rightarrow \infty$ for any fixed $N$. Evaluating the zeroth-order truncation ($N=0$) directly assesses the validity of the leading-order asymptotic approximation ($1+\mathcal{D}(a) \approx 1$) employed in Appendix~\ref{app A} to derive $F_\theta^{\text{asy}}$. Furthermore, examining $J(N)$ for $N=1, 2,$ and $3$ allows us to determine whether finite higher-order corrections can systematically improve accuracy within the practical parameter range $5 \lesssim a \lesssim 12$ covered by the present DNS datasets.

\begin{figure}
    \centering
    \includegraphics[width=0.32\textwidth]{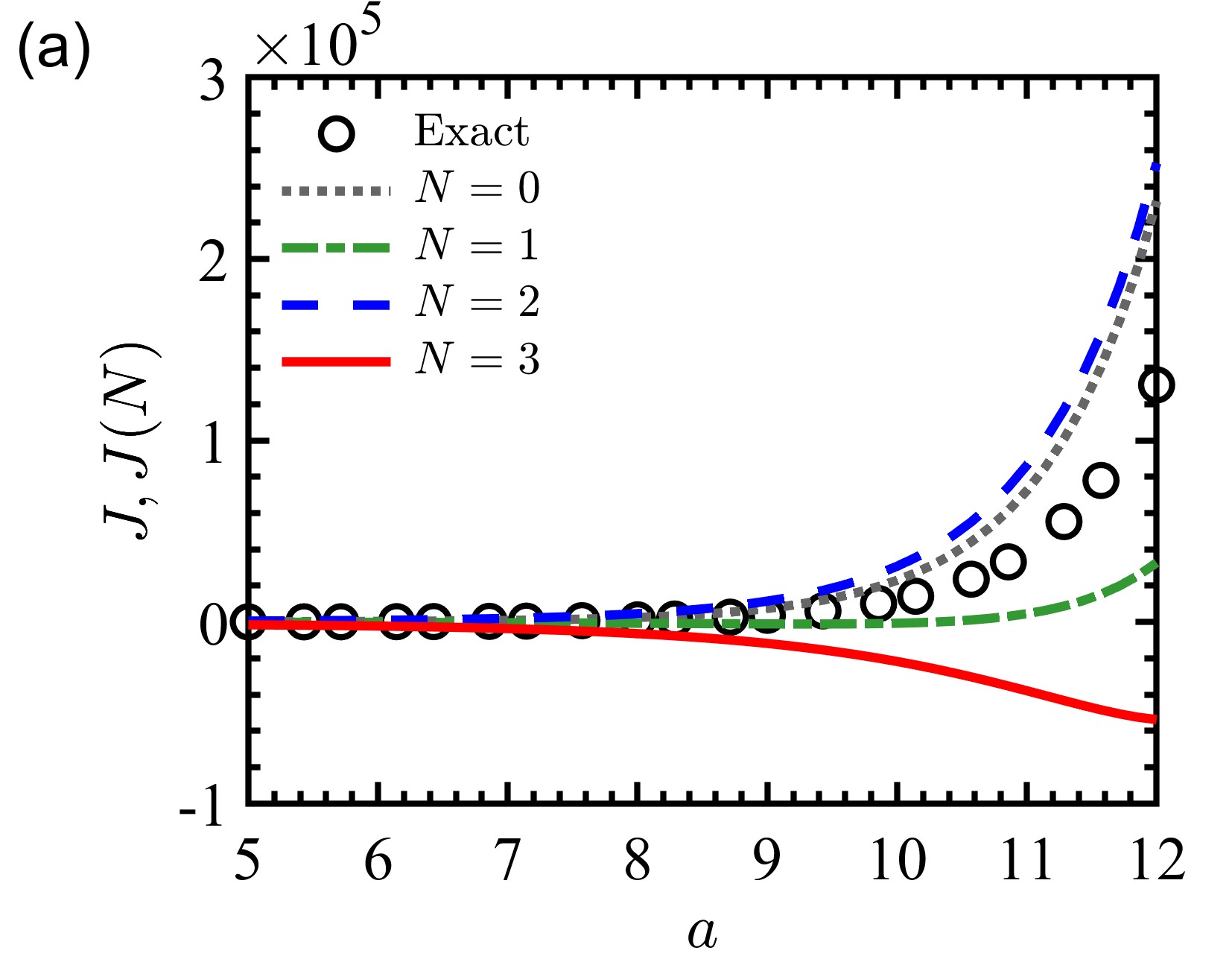}
    \includegraphics[width=0.32\textwidth]{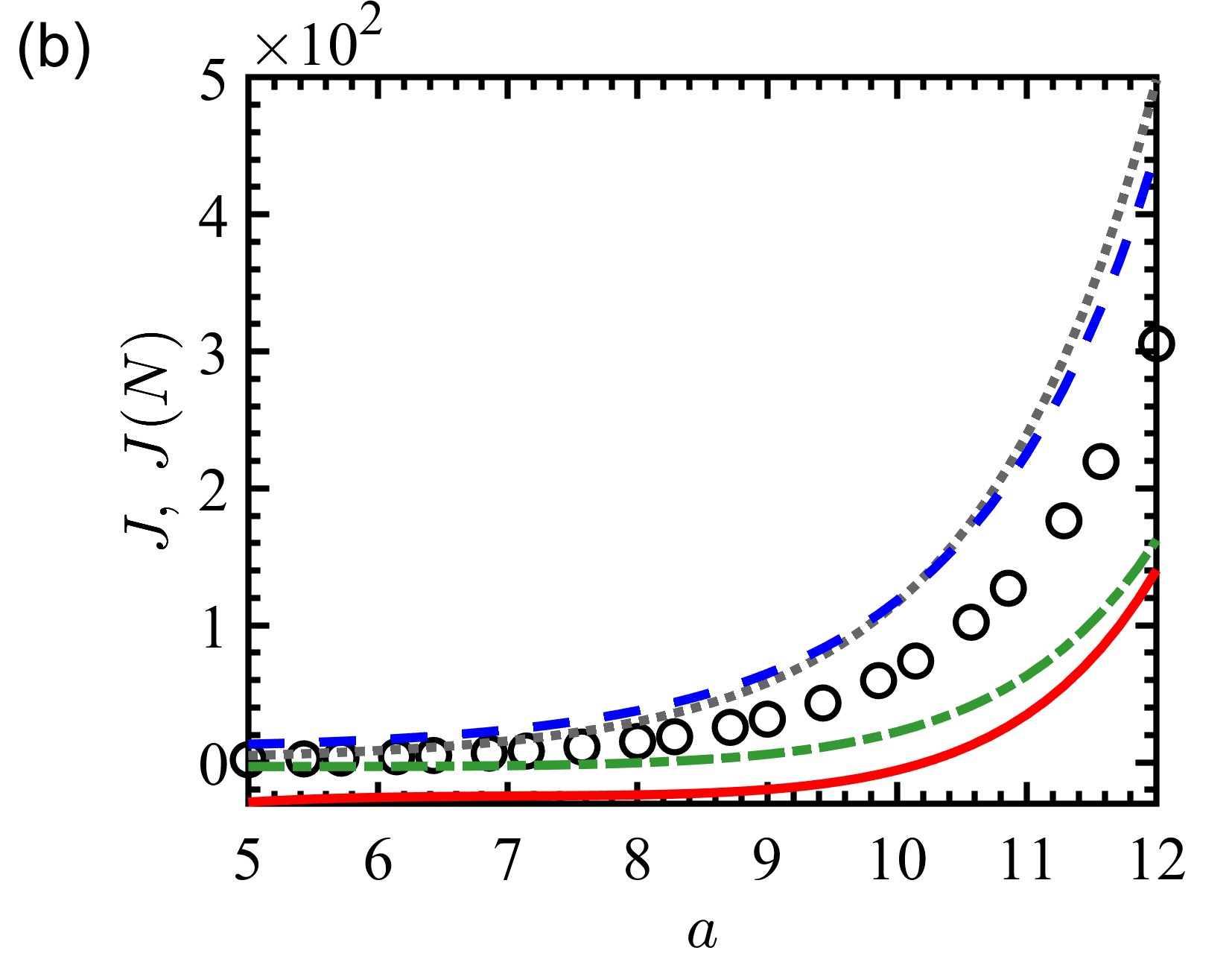}
    \includegraphics[width=0.32\textwidth]{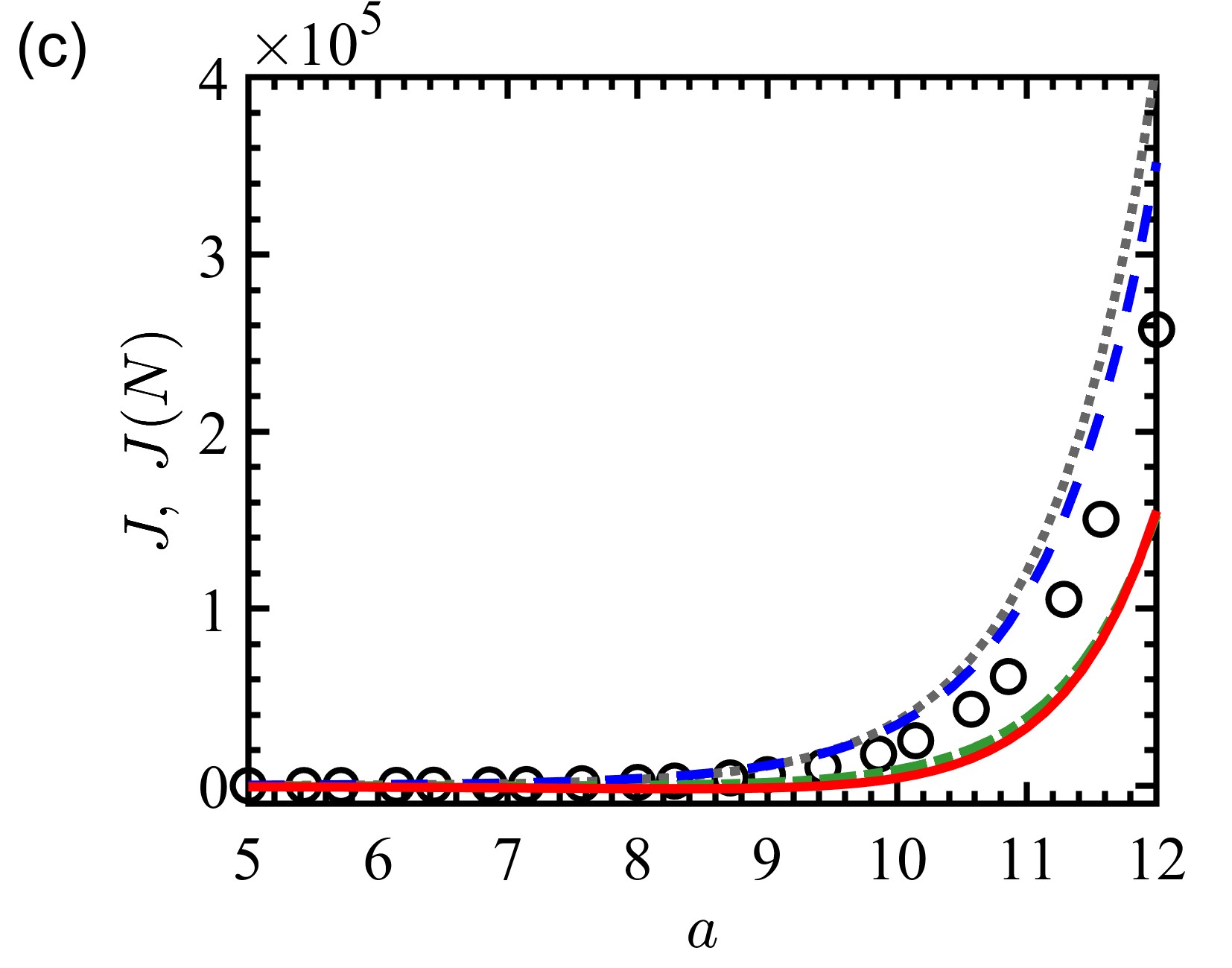}
    \includegraphics[width=0.32\textwidth]{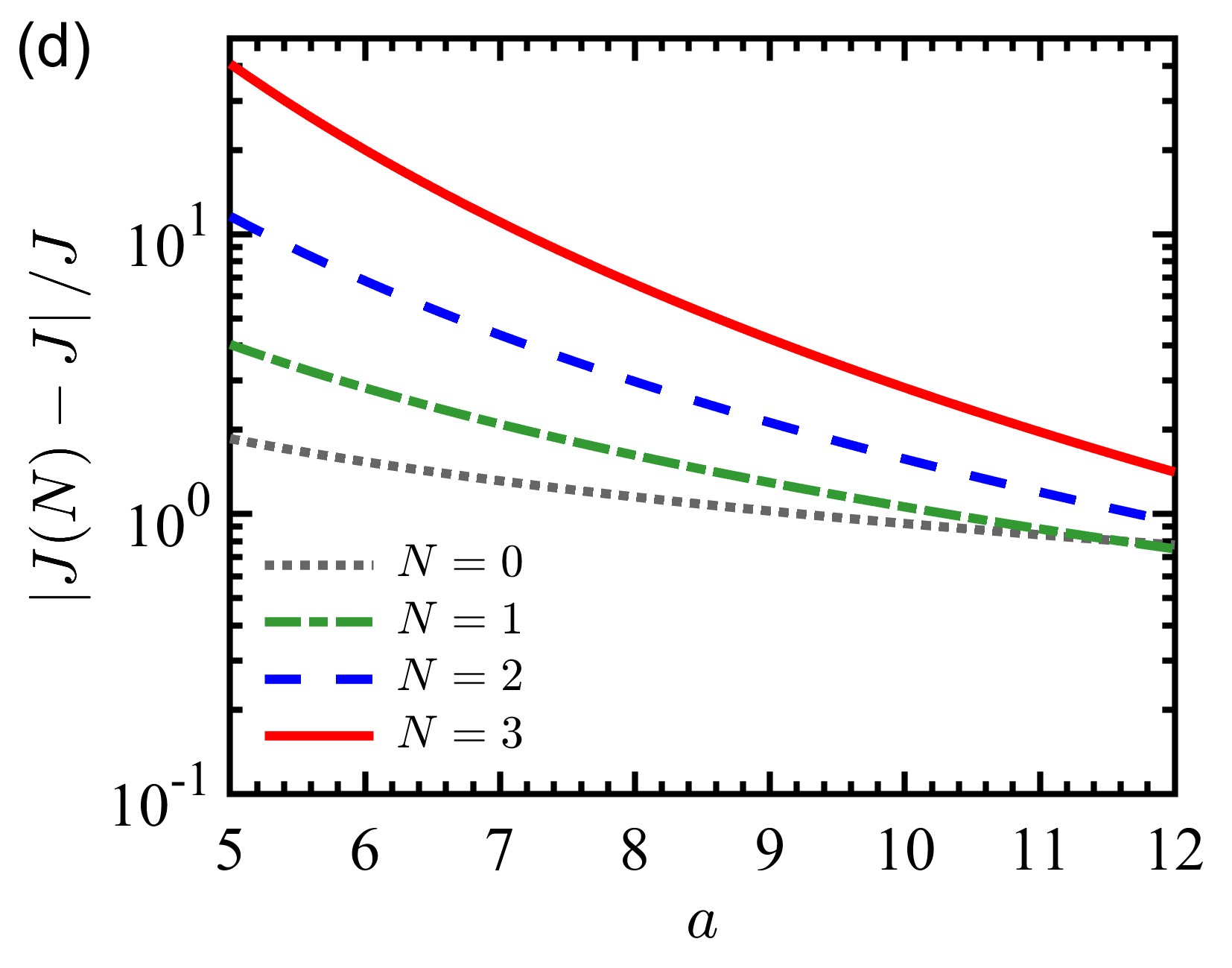}
    \includegraphics[width=0.32\textwidth]{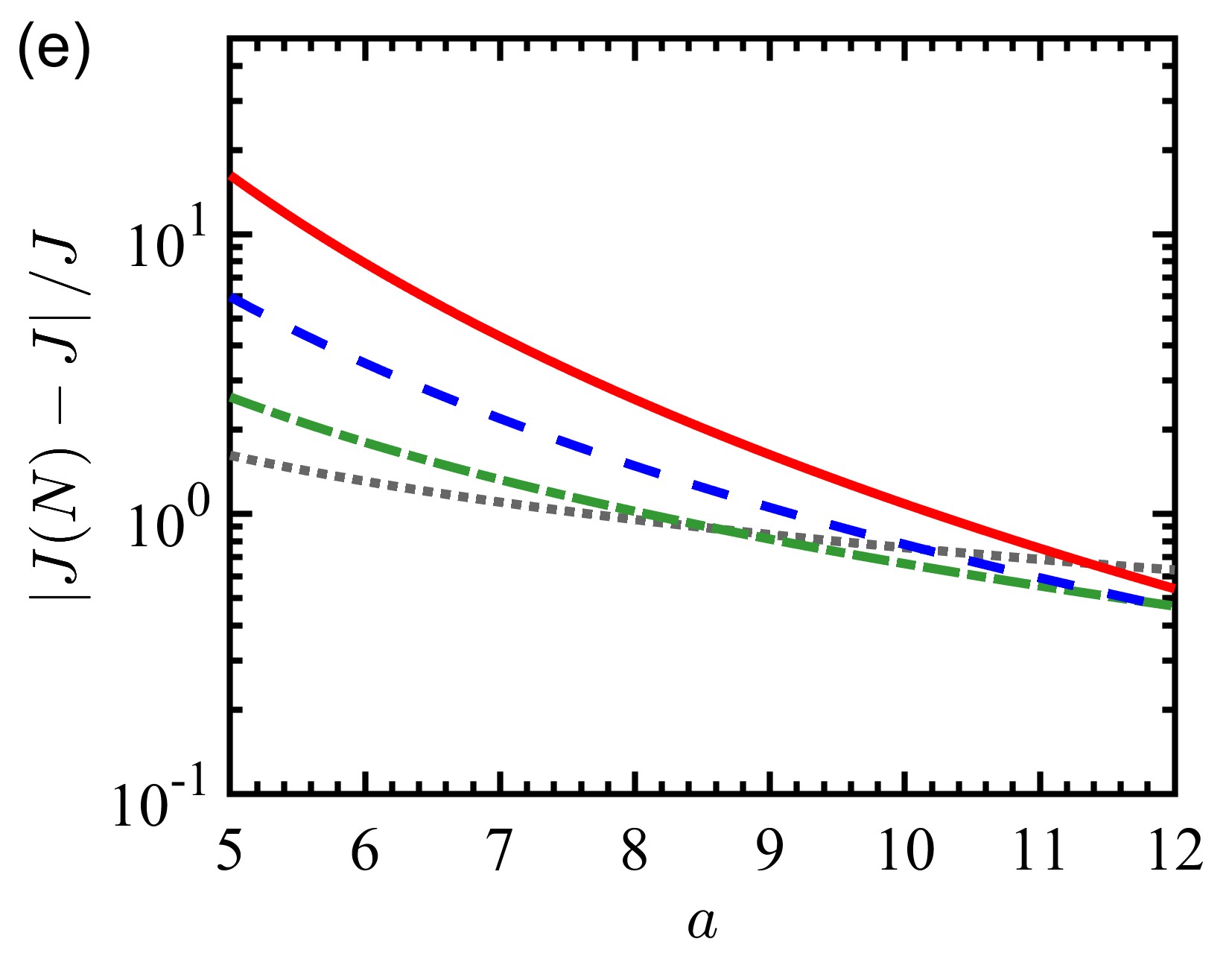}
    \includegraphics[width=0.32\textwidth]{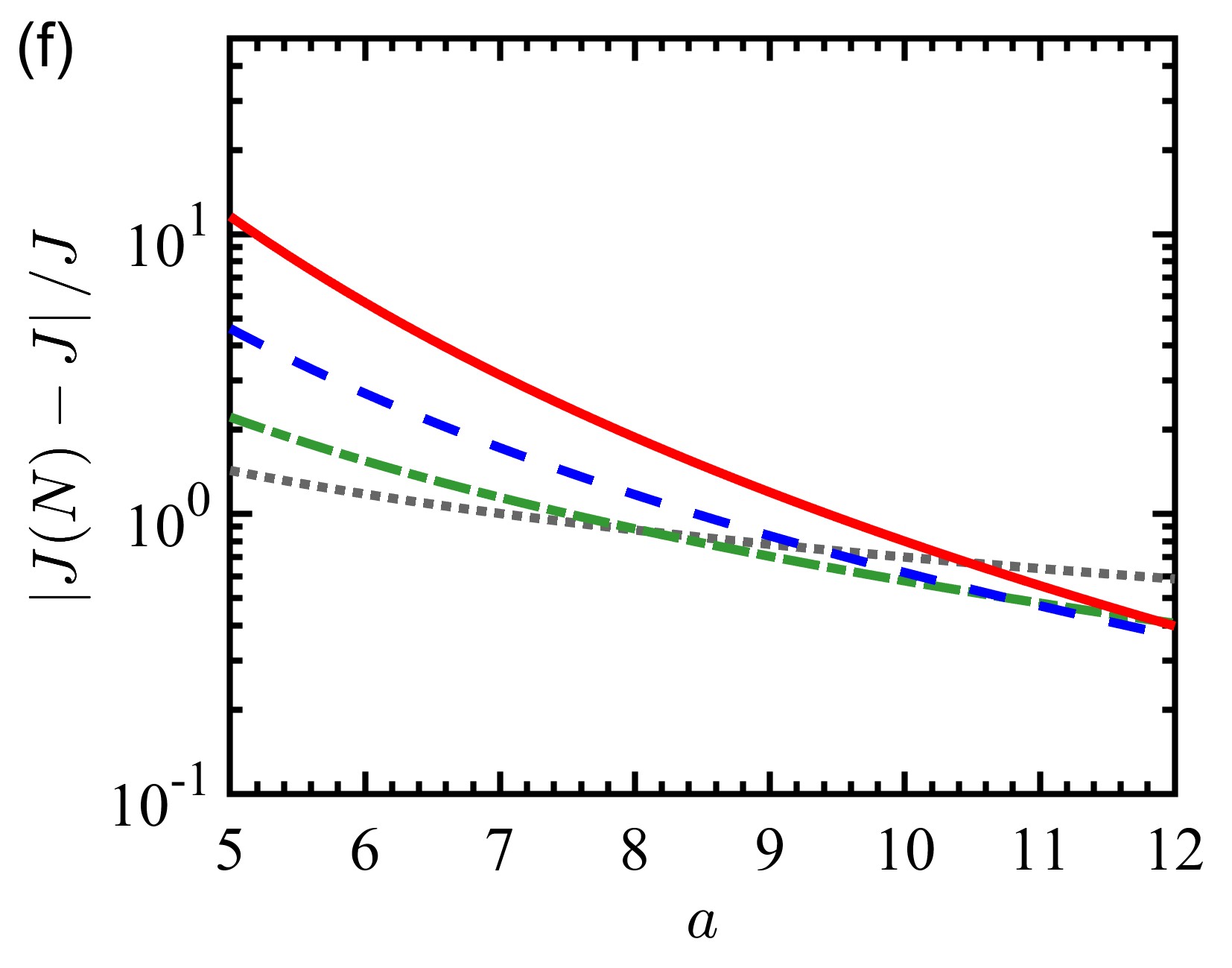}
    \caption{Comparisons between the partial sum $J(N)$ and the exact integral $J$ for the vD~I transformation, plotted against the expansion parameter $a = \kappa_U u_\infty^+$. Panels (a--c) show the exact values of $J$ (open circles) and the partial sums truncated at $N=0, 1, 2, 3$ (lines), while panels (d--f) show the corresponding absolute relative errors, $|J(N)-J|/J$. Each column corresponds to a representative DNS case listed in \hyperref[tab:the ZPG CTBLs database 1]{tables~\ref{tab:the ZPG CTBLs database 1}-\ref{tab:the ZPG CTBLs database 2}}: (a,d) the quasi-adiabatic case with $M_\infty = 8.00$ and $\varTheta = 1.00$; (b,e) the cooled-wall case with $M_\infty = 5.84$ and $\varTheta = 0.13$; and (c,f) the heated-wall case with $M_\infty = 5.00$ and $\varTheta = 2.06$. While the governing parameters ($M_\infty$, $\varTheta$, $Pr$, $T_\infty$) are fixed to match these specific cases, $a$ is treated as a continuous variable spanning the range $5 \le a \le 12$ to represent varying Reynolds numbers.}
    \label{fig:relative error vDI}
\end{figure}

\hyperref[fig:relative error vDI]{Figure~\ref{fig:relative error vDI}} compares the exact integral $J$ with the partial sums $J(N)$ for the vD~I transformation under three representative flow conditions, alongside the corresponding absolute relative errors $|J(N) - J| / J$. Consistent with asymptotic theory, the relative error for any fixed truncation order $N$ is observed to decrease with increasing $a$, reflecting the approach towards the formal asymptotic limit of infinite Reynolds number. Nevertheless, the zeroth-order approximation $J \approx J(0)$, corresponding to the asymptotic solution in classical van Driest's theory, fails to provide acceptable accuracy within the Reynolds number range of interest. As clearly visualised in panels (a--c), this leading-order approximation systematically and substantially overpredicts the exact integral, with the relative error exceeding 50\% even at $a = 12$. Moreover, the asymptotic series exhibits poor convergence properties. Notably, the partial sums $J(N)$ yield unphysical negative values for odd truncation orders ($N = 1$ and $N = 3$) at lower values of $a$. Furthermore, rather than improving accuracy, the inclusion of higher-order terms generally increases the error in the cases examined. This behaviour is likely associated with the alternating structure of the asymptotic series, and the rapid growth in magnitude of the expansion coefficients $w_n$ in \eqref{eq:partial sum of J}.

\begin{figure}
    \centering
    \includegraphics[width=0.32\textwidth]{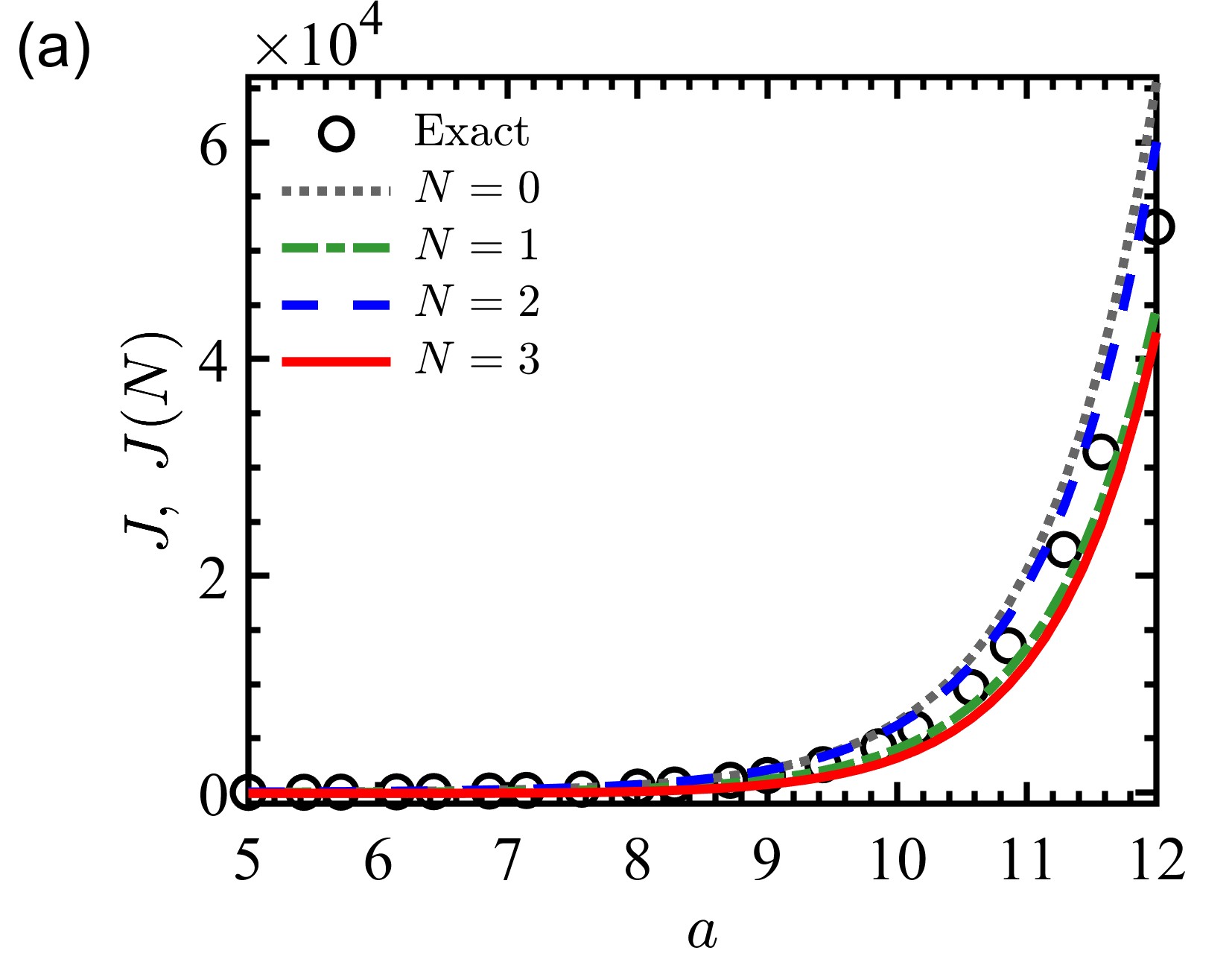}
    \includegraphics[width=0.32\textwidth]{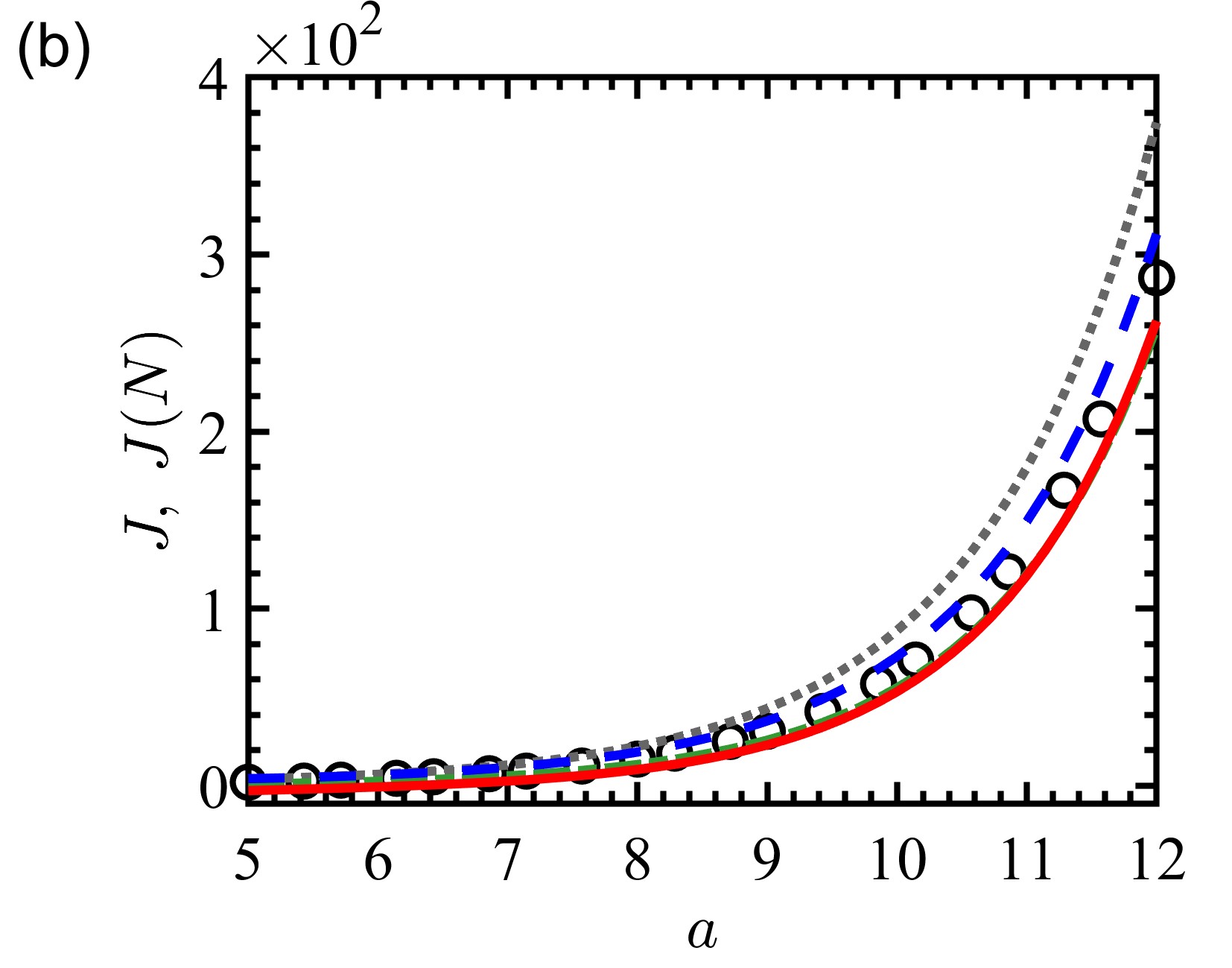}
    \includegraphics[width=0.32\textwidth]{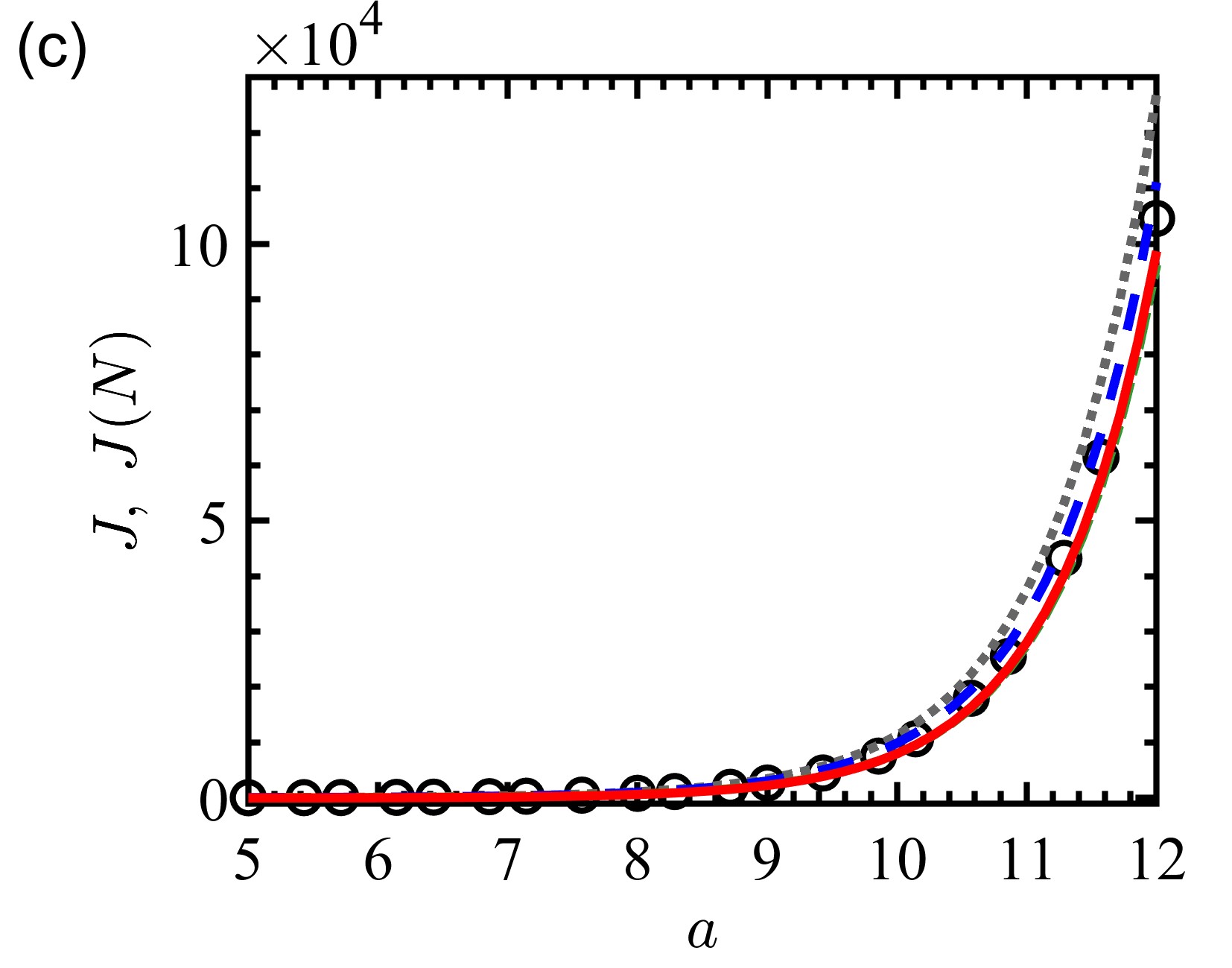}
    \includegraphics[width=0.32\textwidth]{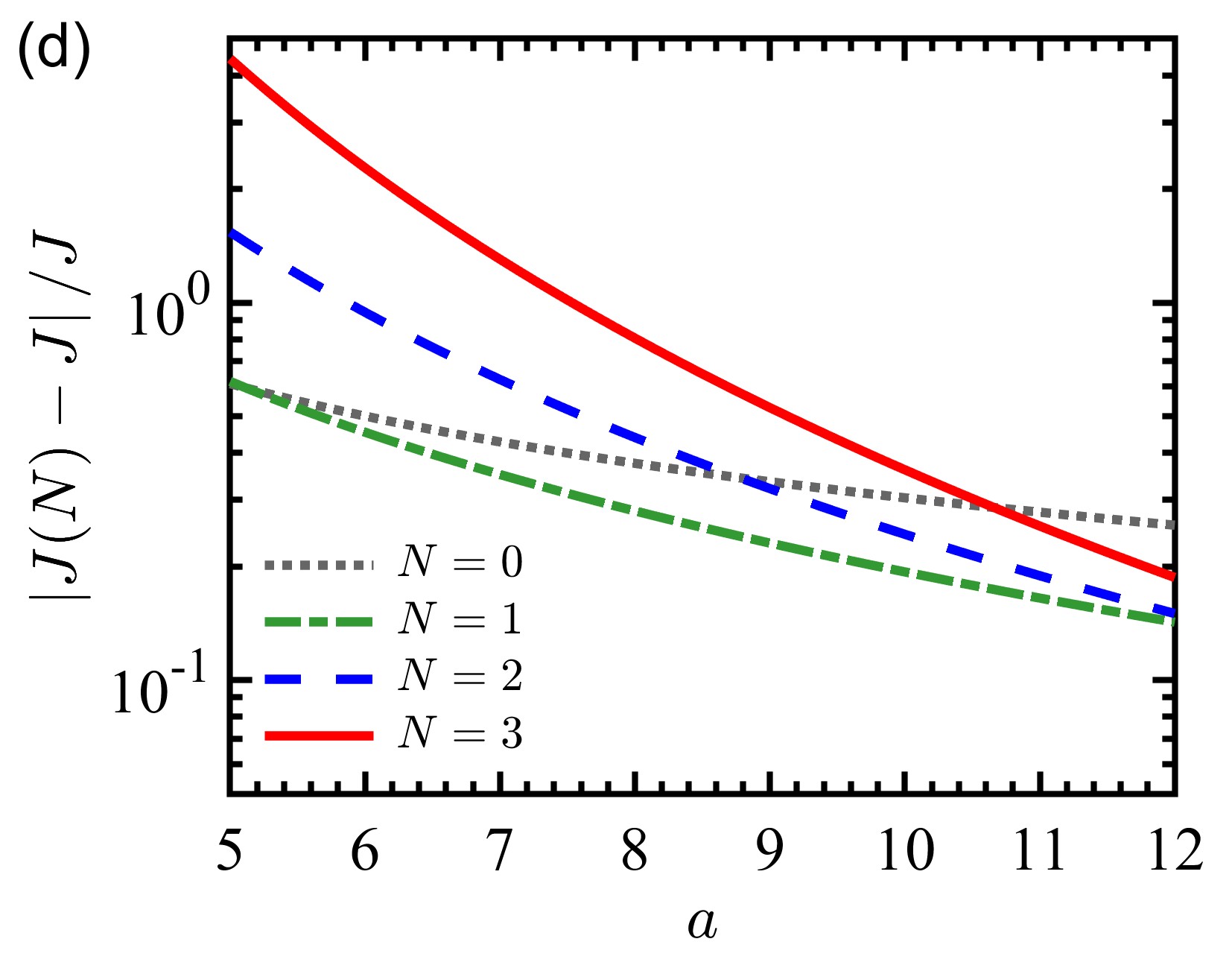}
    \includegraphics[width=0.32\textwidth]{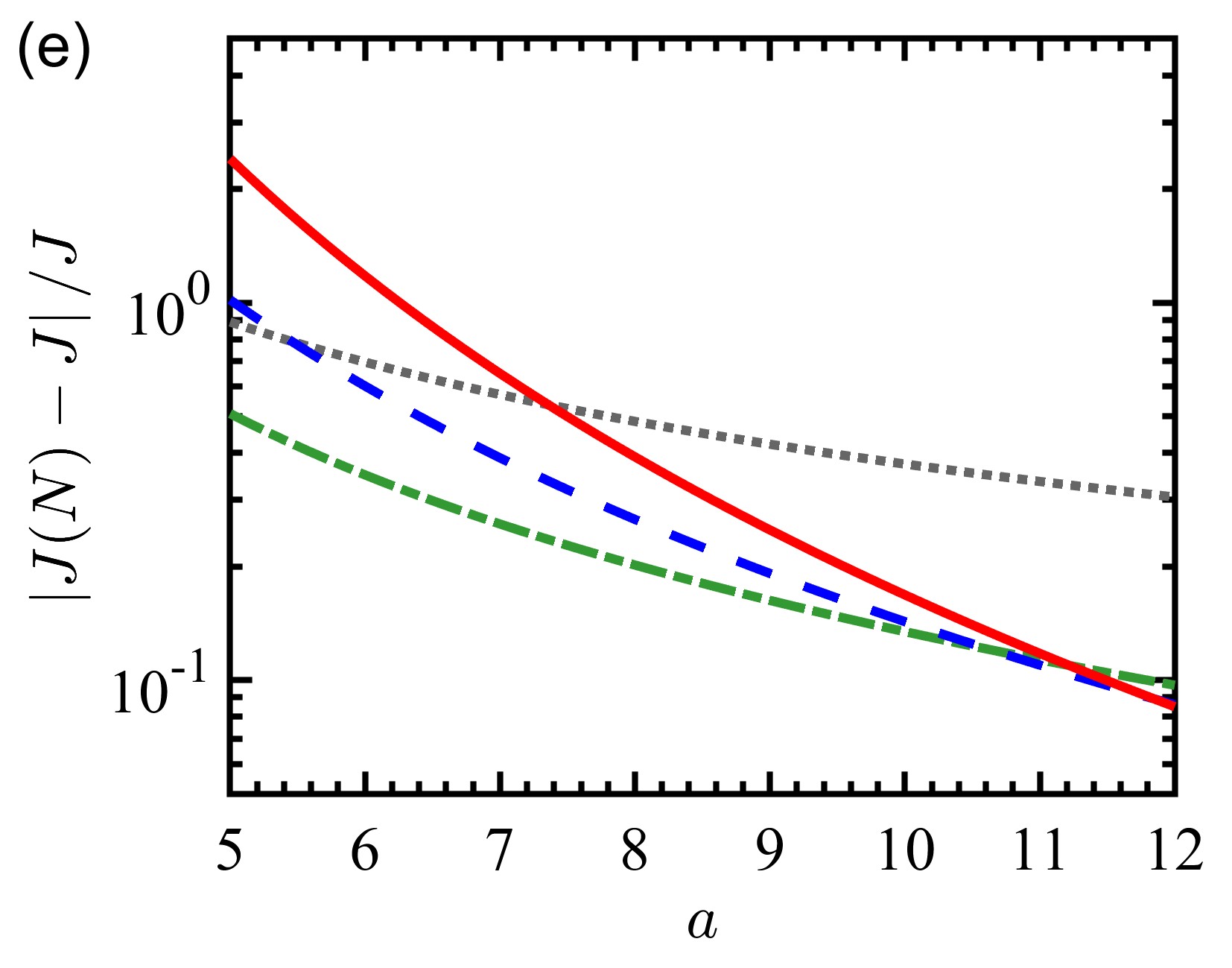}
    \includegraphics[width=0.32\textwidth]{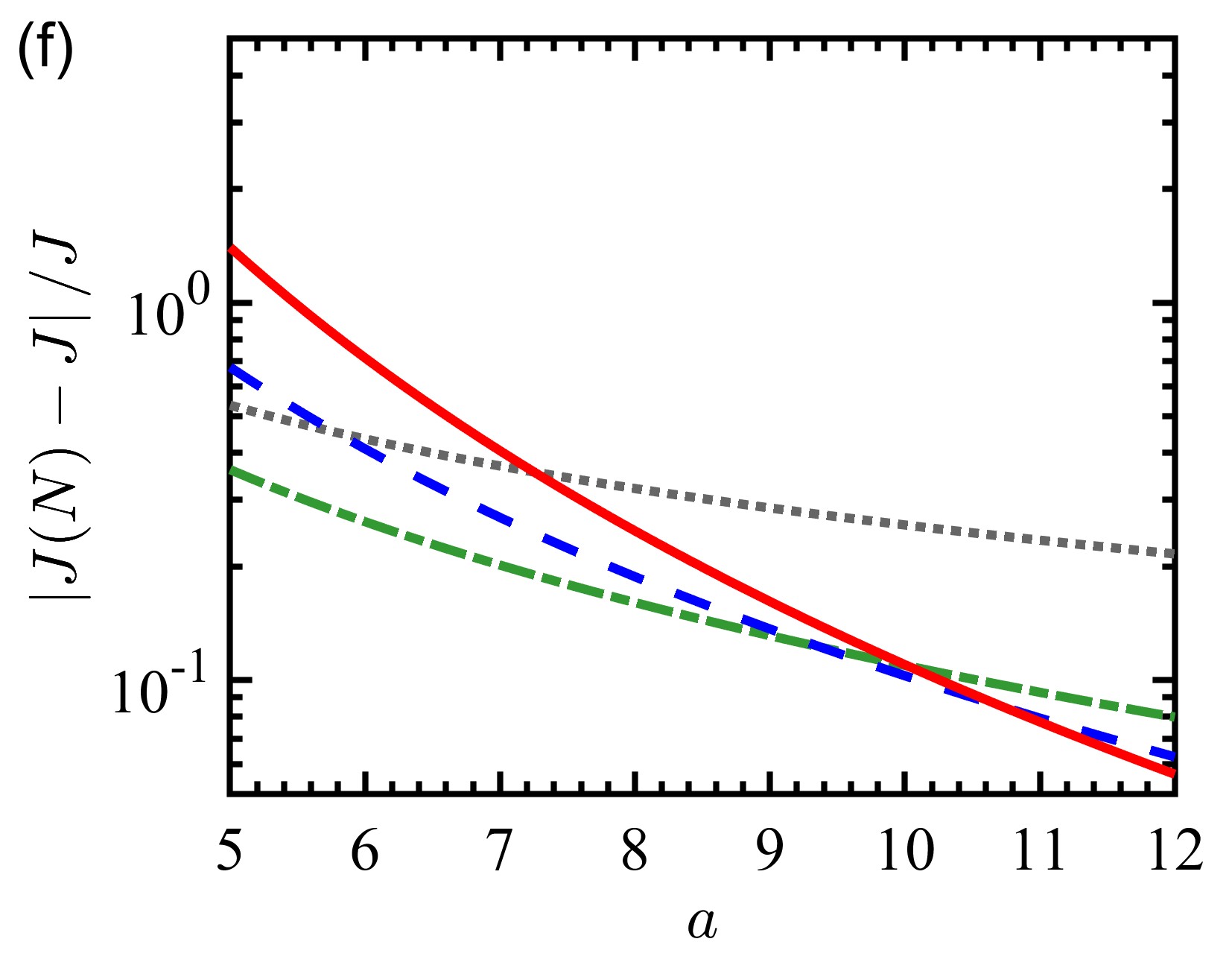}
    \caption{Comparisons between the partial sum $J(N)$ and the exact integral $J$ for the vD~II transformation, plotted against the expansion parameter $a = \kappa_U u_\infty^+$. The panel layouts, flow conditions for each column, as well as the symbols and line styles, are identical to those in \hyperref[fig:relative error vDI]{figure~\ref{fig:relative error vDI}}.}
    \label{fig:relative error vDII}
\end{figure}

The corresponding results for the vD~II transformation are presented in \hyperref[fig:relative error vDII]{figure~\ref{fig:relative error vDII}}. Similarly, the relative error for any fixed truncation order $N$ decreases monotonically with increasing $a$. As shown directly in panels (a--c), the zeroth-order asymptotic approximation $J \approx J(0)$ still systematically overpredicts the exact integral, although the discrepancy is smaller than for vD~I. Nevertheless, the deviation remains significant, with relative errors ranging between 22\% and 31\% at $a=12$. This confirms that the relatively good performance of the classical vD~II skin-friction transformation cannot be attributed to an accurate leading-order approximation of the compressible integral $J$ alone; the associated error-cancellation mechanism is discussed in \textsection~\ref{sec:subsection 4.3}. Regarding the effect of higher-order terms, the series exhibits an evident crossover behaviour. At lower values of $a$, adding terms tends to increase the error, similar to the vD~I case. However, as $a$ becomes sufficiently large, the higher-order approximations begin to outperform their lower-order counterparts. Nevertheless, within the practical range of Reynolds numbers considered, retaining a finite number of terms remains insufficient to yield a precise and reliable approximation.

\begin{figure}
    \centering
    \includegraphics[width=0.32\textwidth]{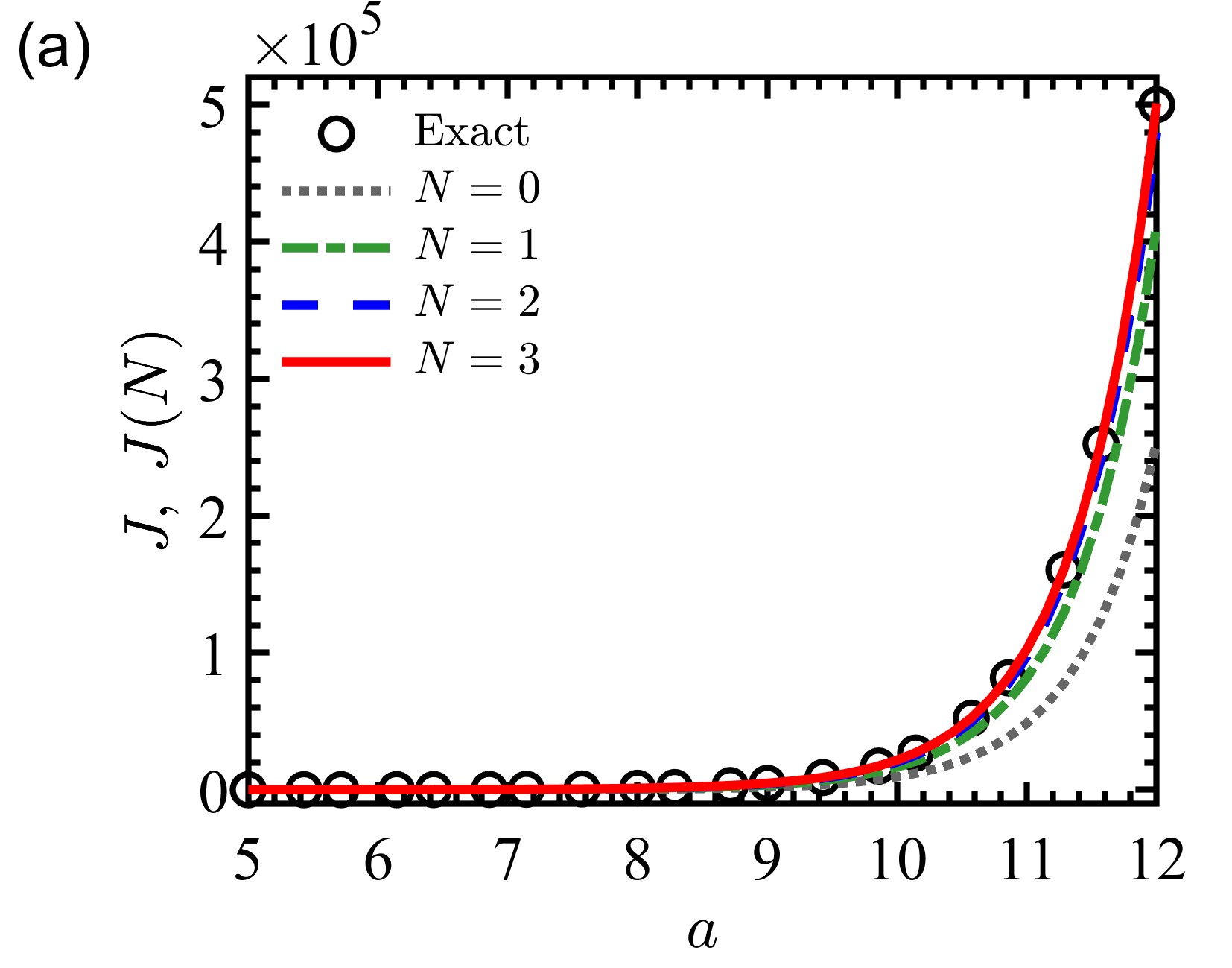}
    \includegraphics[width=0.32\textwidth]{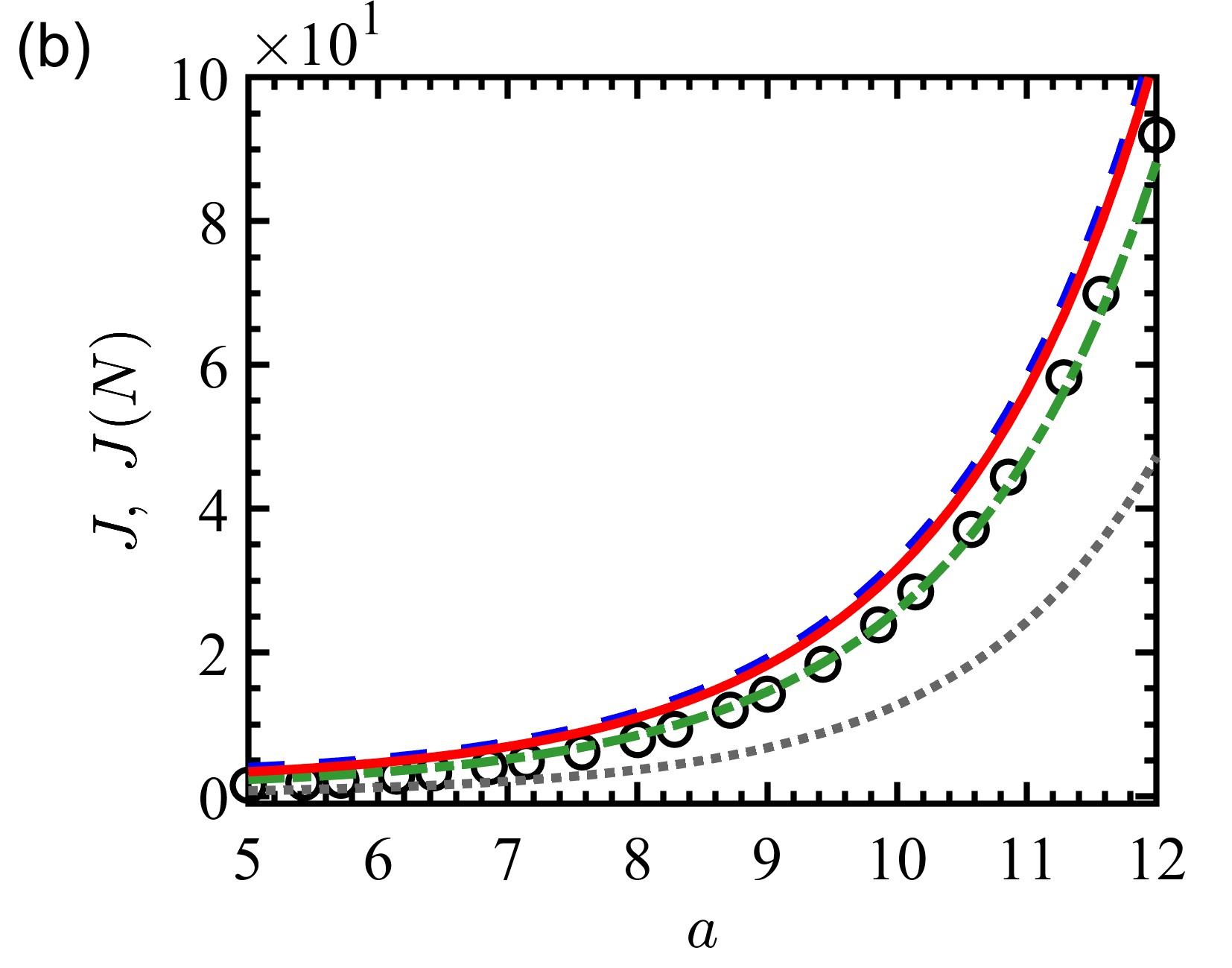}
    \includegraphics[width=0.32\textwidth]{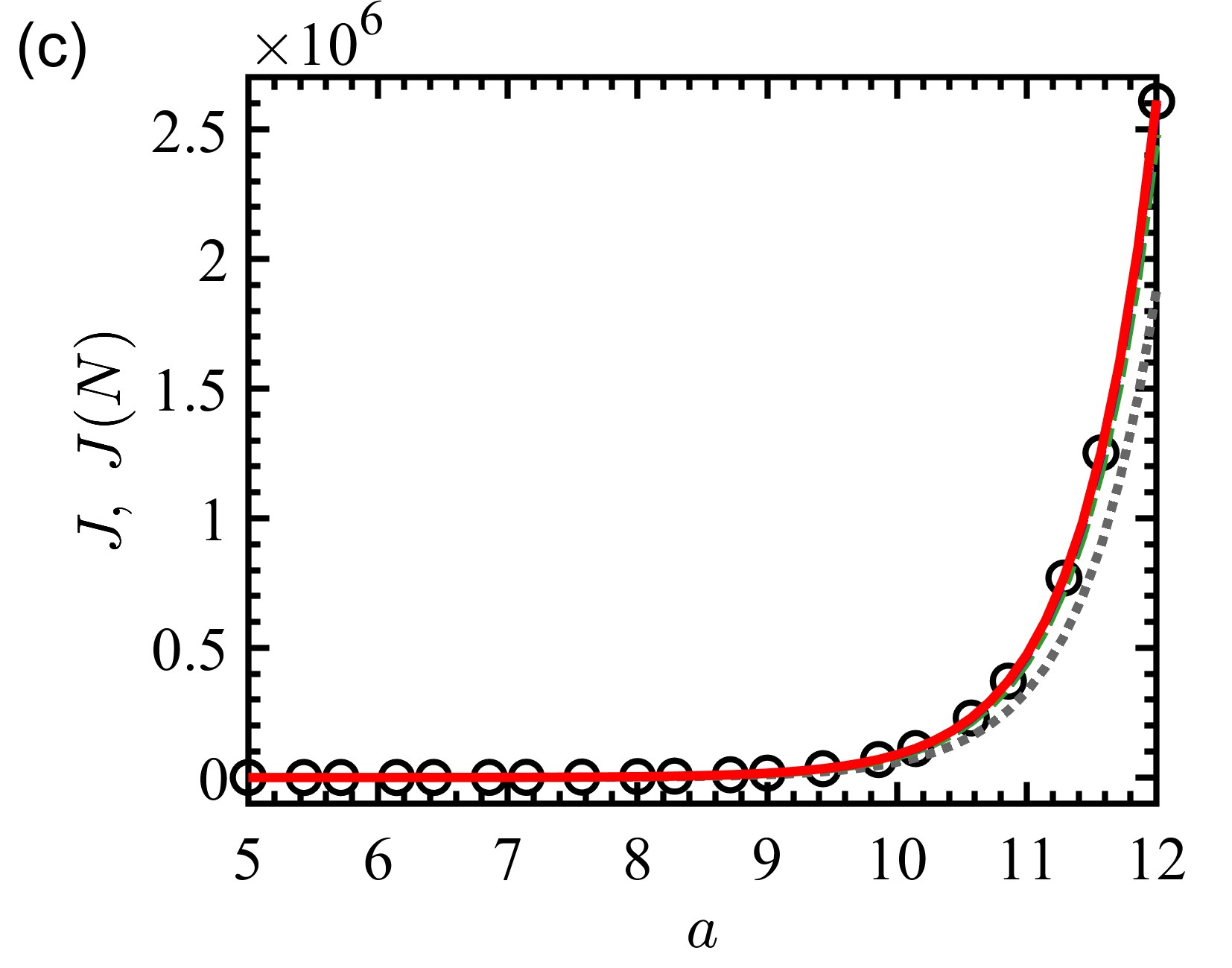}
    \includegraphics[width=0.32\textwidth]{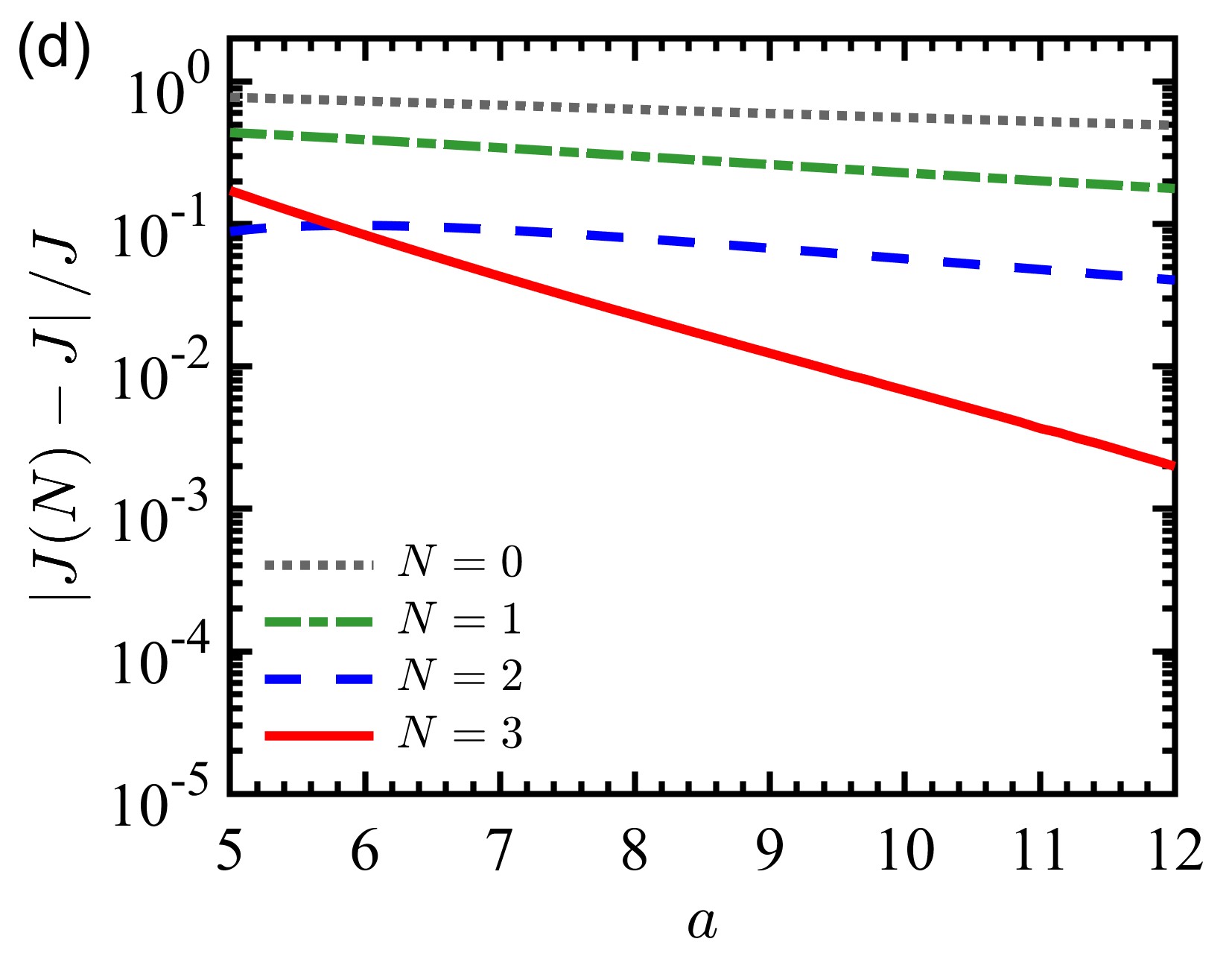}
    \includegraphics[width=0.32\textwidth]{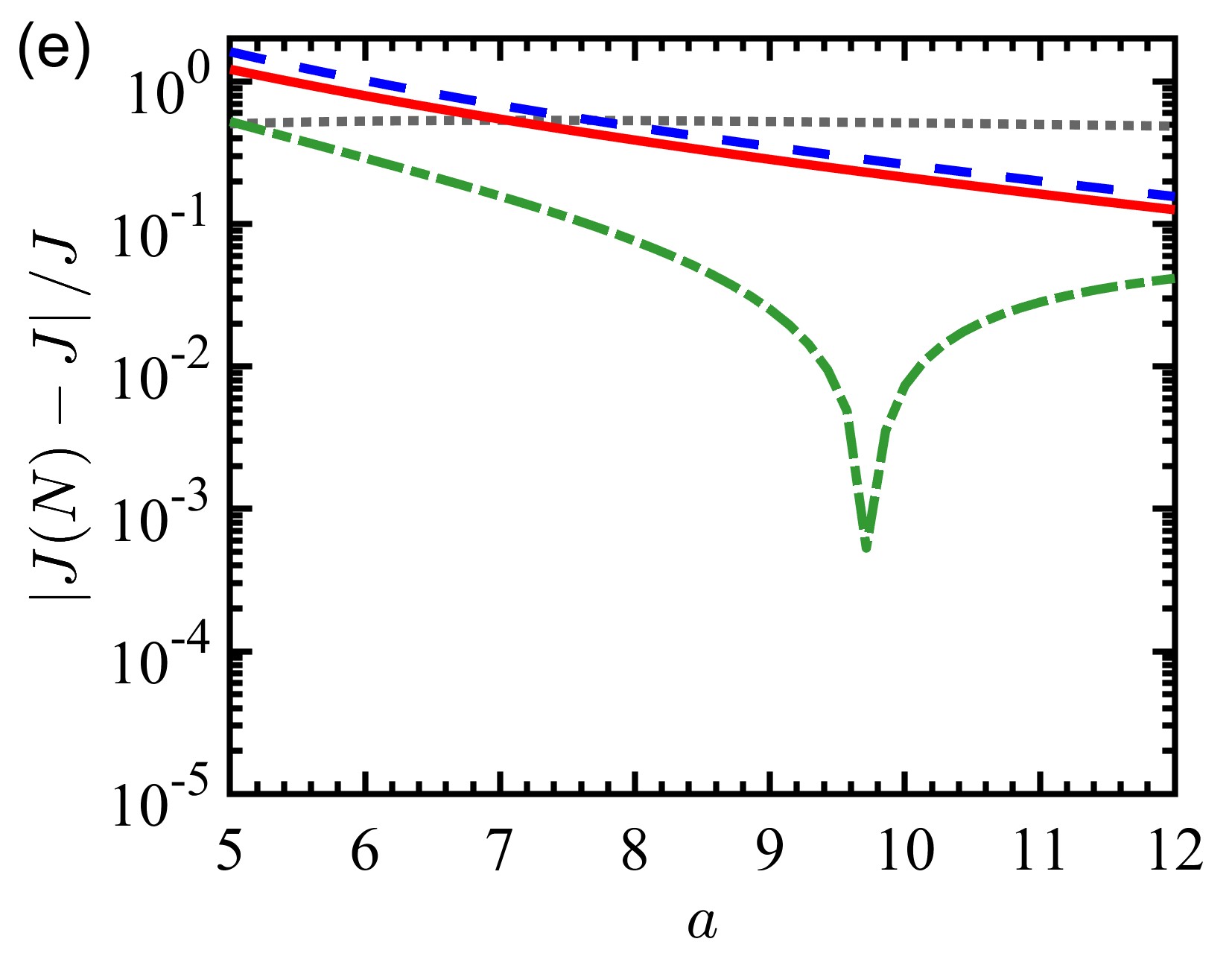}
    \includegraphics[width=0.32\textwidth]{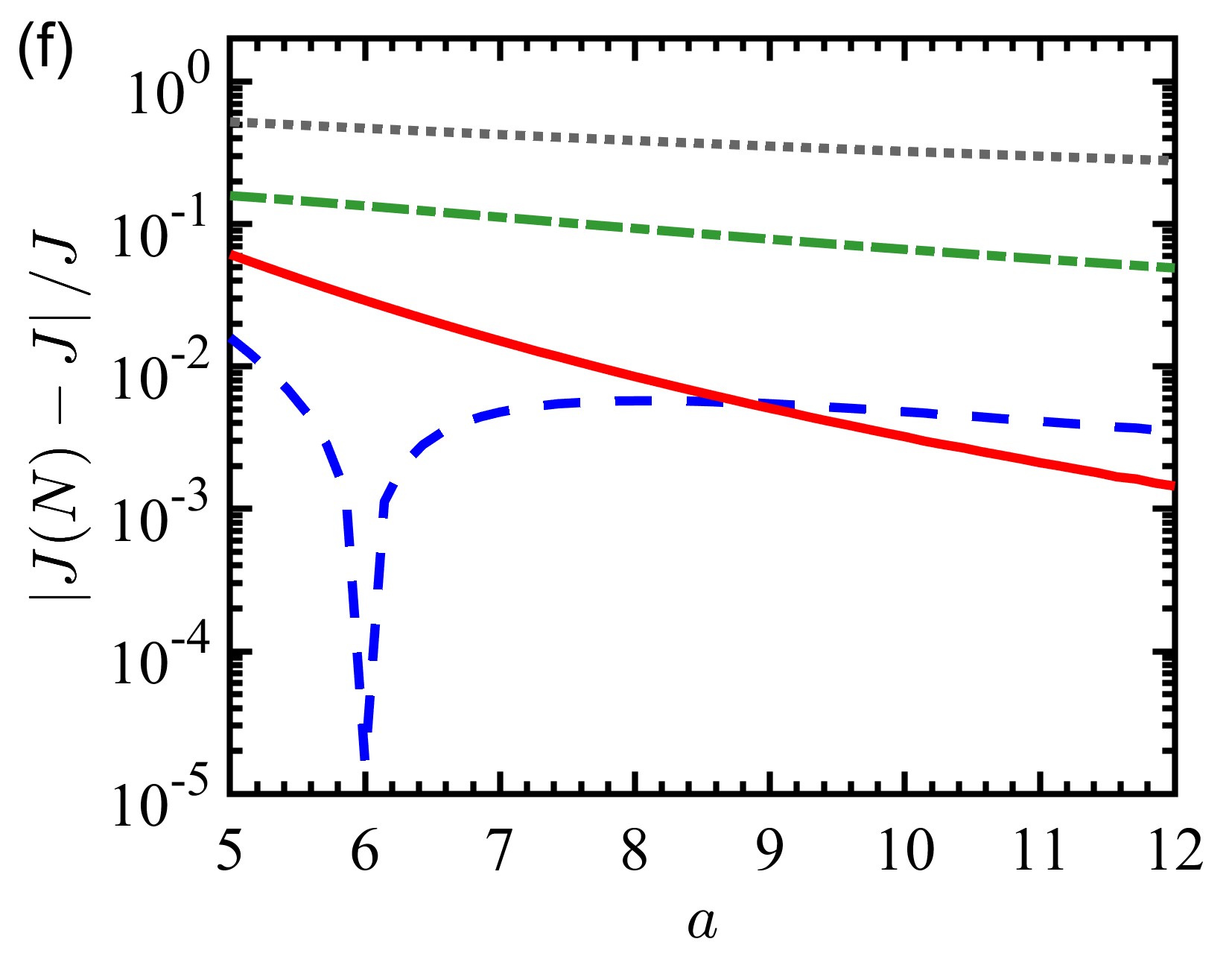}
    \caption{Comparisons between the partial sum $J(N)$ and the exact integral $J$ for the VIPL transformation, plotted against the expansion parameter $a = \kappa_U u_\infty^+$. The panel layouts, flow conditions for each column, as well as the symbols and line styles, are identical to those in \hyperref[fig:relative error vDI]{figure~\ref{fig:relative error vDI}}.}
    \label{fig:relative error VIPL}
\end{figure}

Finally, the convergence behaviour of the VIPL transformation is examined in \hyperref[fig:relative error VIPL]{figure~\ref{fig:relative error VIPL}}. In contrast to the overprediction observed in vD~I and vD~II, panels (a--c) directly show that the asymptotic approximation $J(0)$ substantially underestimates the exact integral $J$, yielding relative errors between 28\% and 49\% at $a=12$. This directly explains the failure of the analytical VIPL skin-friction transformations~\eqref{eq:VIPL skin friction transformation, power law} and \eqref{eq:VIPL skin friction transformation, Sutherland law}, which were derived under the asymptotic limit $a \rightarrow \infty$ (see \hyperref[fig:original VIPL transformation]{figure~\ref{fig:original VIPL transformation}}). Unlike the vD~I and vD~II cases, the asymptotic series exhibits a stronger tendency towards convergence in cases with quasi-adiabatic and heated walls. For instance, as shown in the relative error plots in panels (d) and (f), the inclusion of higher-order terms progressively reduces the error, with relative errors at $a=12$ decreasing to 0.20\% and 0.14\% for $N=3$, respectively. However, no such convergence behaviour is observed for the cooled-wall case. It is also worth noting that the error curves for intermediate orders exhibit non-monotonic behaviours for $N=1$ and $N=2$ in panels (e) and (f), with local minima corresponding to sign changes in the truncation error. Overall, these results demonstrate that retaining a few higher-order terms does not ensure a robust approximation across a wide range of Mach numbers, Reynolds numbers and wall-thermal conditions.

\section{Diagnostic assessment of self-fitted logarithmic closures}\label{app C}
\begin{figure}
  \centering
  \includegraphics[width=0.49\textwidth]{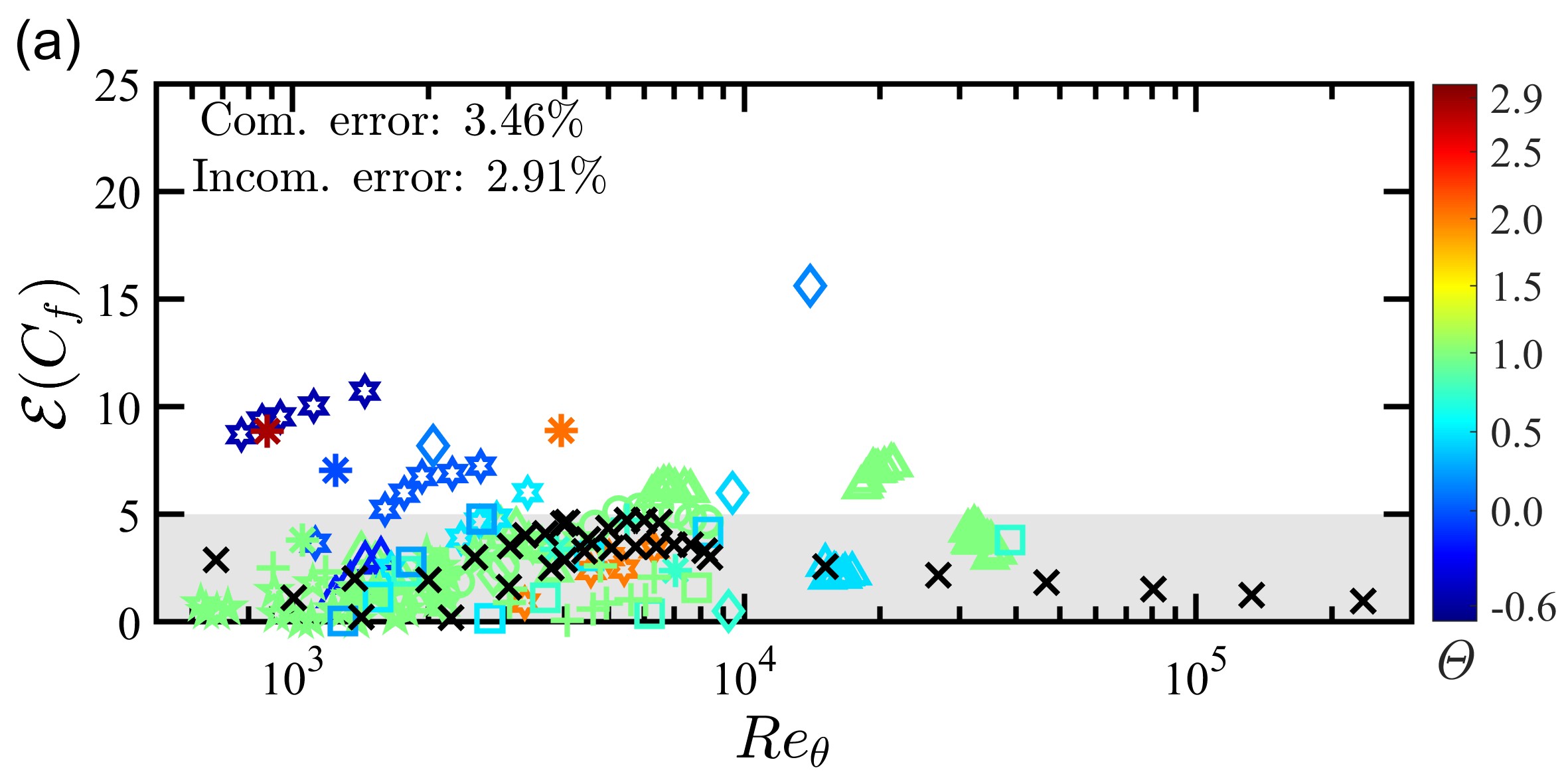}
  \includegraphics[width=0.49\textwidth]{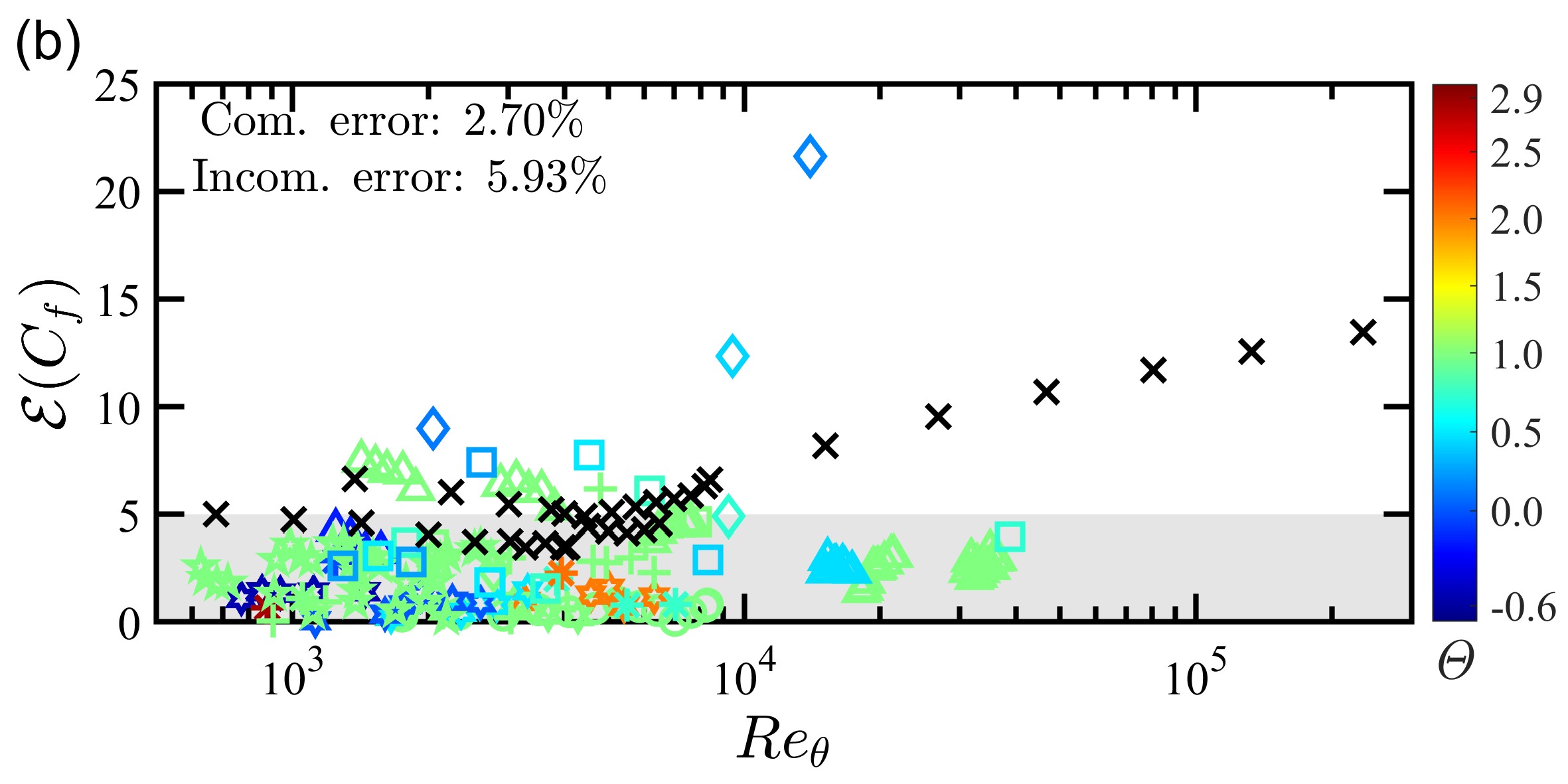}
  \includegraphics[width=0.49\textwidth]{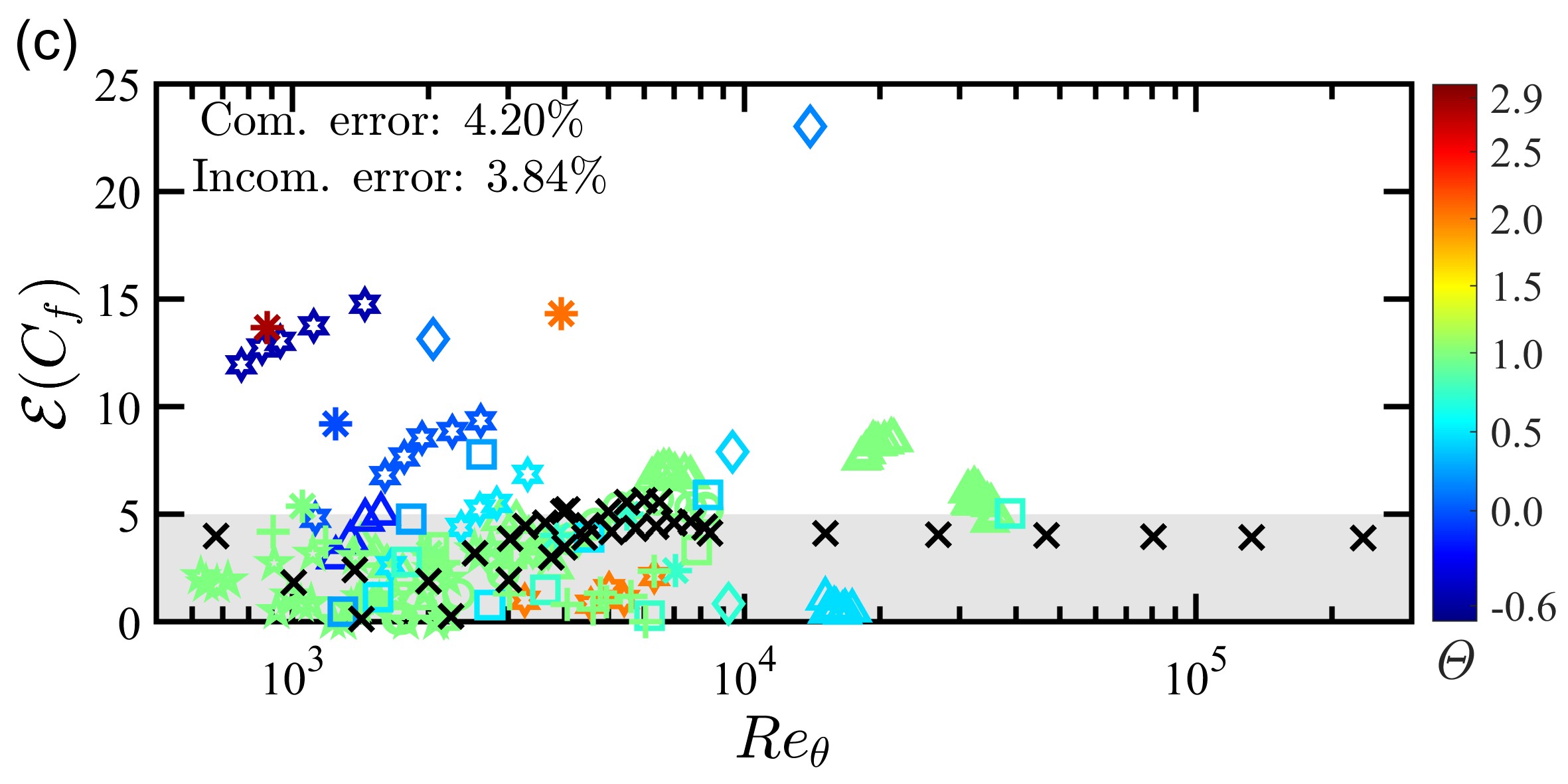}
  \includegraphics[width=0.49\textwidth]{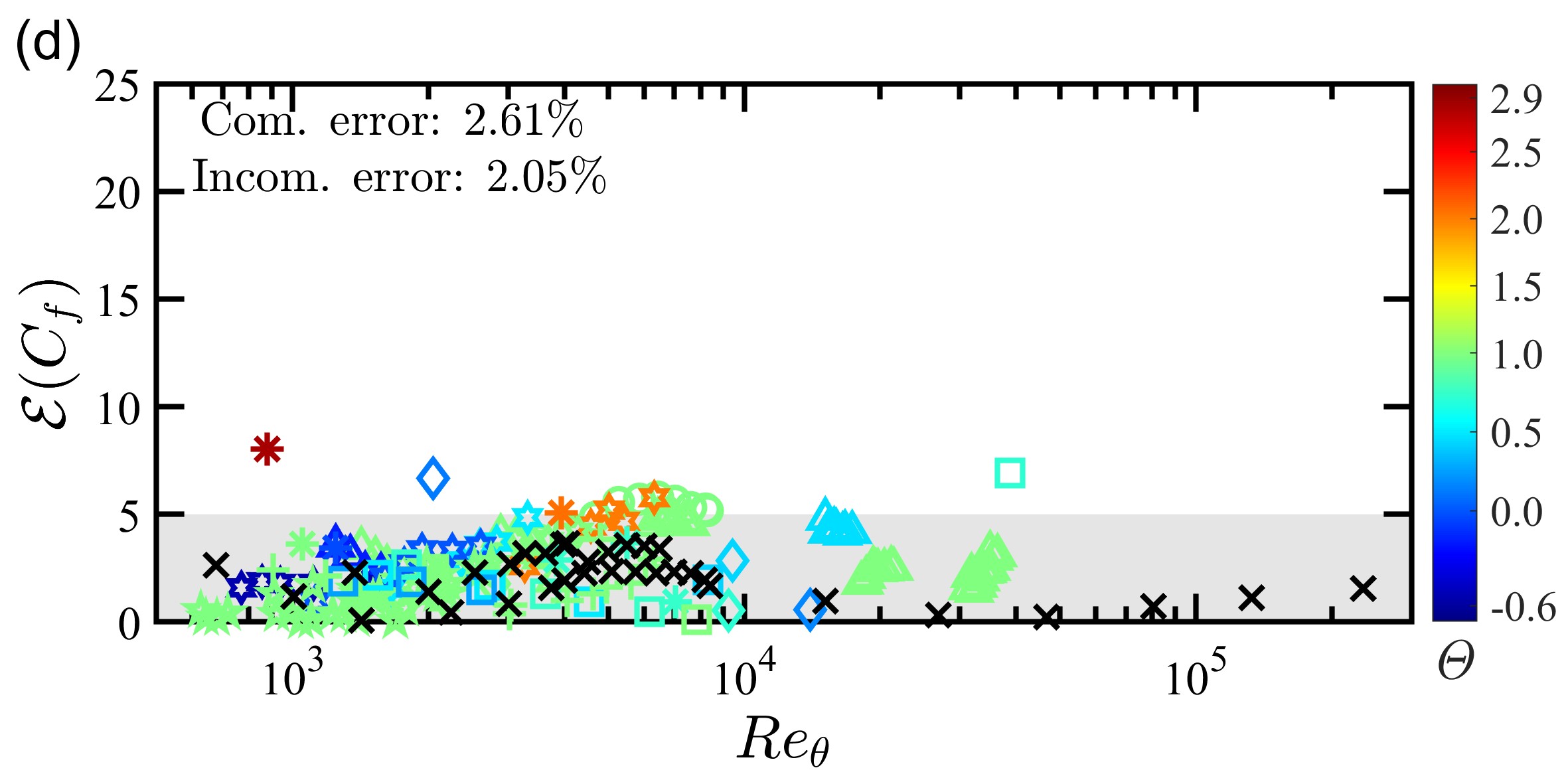}
  \caption{Diagnostic errors obtained when the CF relation~\eqref{eq:Coles-Fernholz relation} is replaced by method-specific logarithmic closures fitted to the transformed compressible data in the \emph{a priori} assessment. Absolute relative errors of the resulting skin-friction coefficient $C_f$, evaluated with respect to the corresponding reference data, are plotted as a function of $Re_\theta$. (a) Baseline results using the classical vD~II transformation~\eqref{eq:vD II skin friction transformation}. (b--d) Results obtained using the modified (b) vD~I, (c) vD~II, and (d) VIPL transformations based on the exact-integral formulation~\eqref{eq:final expression for F_C and F_theta}. For the modified transformations, $\kappa_U$ is dynamically determined from the compatibility condition~\eqref{eq:relation between kappa_u and kappa_theta} together with the corresponding fitted logarithmic closure. Errors over the compressible database represent in-sample diagnostic residuals rather than independent \emph{a posteriori} prediction errors, whereas errors for the incompressible reference cases provide an independent test of physical consistency with the incompressible-limit behaviour. Symbols and colours follow those in \hyperref[fig:original vD I, vD II, SC transformations (logarithmic)]{figure~\ref{fig:original vD I, vD II, SC transformations (logarithmic)}}.}
\label{fig:error of predicted Cf (fitted relation)}
\end{figure}

As a diagnostic extension of the \emph{a posteriori} assessment in \textsection~\ref{sec:section 5}, we examine an alternative closure strategy in which method-specific logarithmic scalings fitted to the transformed compressible DNS data in the \emph{a priori} assessment are used as the baseline `incompressible' correlations, as in \citet{zhao2025revisiting} and \citet{ying2025general}. For the classical vD~II baseline, the fitted logarithmic relation is used directly as the incompressible closure in the compressible extension of the skin-friction relation. For the modified transformations based on the exact-integral formulation, the fitted slope and intercept replace $\kappa_\theta$ and $C_\theta$ in the compatibility relations~\eqref{eq:relation between kappa_u and kappa_theta} and \eqref{eq:relation between C_U and C_theta}, so that $\kappa_U$ is determined with respect to the corresponding fitted closure rather than the CF relation~\eqref{eq:Coles-Fernholz relation}. Since these closures are calibrated using the same transformed compressible database on which the predictions are evaluated, the corresponding errors should be interpreted as in-sample diagnostic residuals rather than independent \emph{a posteriori} prediction errors. In contrast, the incompressible reference cases provide an independent check of whether the fitted closure remains consistent with the expected ZPG ITBL behaviour. The resulting errors are shown in \hyperref[fig:error of predicted Cf (fitted relation)]{figure~\ref{fig:error of predicted Cf (fitted relation)}}.

As expected, replacing the prescribed CF closure with these fitted logarithmic closures reduces the residuals over the compressible database. The mean errors decrease to 3.46\%, 2.70\%, 4.20\% and 2.61\% for the classical vD~II, modified vD~I, modified vD~II and modified VIPL transformations, respectively. This reduction, however, should be interpreted with caution. For the modified vD~I and vD~II transformations, the fitted logarithmic scalings deviate appreciably from the ZPG ITBL reference behaviour, as already evident in \hyperref[fig:modified vD I, vD II, SC transformations (logarithmic)]{figure~\ref{fig:modified vD I, vD II, SC transformations (logarithmic)}(b,c)}. Consequently, when these fitted closures are applied to incompressible reference cases, the mean errors increase substantially to 5.93\% and 3.84\%, respectively. This confirms that organizing transformed compressible data around a database-fitted logarithmic trend is insufficient unless the resulting closure also recovers the expected ZPG ITBL behaviour in the incompressible limit. Without this physical consistency, the apparent improvement over the compressible database mainly reflects in-sample fitting rather than a transferable skin-friction closure.

The situation is more favourable for the classical vD~II and modified VIPL transformations, whose fitted logarithmic scalings remain closer to the ZPG ITBL reference behaviour, as shown in \hyperref[fig:modified vD I, vD II, SC transformations (logarithmic)]{figure~\ref{fig:modified vD I, vD II, SC transformations (logarithmic)}(a,d)}. The corresponding mean errors for the incompressible reference cases are 2.91\% and 2.05\%, respectively. Among these fitted closures, the modified VIPL case yields the smallest residuals over the compressible database, with a mean error of 2.61\% and a maximum error of approximately 8.01\%, while retaining good agreement with the incompressible reference data. Because this closure is calibrated on the same transformed compressible database, these residuals should not be regarded as independent standalone prediction errors. Nevertheless, the relatively small error for the independent incompressible reference cases indicates that the fitted scaling associated with the modified VIPL transformation remains close to the adopted ZPG ITBL reference scaling over the Reynolds-number range considered. Thus, the fitted-closure variant provides diagnostic evidence that the modified VIPL transformation can support a calibrated logarithmic closure with improved in-sample accuracy and reasonable consistency with the independent incompressible reference cases. This suggests potential utility as a calibrated practical variant over parameter ranges similar to those considered here, but it should not be interpreted as the primary standalone predictive closure adopted in the present study.


\bibliographystyle{jfm}
\bibliography{jfm}

\end{document}